\catcode`\@=11
\def\citen#1{\if@filesw \immediate\write \@auxout {\string\citation{#1}}\fi%
\@tempcntb\m@ne \let\@h@ld\relax \def\@citea{}%
\@for \@citeb:=#1\do {\@ifundefined {b@\@citeb}%
    {\@h@ld\@citea\@tempcntb\m@ne{\bf ?}%
    \@warning {Citation `\@citeb ' on page \thepage \space undefined}}%
    {\@tempcnta\@tempcntb \advance\@tempcnta\@ne
    \setbox\z@\hbox\bgroup\ifcat0\csname b@\@citeb \endcsname \relax
    \egroup \@tempcntb\number\csname b@\@citeb \endcsname \relax
    \else \egroup \@tempcntb\m@ne \fi \ifnum\@tempcnta=\@tempcntb
    \ifx\@h@ld\relax \edef \@h@ld{\@citea\csname b@\@citeb\endcsname}%
    \else \edef\@h@ld{\hbox{--}\penalty\@highpenalty
    \csname b@\@citeb\endcsname}\fi
    \else \@h@ld\@citea\csname b@\@citeb \endcsname \let\@h@ld\relax \fi}%
\def\@citea{,\penalty\@highpenalty\hskip.13em plus.13em minus.13em}}\@h@ld}
\def\@citex[#1]#2{\@cite{\citen{#2}}{#1}}%
\def\@cite#1#2{\leavevmode\unskip\ifnum\lastpenalty=\z@\penalty\@highpenalty\fi%
  \ [{\multiply\@highpenalty 3 #1%
  \if@tempswa,\penalty\@highpenalty\ #2\fi}]}   %
\makeatother 
\catcode`\@=12

\documentclass[12pt]{amsart}
\usepackage{amssymb}
\usepackage[dvips]{graphics}
\newcommand{\Z}{{\mathbb Z}}

\newcommand{\R}{{\mathbb R}}
\newcommand{\C}{{\mathbb C}}

\newcommand{\I}{{\mathrm{i}}}
\newcommand{\one}{{\mathbf 1 }}
\newcommand{\Hom}{{\mathrm{Hom}}}
\newcommand{\M}{{\mathcal{M}}}
\newcommand{\HH}{{\mathcal{H}}}

\newcommand{\Ref}[1]{{(\ref{#1})}}
\newcommand{\dontprint}[1]{\relax}
\newcommand{\Otimes}{\,{\otimes}\,}
\newcommand{\OTimes}{{\otimes}}
\newcommand{\Times}{\,{\times}\,}
\newcommand{\TImes}{{\times}}
\newcommand{\Sqcup}{\,{\sqcup}\,}
\newcommand{\In}{\,{\in}\,}
\newcommand{\To}{\,{\to}\,}
\newcommand{\eq}{\,{=}\,}
\newcommand\Frac[2]{\mbox{\large$\frac{#1}{#2}$}}
\newcommand{\zero}{\one} 
\newcommand{\sixj}[8]
{\left\{
\begin{array}{ccc}
#1&#2&#3\\
#4&#5&#6
\end{array}
\right\}
^{\!#7}_{\!#8}
}
\newcommand{\barsixj}[8]
{
\overline{
\left\{
\begin{array}{ccc}
#1&#2&#3\\
#4&#5&#6
\end{array}
\right\}}
^{\,#7}_{\,#8}
}
\newcommand{\ST}{s}
\newcommand{\SR}{S} 
\newcommand{\T}{\hat T}      
\newcommand{\id}{\mathrm{id}}
\newcommand{\funo}
{\begin{picture}(3,3)(0,0) 
\scalebox{.25}{\includegraphics{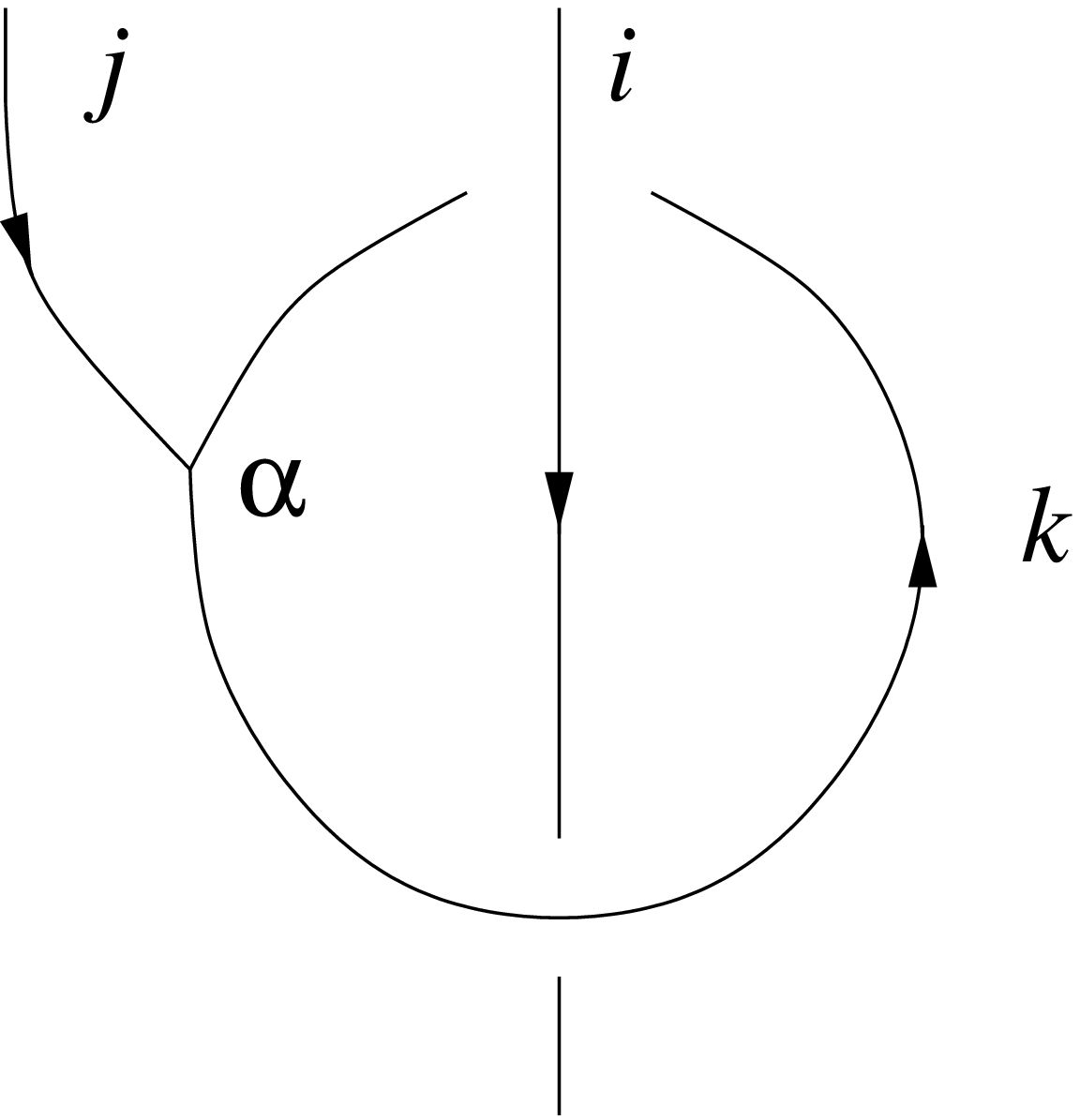}}
\end{picture}}
\newcommand{\fdue}
{\begin{picture}(3,3)(0,-0.5) 
\scalebox{.25}{\includegraphics{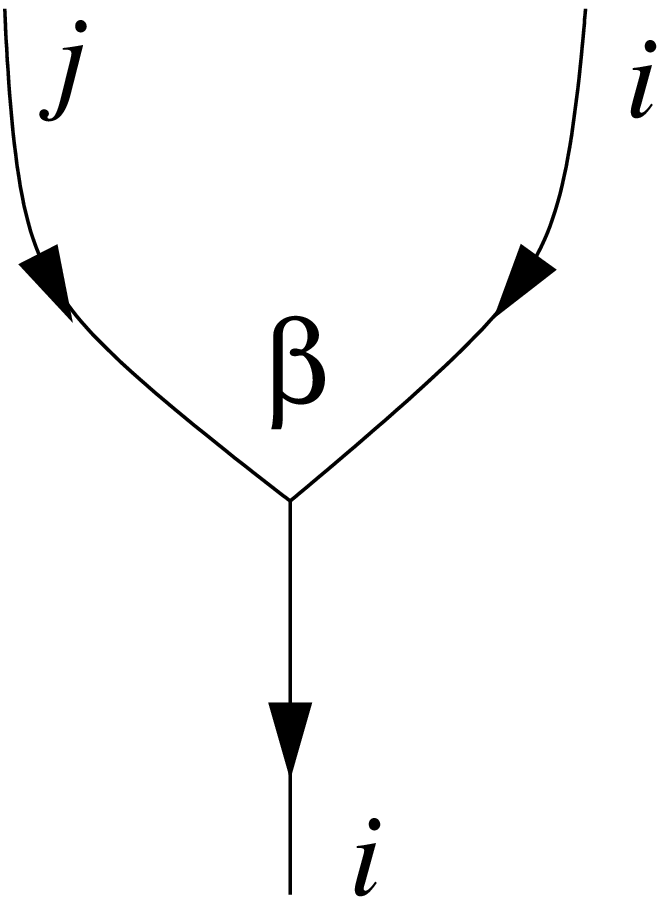}}
\end{picture}}
\newcommand{\ftre}
{\begin{picture}(3,3.5)(0,0) 
\scalebox{.25}{\includegraphics{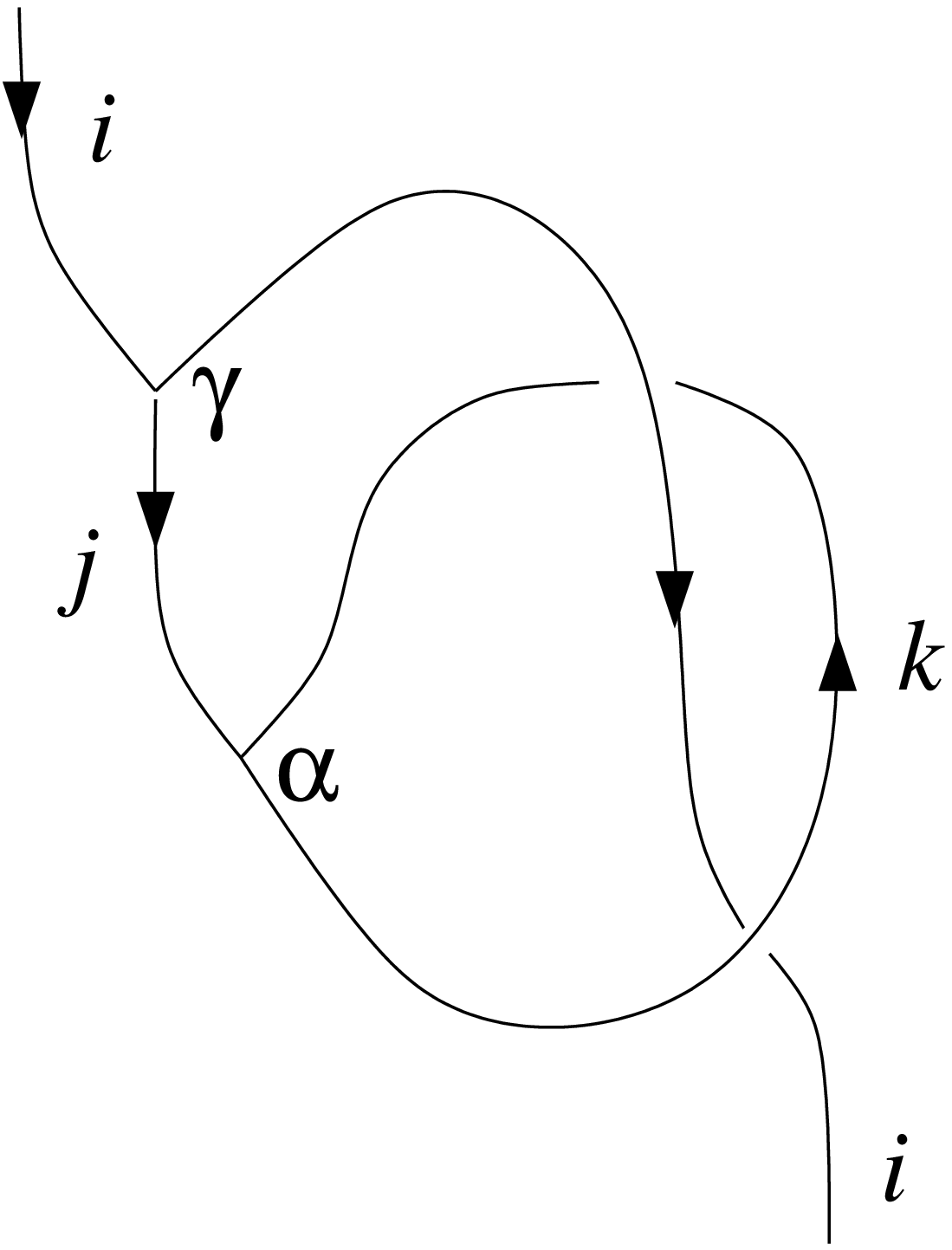}}
\end{picture}}
\newcommand{\fquattro}
{\begin{picture}(1.5,3)(0,-0.5) 
\scalebox{.25}{\includegraphics{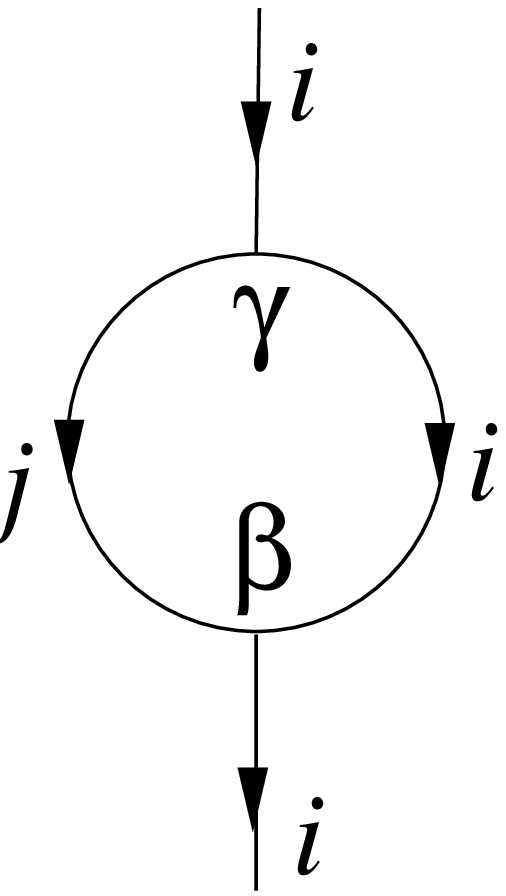}}
\end{picture}}
\newcommand{\fcinque}
{\begin{picture}(1,3)(0,-0.5) 
\scalebox{.25}{\includegraphics{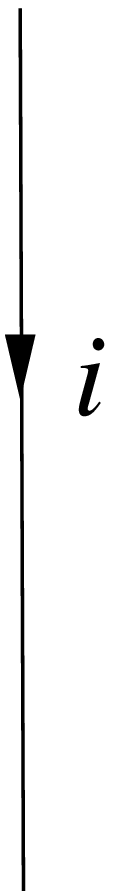}}
\end{picture}}
\newcommand{\fsei}
{\begin{picture}(3,3)(0,0) 
\scalebox{.3}{\includegraphics{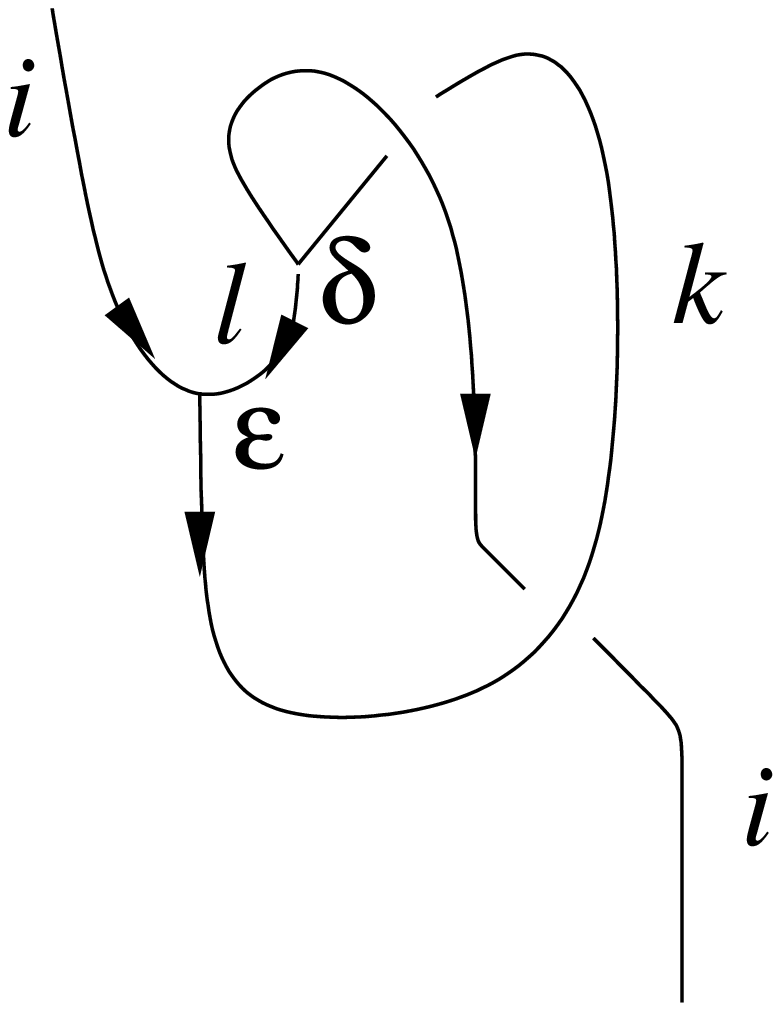}}
\end{picture}}
\newcommand{\fsette}
{\begin{picture}(3,3)(0,0) 
\scalebox{.25}{\includegraphics{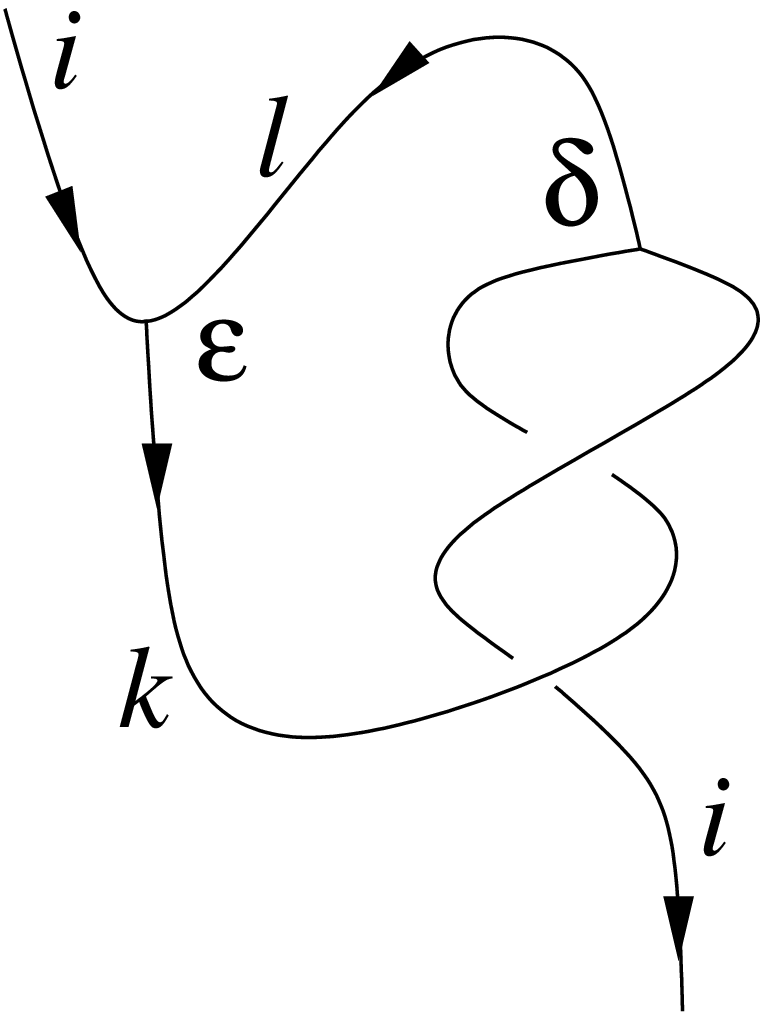}}
\end{picture}}
\newcommand{\fotto}
{\begin{picture}(3,3)(0,0) 
\scalebox{.25}{\includegraphics{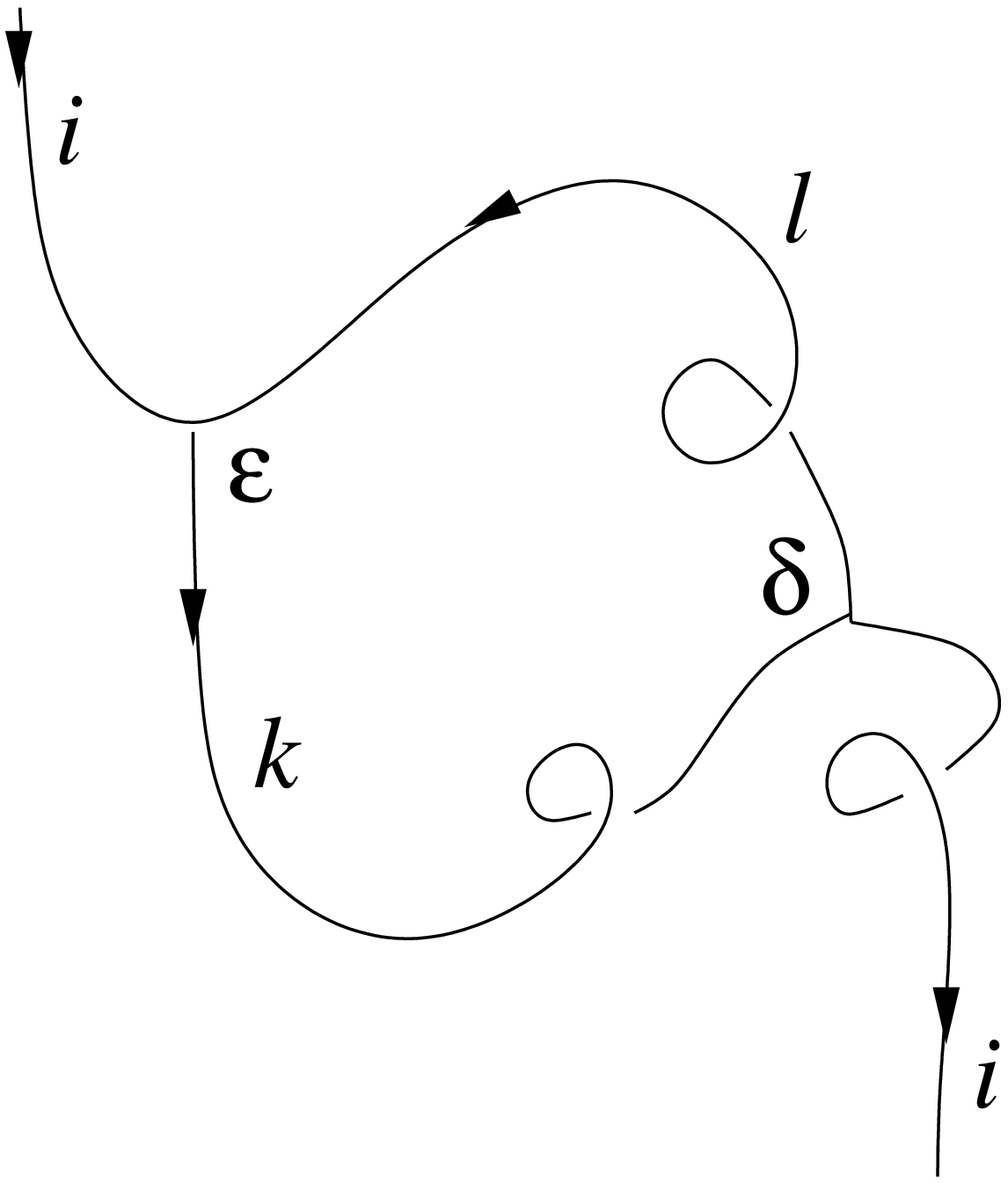}}
\end{picture}}
\newcommand{\fnove}
{\begin{picture}(3,3)(0,0) 
\scalebox{.25}{\includegraphics{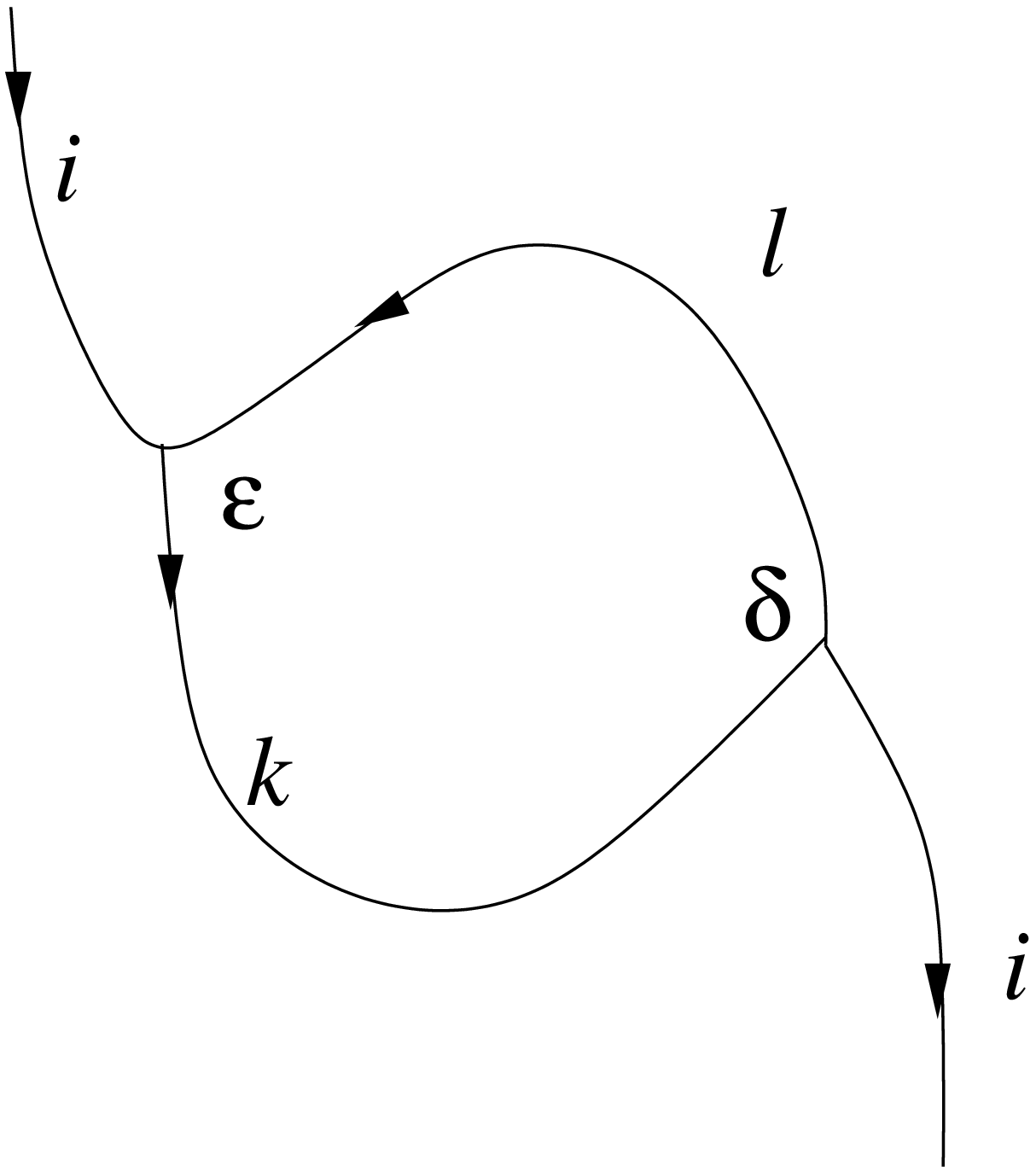}}
\end{picture}}
\newcommand{\fundici}
{\begin{picture}(2.1,3)(0,0) 
\scalebox{.25}{\includegraphics{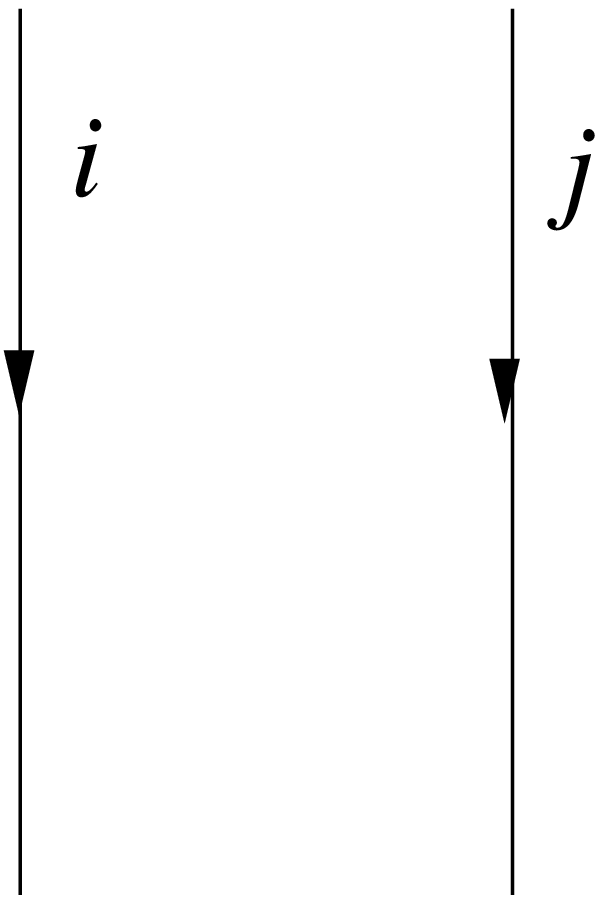}}
\end{picture}}
\newcommand{\fdodici}
{\begin{picture}(2.5,3)(0,0) 
\scalebox{.25}{\includegraphics{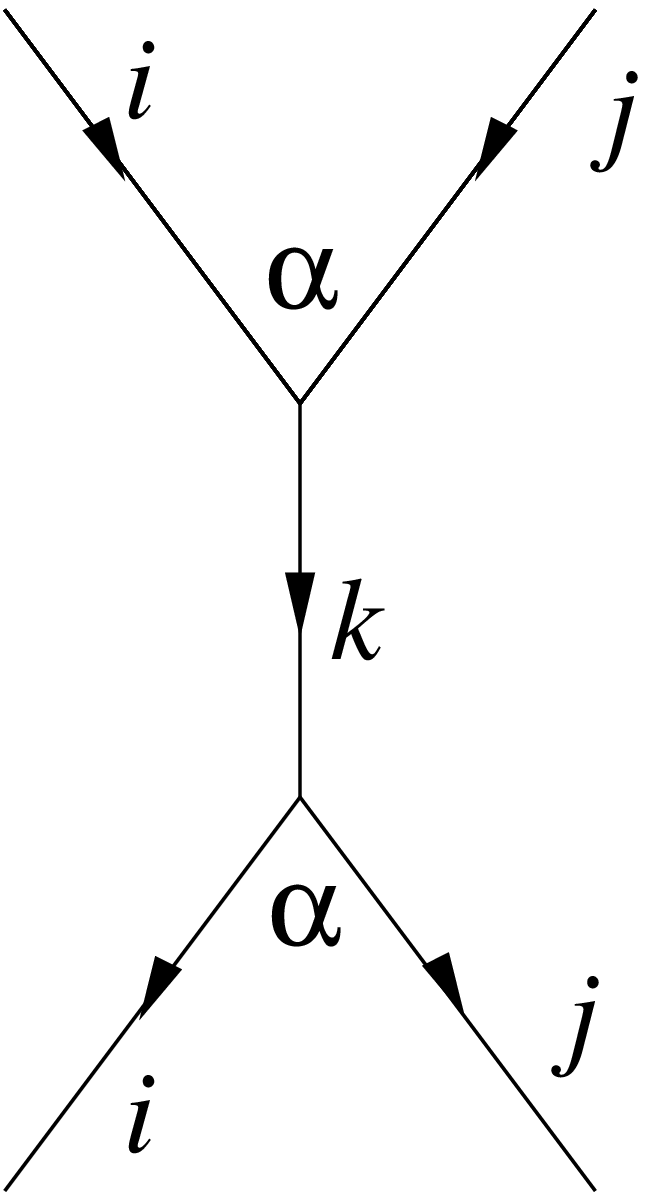}}
\end{picture}}
\newcommand{\ftredici}
{\begin{picture}(1.8,3)(0,0) 
\scalebox{.25}{\includegraphics{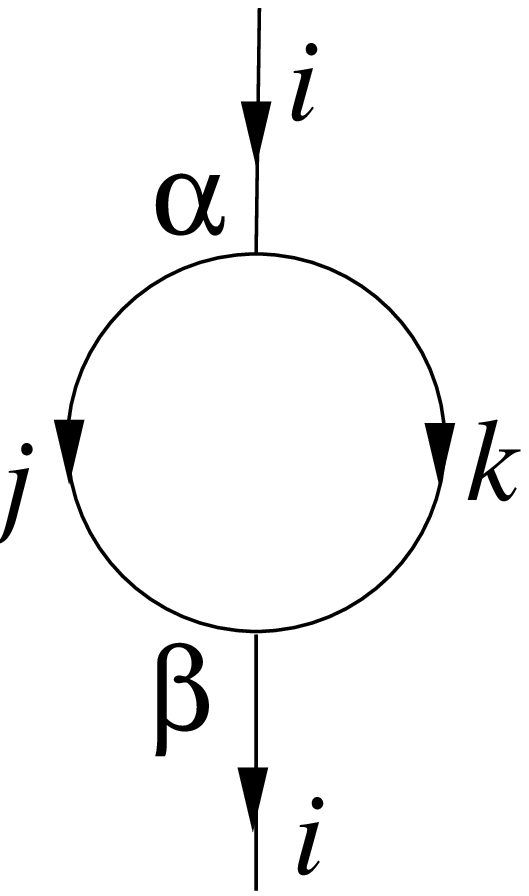}}
\end{picture}}
\newcommand{\fquattordici}
{\begin{picture}(1,3)(0,0)
\scalebox{.25}{\includegraphics{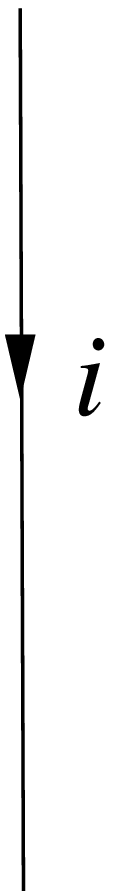}}
\end{picture}}
\newcommand{\fventuno}
{\begin{picture}(3,3)(0,0) 
\scalebox{.25}{\includegraphics{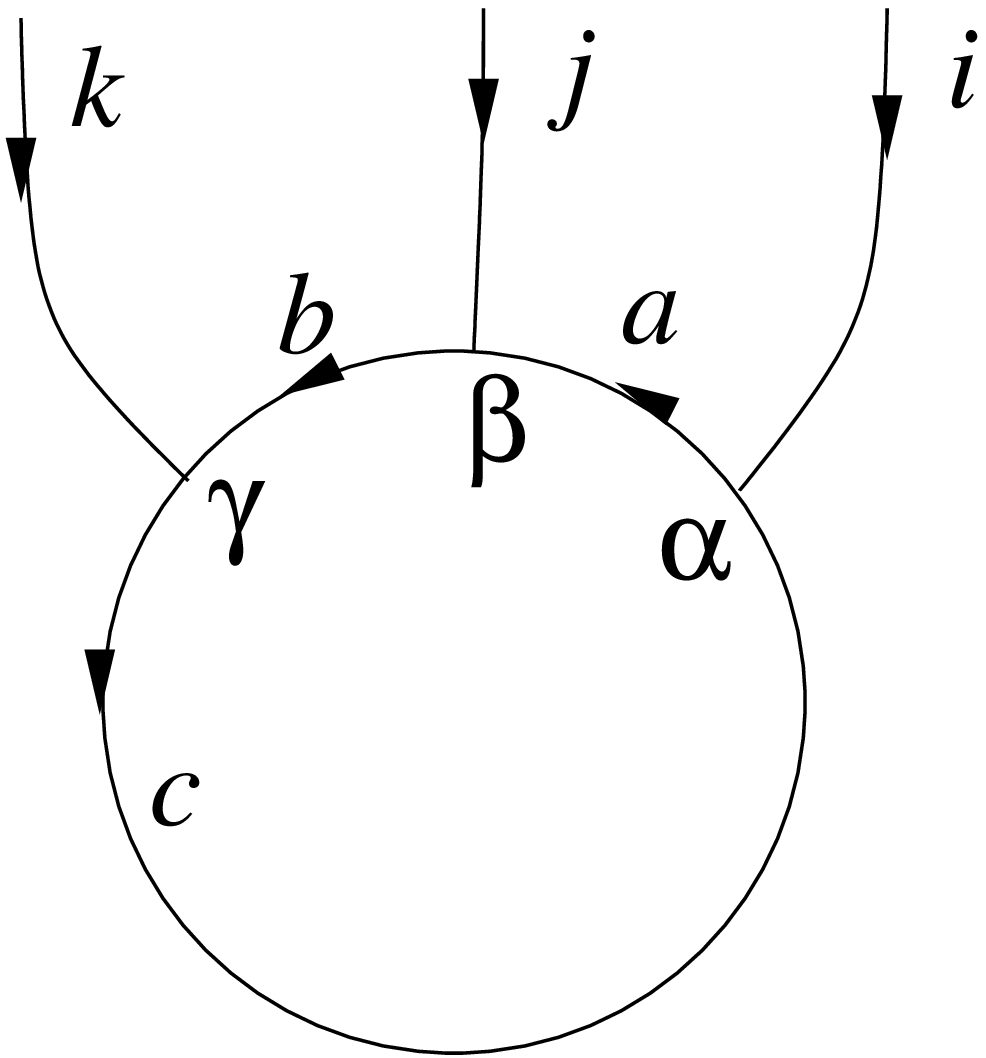}}
\end{picture}}
\newcommand{\fventidue}
{\begin{picture}(3,3)(0,0) 
\scalebox{.25}{\includegraphics{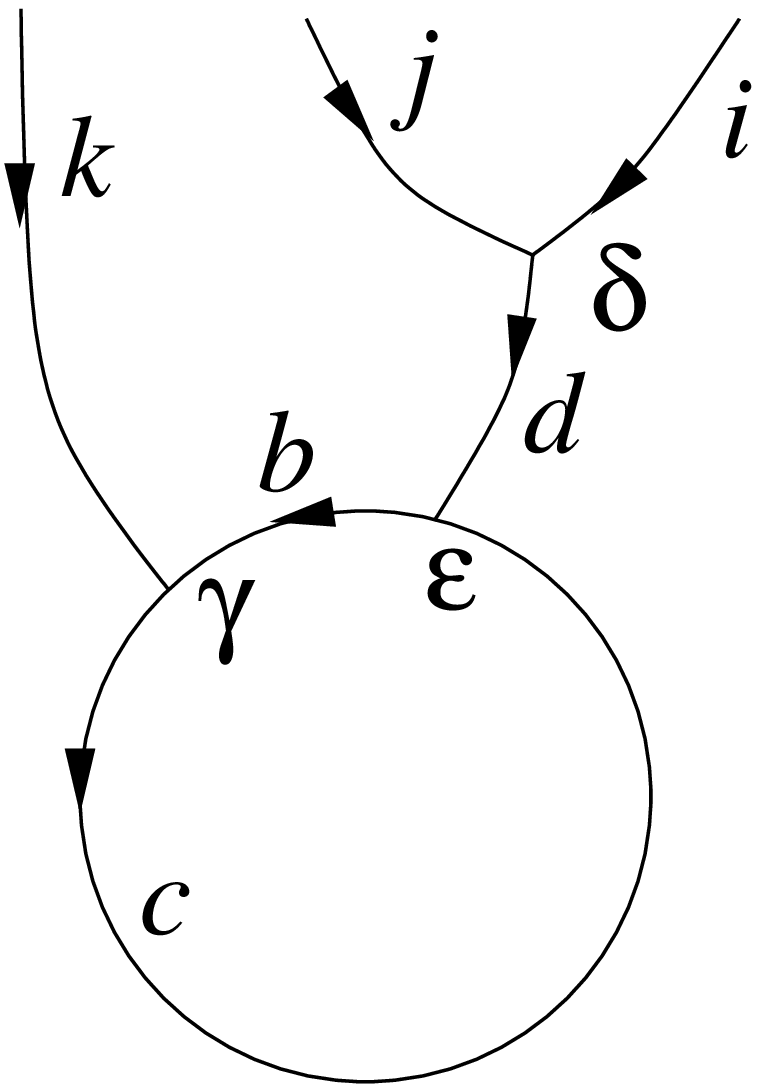}}
\end{picture}}
\newcommand{\fventitre}
{\begin{picture}(3,3)(0,0) 
\scalebox{.25}{\includegraphics{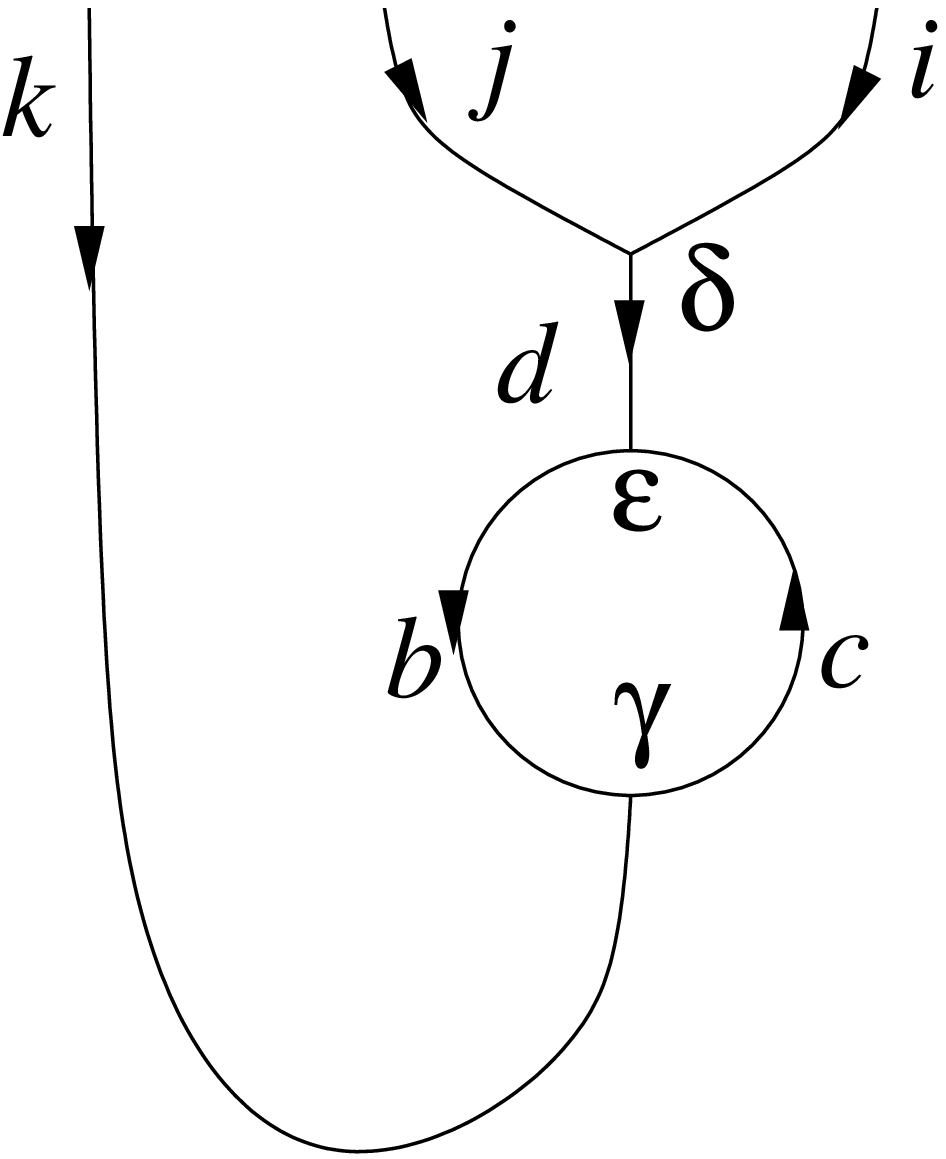}}
\end{picture}}
\newcommand{\fventiquattro}
{\begin{picture}(3,3)(0,-0.5) 
\scalebox{.25}{\includegraphics{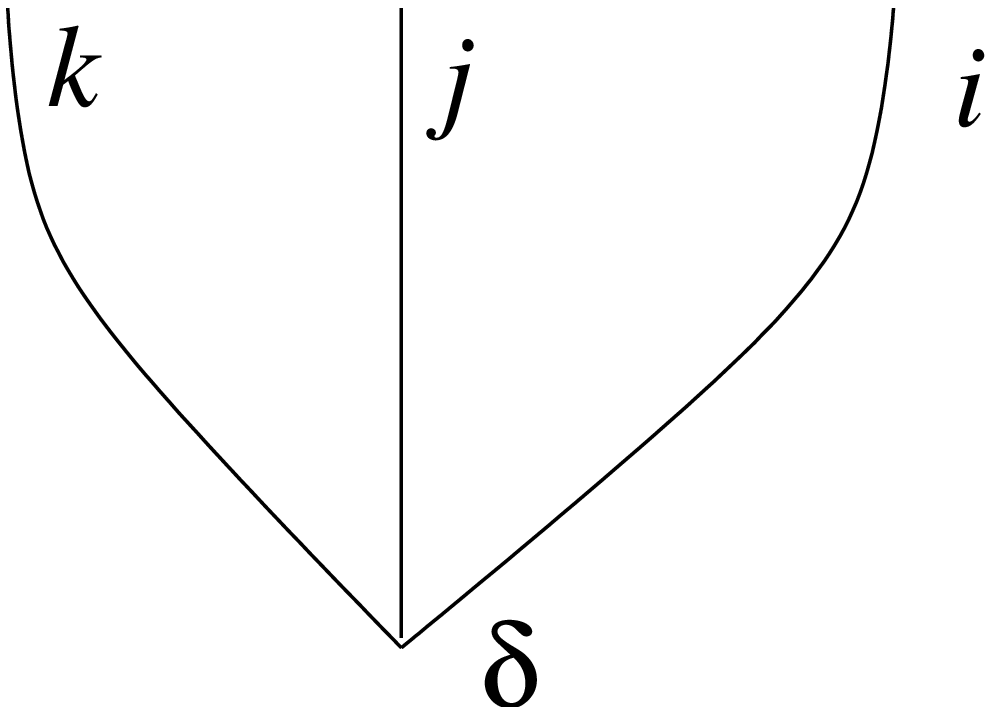}}
\end{picture}}
\newcommand{\ftrentadue}
{\begin{picture}(3,4)(0,-0.5) 
\scalebox{.25}{\includegraphics{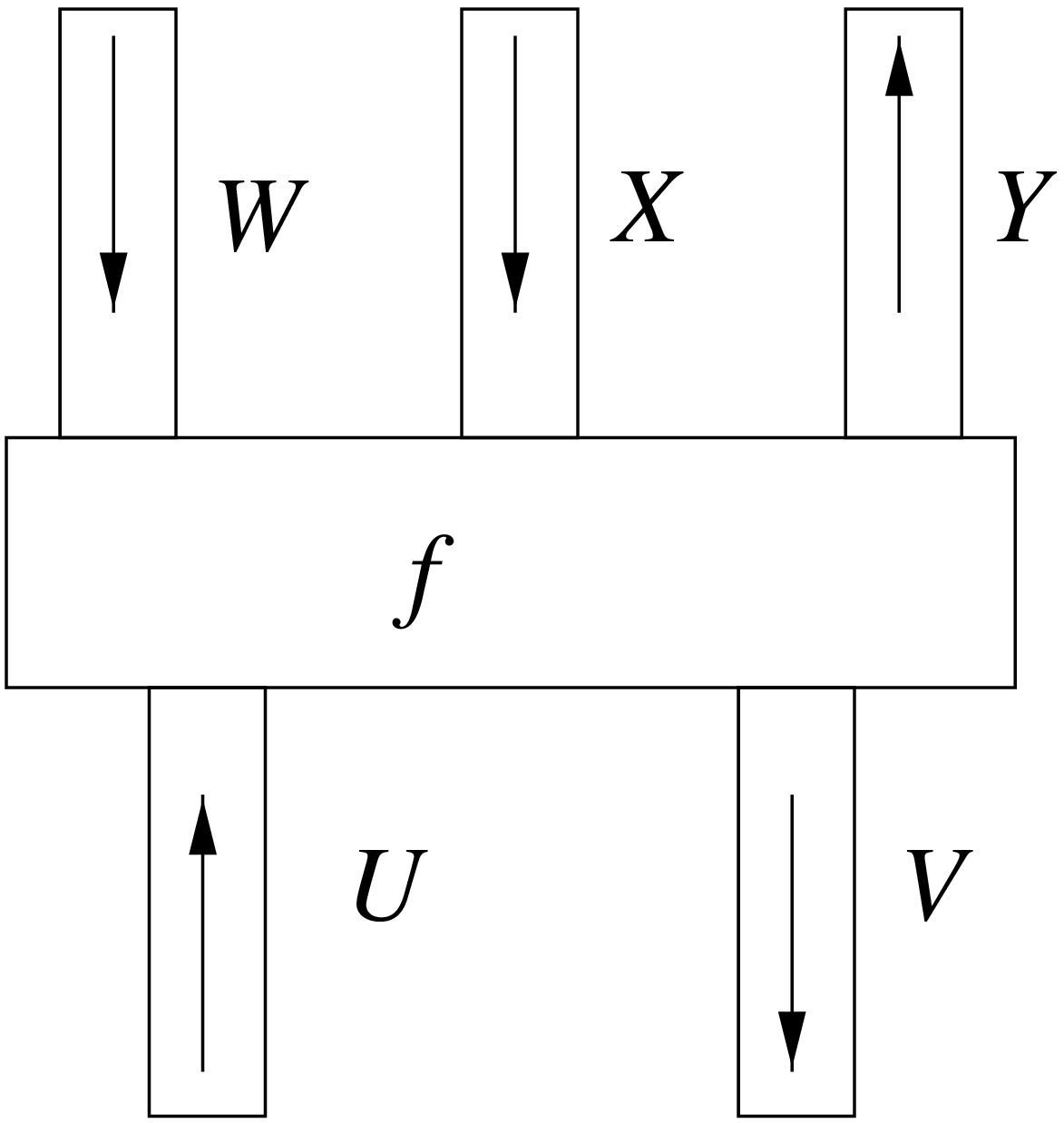}}
\end{picture}}
\newcommand{\ftrentatre}
{\begin{picture}(3,4)(0,-0.5) 
\scalebox{.30}{\includegraphics{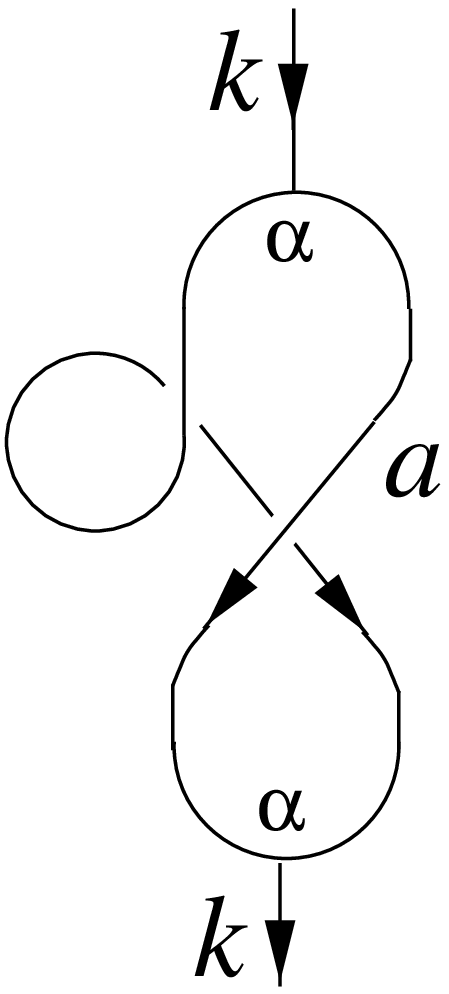}}
\end{picture}}

\newcommand{\fduecentouno}
{\begin{picture}(3,3)(0,0) 
\scalebox{.3}{\includegraphics{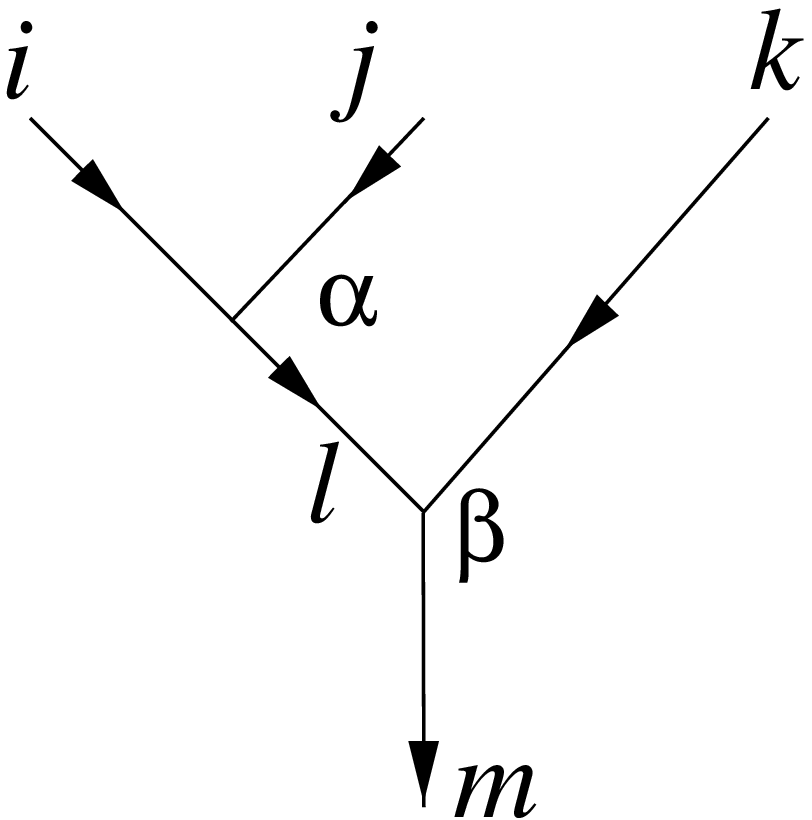}}
\end{picture}}
\newcommand{\fduecentodue}
{\begin{picture}(3,3)(0,0) 
\scalebox{.3}{\includegraphics{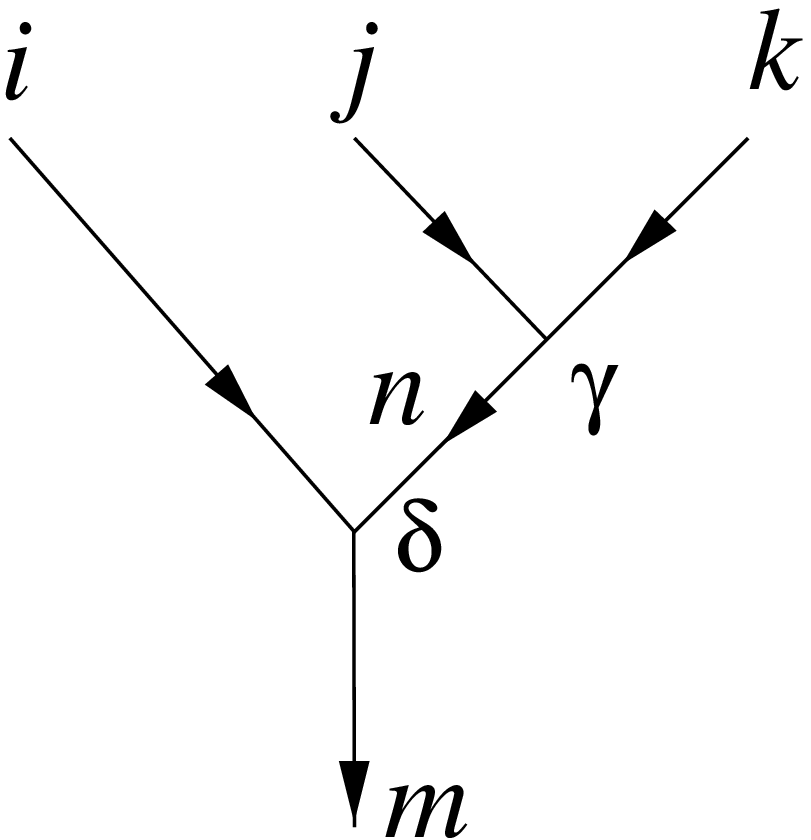}}
\end{picture}}
\newcommand{\ftrecentodue}
{\begin{picture}(3,3)(0,0) 
\scalebox{.20}{\includegraphics{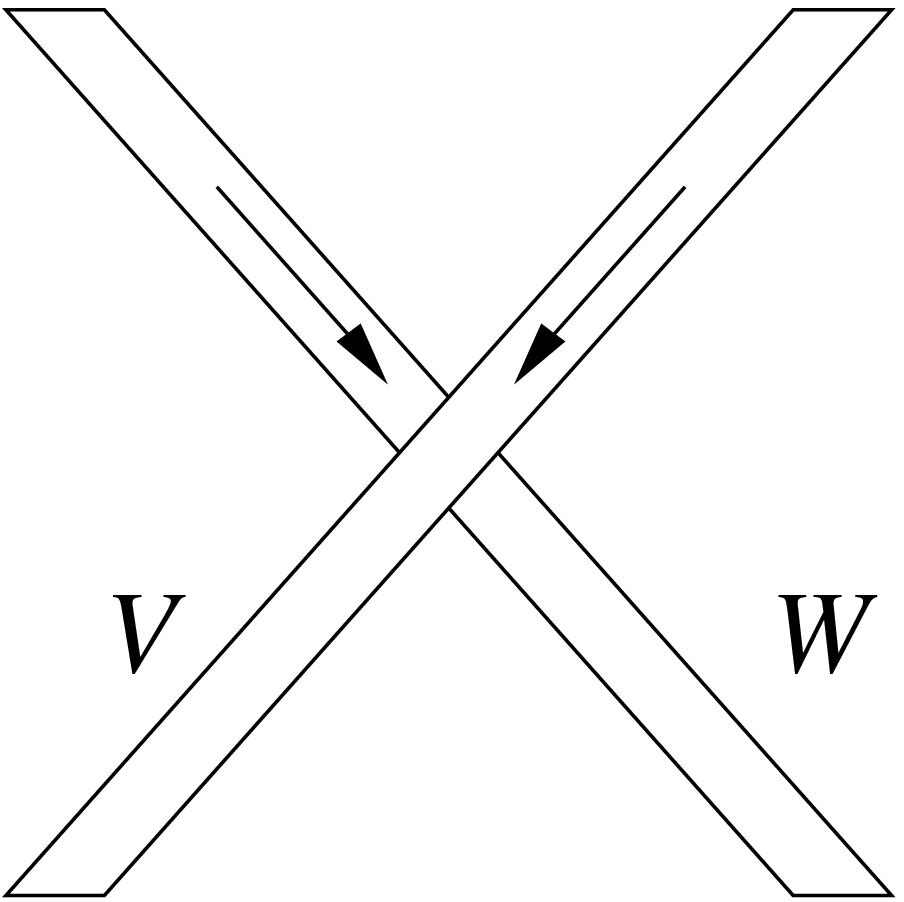}}
\end{picture}}
\newcommand{\ftrecentotre}
{\begin{picture}(3,3)(0,0) 
\scalebox{.23}{\includegraphics{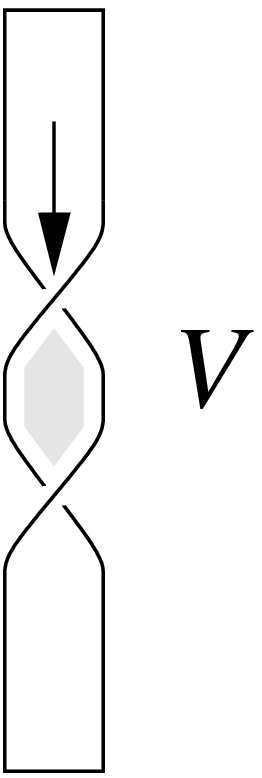}}
\end{picture}}
\newcommand{\ftrecentoquattro}
{\begin{picture}(3,0.01)(0,0) 
\scalebox{.20}{\includegraphics{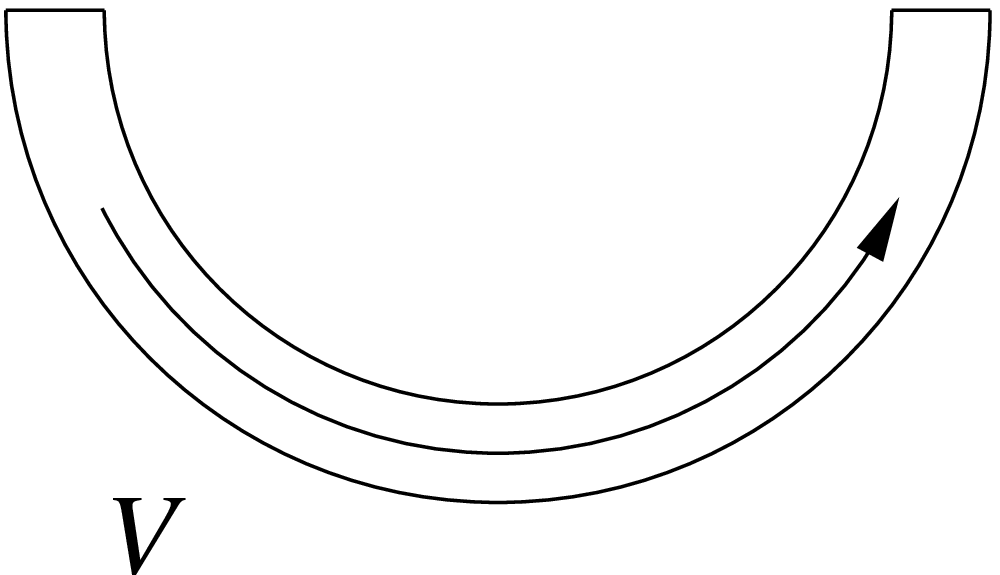}}
\end{picture}}
\newcommand{\ftrecentocinque}
{\begin{picture}(3,2)(0,0) 
\scalebox{.20}{\includegraphics{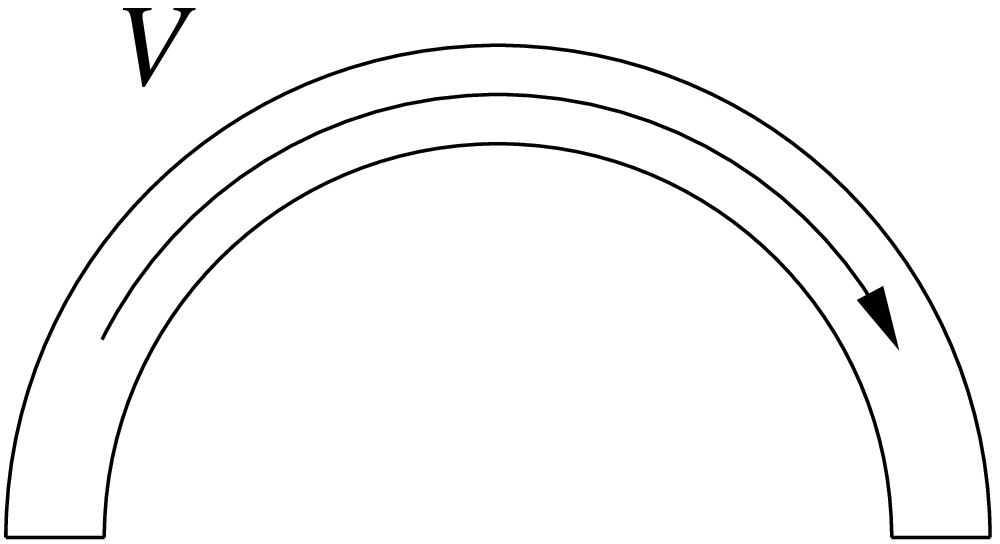}}
\end{picture}}

\newcommand{\fquattrocentouno}        
{\begin{picture}(3,4)(2,0) 
\scalebox{.34}{\includegraphics{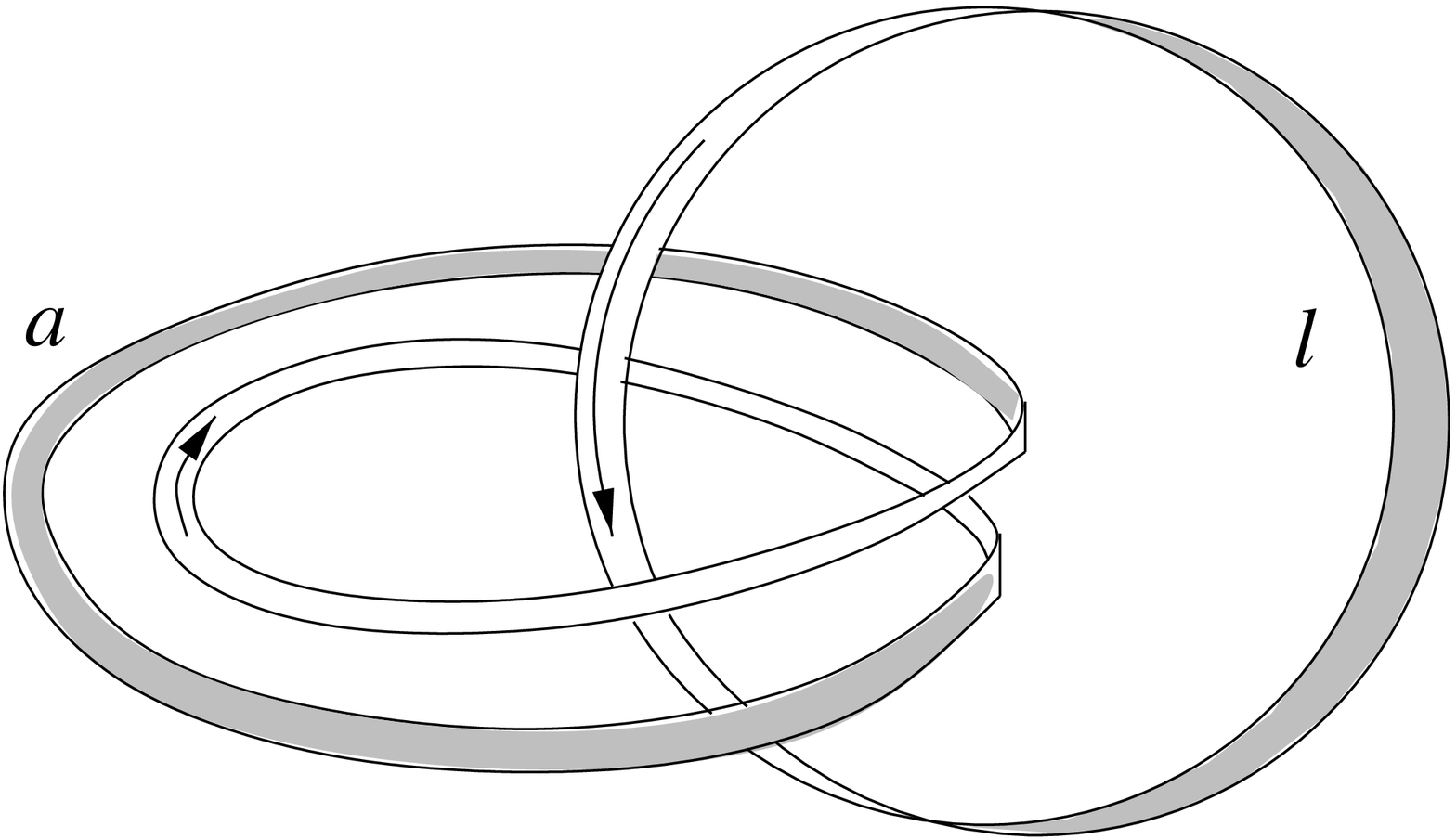}}
\end{picture}}
\newcommand{\fquattrocentodue}
{\begin{picture}(3,3)(0,0) 
\scalebox{.25}{\includegraphics{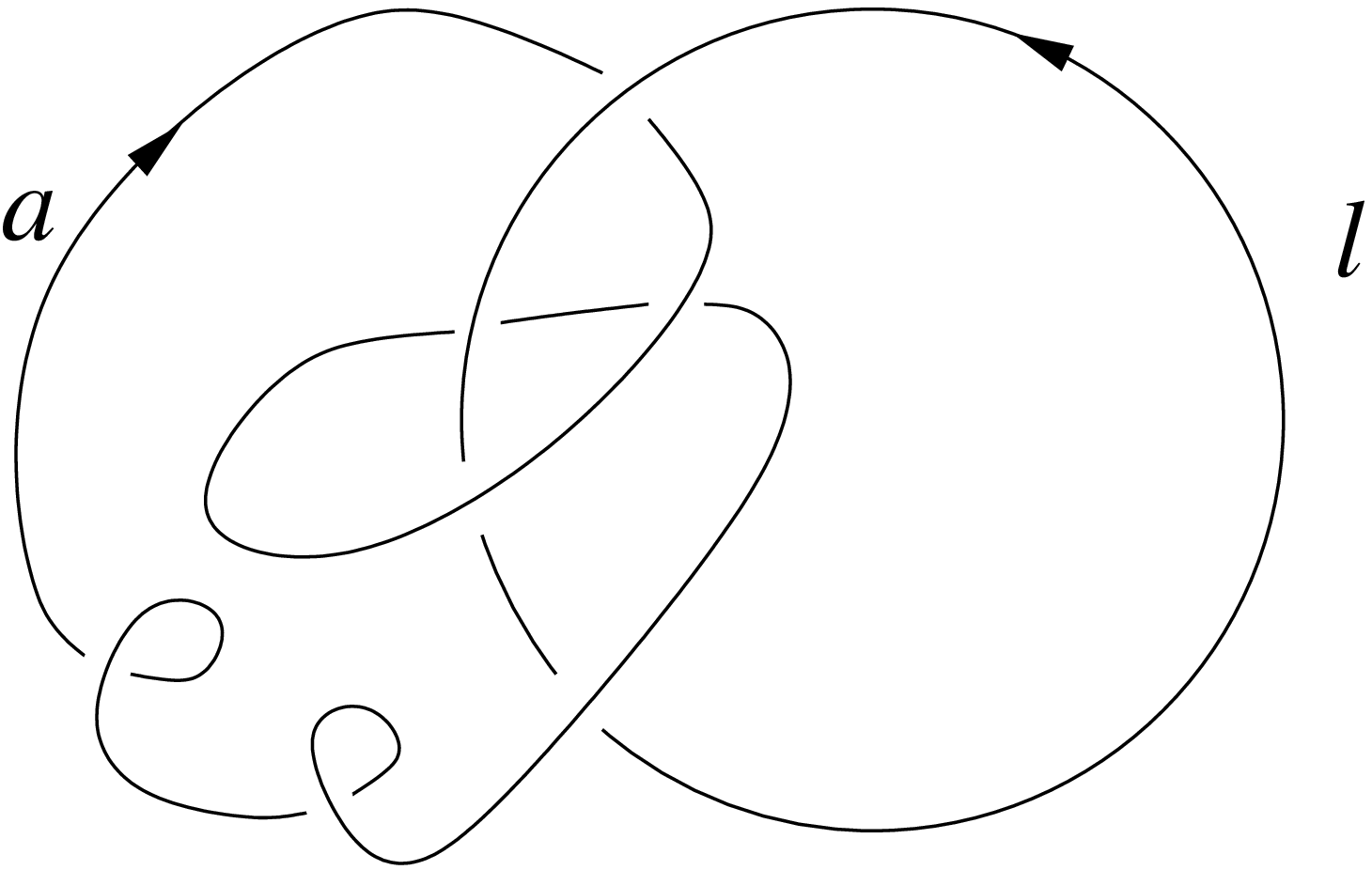}}
\end{picture}}
\newcommand{\fquattrocentotre}
{\begin{picture}(3,3)(0,0) 
\scalebox{.25}{\includegraphics{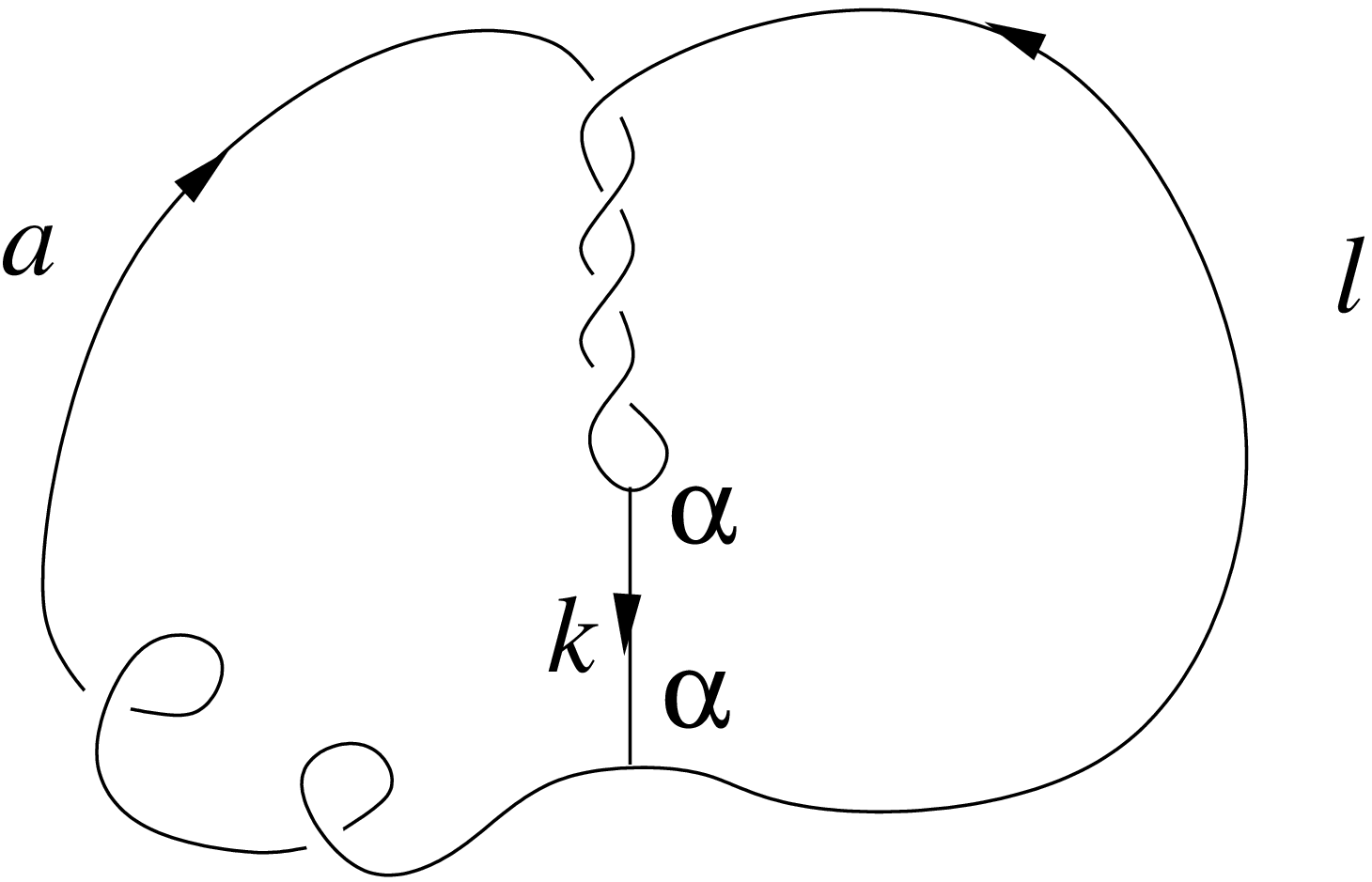}}
\end{picture}}
\newcommand{\fquattrocentoquattro}
{\begin{picture}(3,3)(0,0) 
\scalebox{.20}{\includegraphics{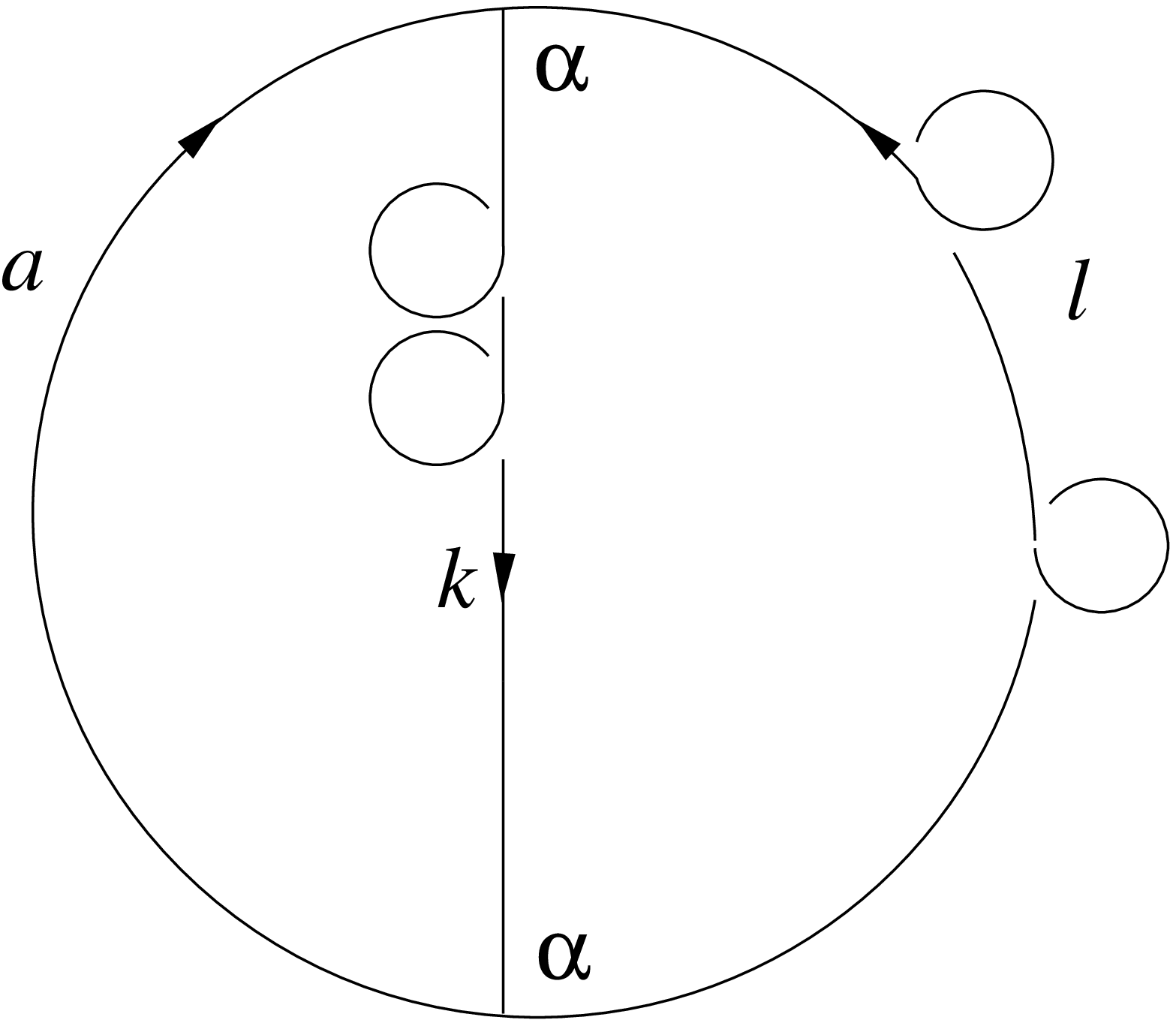}}
\end{picture}}
\newcommand{\fvuno} 
{\begin{picture}(3,3)(0,0) 
\scalebox{.3}{\includegraphics{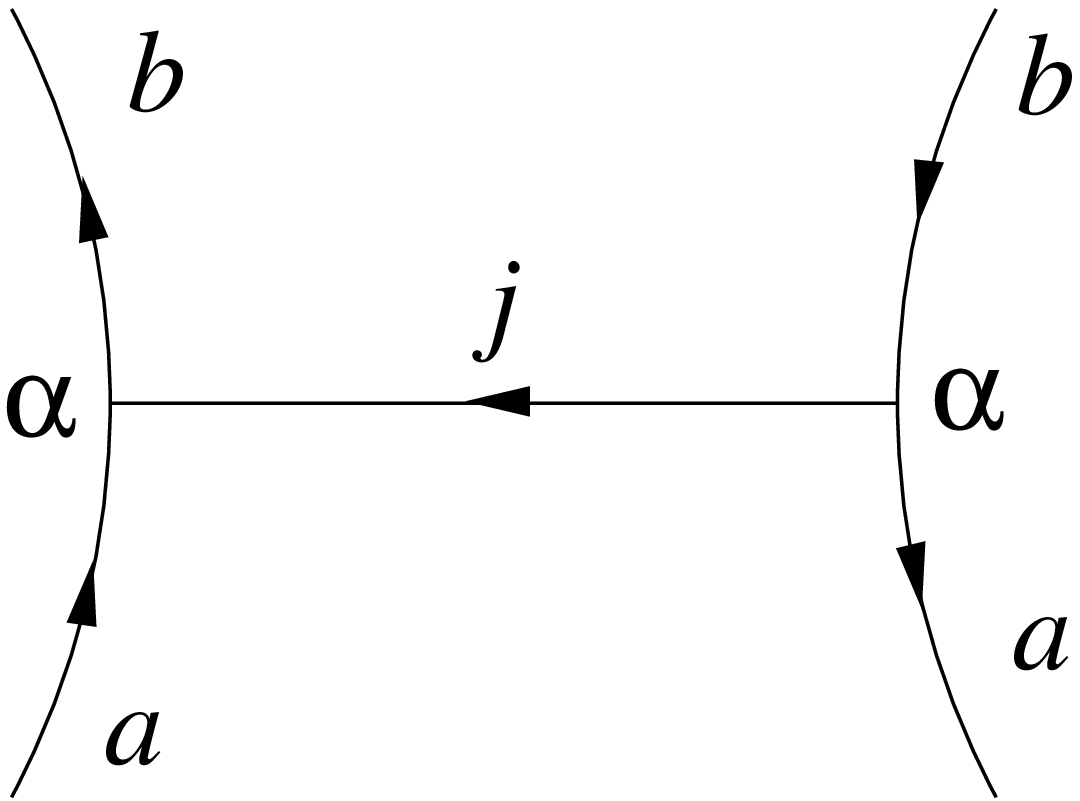}}
\end{picture}}
\newcommand{\fmilleuno} 
{\begin{picture}(3,3)(0,0) 
\scalebox{.3}{\includegraphics{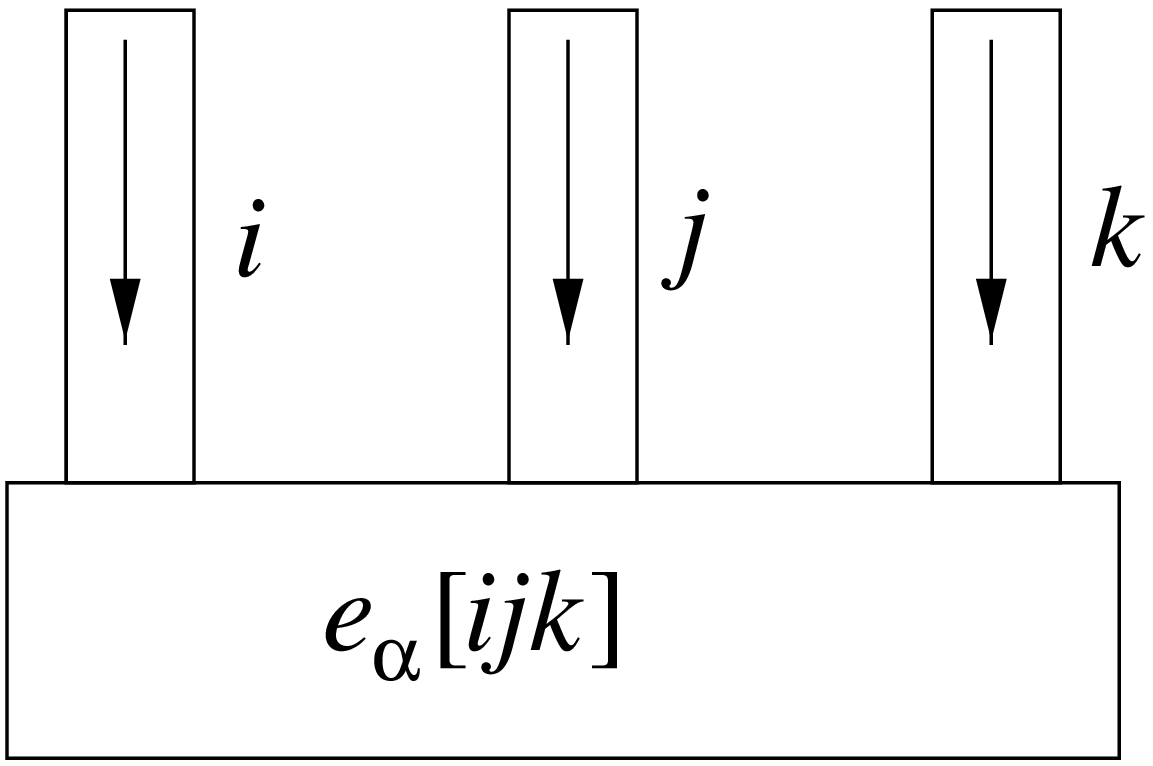}}
\end{picture}}
\newcommand{\fmilledue} 
{\begin{picture}(3,3)(0,0) 
\scalebox{.3}{\includegraphics{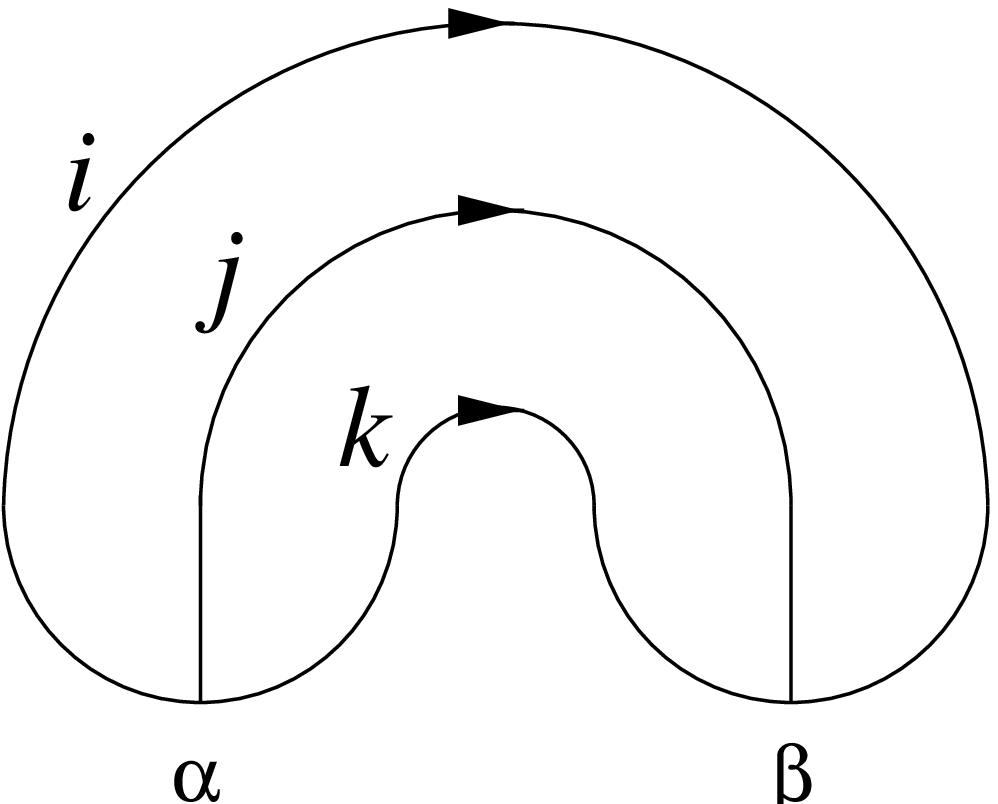}}
\end{picture}}
\newcommand{\fmilletre} 
{\begin{picture}(3,3)(0,0) 
\scalebox{.3}{\includegraphics{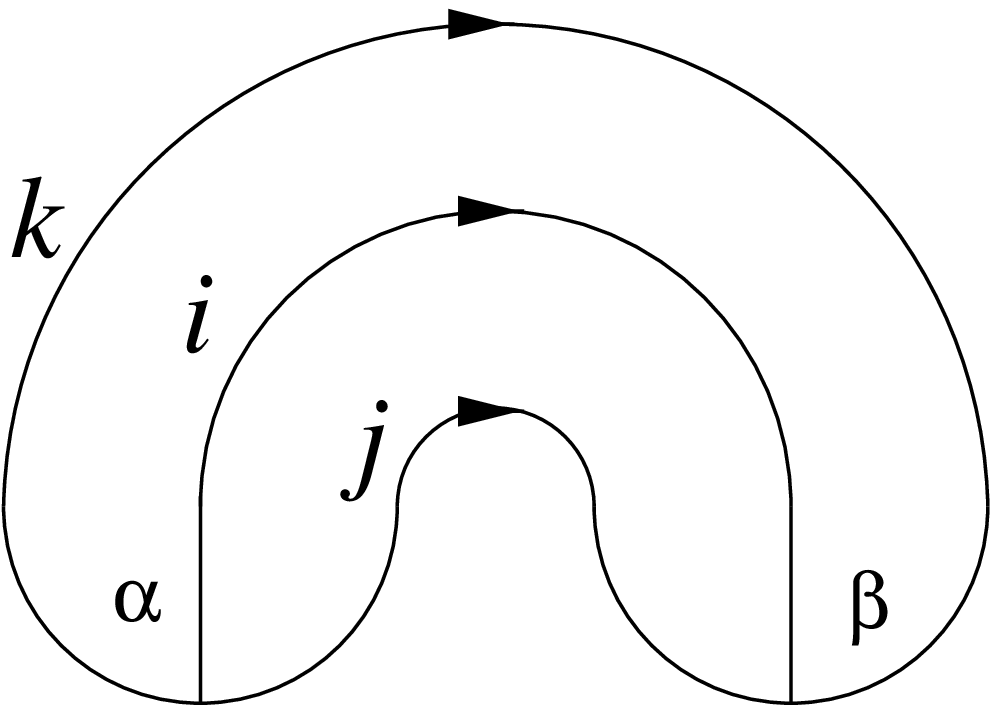}}
\end{picture}}
\newcommand{\fmillequattro} 
{\begin{picture}(3,3)(0,0) 
\scalebox{.3}{\includegraphics{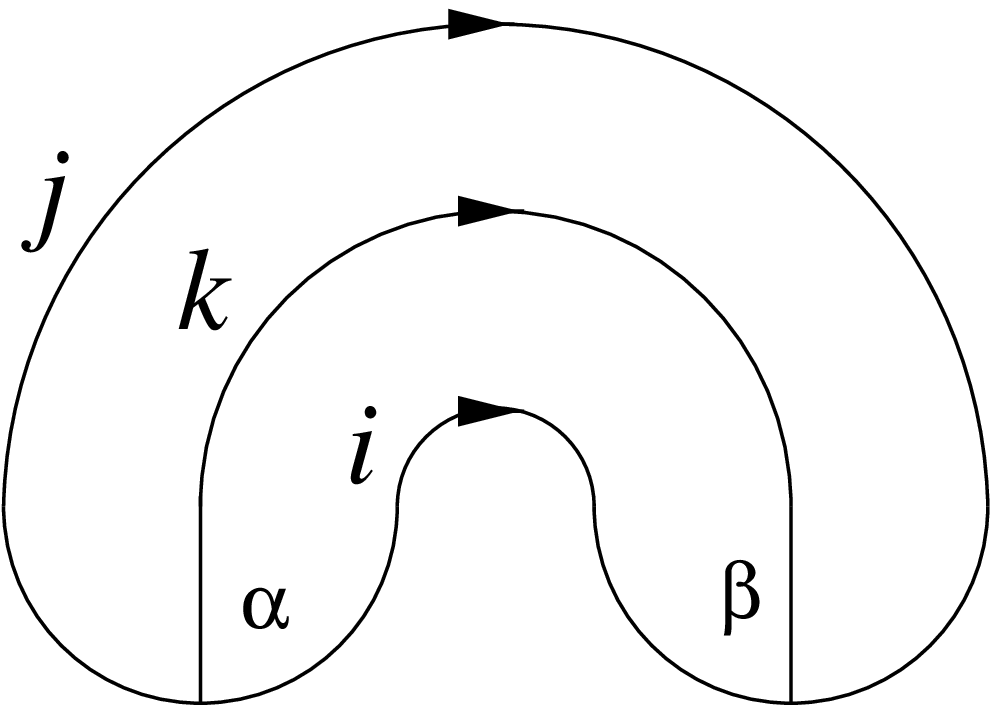}}
\end{picture}}
\newcommand{\alza}[2]{\raisebox{#1}{$\displaystyle{#2}$}}

\newtheorem%
{thm}{Theorem}[section]
\newtheorem%
{proposition}[thm]{Proposition}
\newtheorem%
{lemma}[thm]{Lemma}
\newtheorem%
{lemmadef}[thm]{Lemma-Definition}
\newtheorem%
{corollary}[thm]{Corollary}
\newtheorem%
{conjecture}[thm]{Conjecture}

\title[Correlation functions in RCFT and 3D topology]
{Correlation functions and boundary conditions\\ in rational
conformal field theory
and three-dimensional topology}
\author[G. Felder, J. Fr\"ohlich, J. Fuchs and C. Schweigert]
{Giovanni Felder, J\"urg Fr\"ohlich, J\"urgen Fuchs  and 
Christoph Schweigert}
\address{G. F.: Department of mathematics, ETH-Zentrum, CH-8092 Z\"urich}
\address{
J. F.: Institute for theoretical physics, ETH-H\"onggerberg, CH-8093 Z\"urich}
\address{
J. F.: Institute for theoretical physics, ETH-H\"onggerberg, CH-8093 Z\"urich}
\address{C. S.:
LPTHE, Universit\'e Paris VI, 4, place Jussieu, F-75\,252 Paris CEDEX 05}
\date{}
\begin{document}
\begin{abstract}
We give a general construction of
correlation functions in rational conformal field theory
on a possibly non-orientable surface with boundary in terms
of 3-dimensional topological field theory. The
construction applies to any modular category in the sense of Turaev.
It is proved that these correlation functions
obey modular invariance
and factorization rules. Structure constants are calculated and expressed
in terms of the data of the modular category.
\end{abstract}
\begin{flushright} {~}\\[-2em] {\sf hep-th/9912239} \\[1mm]
{\sf ETH-TH/99-30}\\[1mm]{\sf PAR-LPTHE 99-45}\\[2em] \end{flushright}

\maketitle

\section{Introduction}
In this paper, we study correlation functions in conformal
field theory from the point of view of three-dimensional topological
field theory.

The
problem of constructing correlation functions
in rational conformal field theory has two parts. 
The first part of the problem is to determine the
space of conformal blocks. The second part is to 
use this space to construct correlation functions.

The first part of the problem is by now well understood in mathematical
terms. One approach, suggested by Witten's paper
\cite{W} on Chern--Simons theory, is given in terms of 3-dimensional
topological field theory (TFT). Such  a theory assigns to
every ``extended surface'' 
 --- an oriented 2-manifold $X$ with marked points carrying labels,
and certain additional data related to framing ---
a finite dimensional complex vector space $\mathcal H(X)$, the space of 
conformal blocks, or of the states of the TFT, and
to every 3-manifold $M$  bounded by $X$, containing a ``ribbon graph'',
a vector $Z(M)$ in $\mathcal H(X)$.  The ribbon graph is an embedded graph 
ending at the marked points, with some additional structure. 
The spaces $\mathcal{H}(X)$ and the vectors $Z(M)$ 
are supposed to obey a number
of natural axioms relating to homeomorphisms and cutting and pasting.
Turaev showed in \cite{T} how every {\em modular category} produces a 
TFT and, in particular, a space of conformal blocks associated to
every extended surface.

The purpose of this paper is to give a precise meaning in
the same terms to the second part of the 
problem, the construction of correlation
functions out of conformal blocks, and to present a solution.
Let us first describe the data of the problem. First, one requires the 
chiral data of a rational conformal field theory, which for us are
the data of a modular category. In particular there is
a set $I$ of distinguished simple objects. These data are what is
needed to define conformal blocks. The conformal field theory itself
is given by a system of {\em correlation functions} obeying certain axioms.
Each construction of such correlation functions satisfying the
axioms gives a different conformal field theory
with the same underlying modular category.

We proceed to describe what 
     the axioms are
in the simplest situation (the ``Cardy
case'') to which we restrict our attention in this paper. To formulate
the axioms we introduce the notion of a {\em labeled surface}.
A conformal field theory is then an assignment of a
correlation function $C(X)$ 
to each  labeled surface $X$.
 A labeled surface $X$ consists of a (not necessarily orientable)
compact 2-dimensional manifold
with (possibly empty) oriented
boundary, a set of marked points on it, and ``boundary conditions''.
The marked points all carry a label from $I$ and certain local data. The boundary
conditions are a coloring by $I$ of the 
boundary arcs between marked boundary  points.
For example, if $X$
is a disk with $m$ marked points on its boundary, the boundary
conditions are a labeling of the $m$ 
arcs between neighboring points by elements
of $I$.

The correlation function associated to these data is then a linear map
$W_{\partial X}\to\mathcal{H}(\hat X)$ from a ``multiplicity space'',
$W_{\partial X}$, associated
to the boundary  
($W_{\partial X}=\C$ if $\partial X=\emptyset$) 
to the space of
conformal blocks,
$\mathcal{H}(\hat X)$, of the {\em double} $\hat X$ of $X$. The
double of a compact surface is an oriented closed surface with an orientation 
reversing involution $\sigma$ so that $X$ is obtained from $\hat X$ by
identifying pairs of points related by $\sigma$.
For example, the double of a disk is a sphere, and the double
of the projective plane is also a sphere, but with a different involution.
The torus is the double of the annulus, the M\"obius band
 and the Klein bottle. 
The double of a closed orientable surface is the disjoint union of two copies
of the surface, with opposite orientations. The definition of the double
also
applies  naturally to surfaces with marked points, so that the
double of a labeled 
surface is an extended surface in the sense of Turaev \cite{T}.

The correlation functions are supposed to behave
naturally under homeomorphisms (the modular invariance of
correlation functions) and gluing (the factorization properties
of correlation functions). There are two types of gluing properties:
a (possibly disconnected) surface may be glued by identifying two
arcs on its boundary or by cutting out two disks in the interior
and identifying the boundaries of the resulting holes. In both cases
these operations induce gluing operations on the double of the surface,
which in turn, by the rules of TFT, induce homomorphisms between the corresponding
spaces of conformal blocks. The requirement of factorization means that
the correlation functions 
behave naturally under these homomorphisms.

Our main result is the
construction of an assignment 
$X\mapsto C(X)\in\Hom_\C(W_{\partial X}, \mathcal H(\hat X))$ of
a correlation function to every labeled surface
and a proof that it obeys the required modular and factorization properties.
The basic idea is that the correlation function $C(X)$ is the vector
$Z(M_X)\in\mathcal H(\hat X)$ associated to a 3-manifold $M_X$ with a ribbon
graph, the {\em connecting 3-manifold of $X$}, 
whose boundary is the double $\hat X$ of $X$. The vector $Z(M_X)$ depends 
linearly on the colorings of the vertices of the ribbon graphs
by morphisms of the modular category. This space of colorings is identified
with $W_{\partial X}$.
The connecting manifold was first considered by Ho\v{r}ava \cite{H} 
in his study of Chern--Simons theory on $\Z_2$-orbifolds.

\begin{figure}
\scalebox{0.4}{\includegraphics{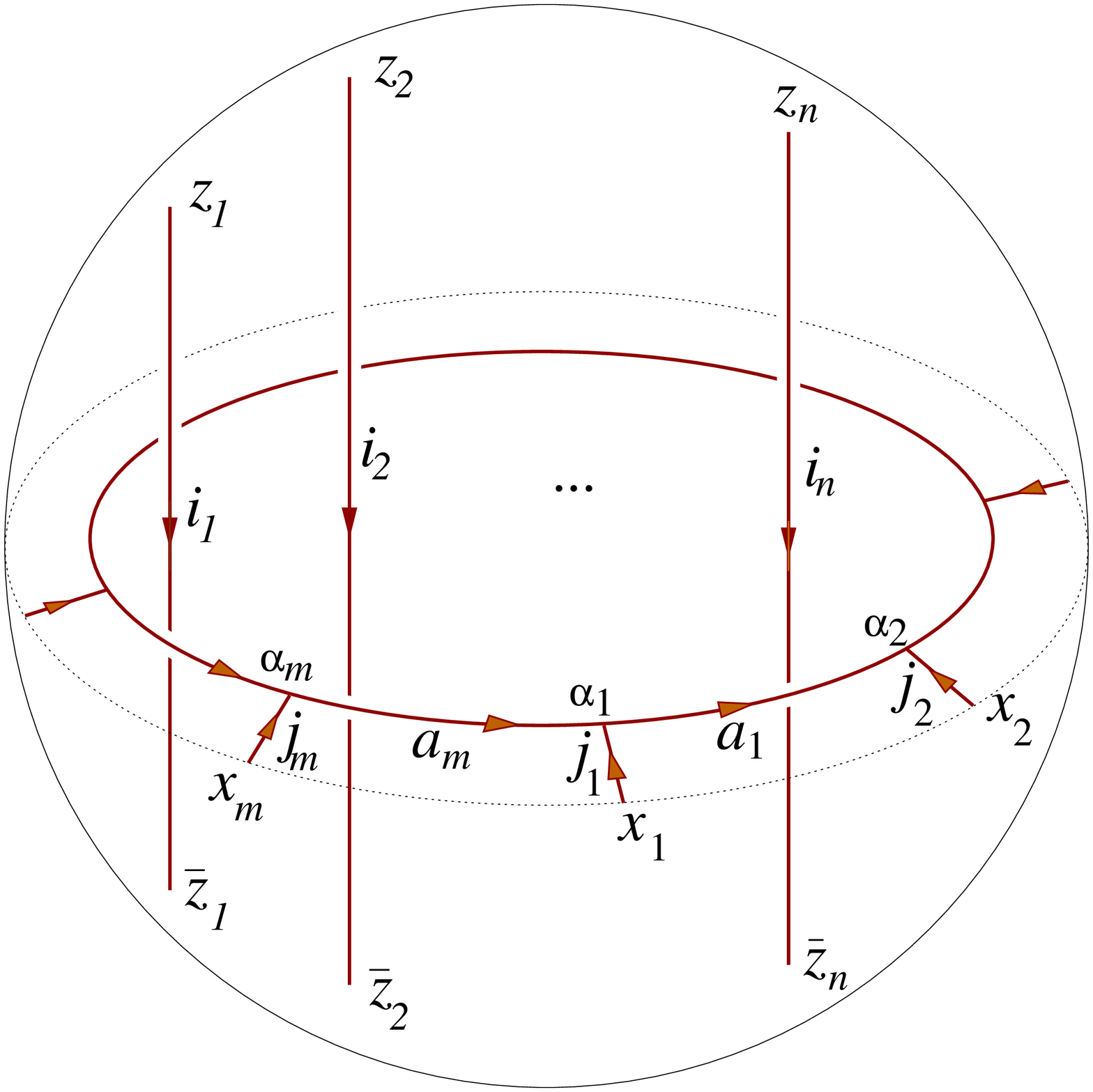}}
\caption{}\label{fig1}\end{figure}

Let us first describe our construction 
in a simple example, suppressing
for the
moment the framing. 
Let $X$ be a disk with $n$ marked points in its interior
with labels $i_1,\dots,i_n\In I$ and $m$ points on
the boundary, with labels $j_1,\dots,j_m\In I$. Let the boundary condition
on the arcs between the $k$th and $k{+}1$st boundary points be
labeled by $a_k\In I$. Then $M_X$ is a 3-ball and the correlation function $C(X)$
is the conformal block on the sphere associated to the 
ribbon graph  depicted in Fig.\ \ref{fig1}.
The points $z_k$, $\bar z_k$ project to the $k$th interior
point on the disk, and $x_l$ projects to the $l$th 
boundary point.

In the general case, the main property of the connecting
manifold is that it comes with a projection
$p{:}\ M_X\to X$ whose fibers over interior points are 
closed intervals, and an inclusion  $i{:}\ X\to M_X$,
which is a homotopy equivalence such that $p\,{\circ}\,i\eq\mathrm{id}_X$. The
ribbon graph  then consists
of fibers of $p$ over the marked points in the interior,
and a loop running close to $i(\partial X)$ and connected by
short lines to the marked boundary points.

After formulating and proving  the modular invariance and
factorization properties of our correlation functions,
we compute ``structure constants'', namely the correlation
functions for elementary building blocks: the disk with three
boundary points, the disk with one interior and one boundary point,
and the projective plane with one point. General correlation functions
are then in principle obtainable from these using the factorization 
theorems. The structure constants are given in terms 
of the data (fusing matrices, modular $S$-matrix) of the modular category.

We also give formulae
for the annulus, the Klein bottle and the M\"obius strip
with no marked points. In these
cases the double is a torus; we show that the coefficients
of correlation functions with respect to a natural basis of the
space of conformal blocks are integers.

The correlation functions are thus given in our approach
as states of a TFT. To get actual functions of the position of
the marked points
and moduli, say for WZW models, one uses the modular category of
integrable modules of an affine Lie algebra \cite{KL,F}. The dependence on 
the moduli should then be
obtained by integrating the Knizhnik--Zamolodchikov connection.
In the case of the sphere and general WZW models
such a construction might be possible along the lines of \cite{K}.

The paper is organized as follows.
We begin by giving a review of three-dimensional
topological field theory, following
\cite{T}, in Sect.~\ref{s-2}. In this section
modular categories, TFT and modular functors
are introduced, and subtleties such as the framing anomaly are explained.
In Sect.~\ref{s-3} we present our proposal
for correlation functions. We also formulate and prove the factorization
and modular properties they obey. These
properties imply in particular that correlation
functions on general surfaces may be expressed
in terms of basic correlation functions.
We compute these basic correlation functions
in Sect.~\ref{s-4}. In Sect.~\ref{s-5} we compute
correlation functions in the cases where the
double has genus one, and prove integrality results.
 
In  Appendix \ref{a-mod} we give the definition of
modular categories and in Appendix \ref{a-surg}
we describe how to obtain
the real projective space by surgery on the
unknot, a result needed to compute correlation
functions on the projective plane.

Some of these results were announced in \cite{FFFS2}.
\medskip

\noindent{\bf Acknowledgements.}
G.\ F.\ is thankful to V.\ Turaev for useful correspondence.
We thank Ph. Mathey for careful reading the manuscript
and for corrections.

\section{Modular categories and 3-dimensional TFT}
\label{s-2}

\subsection{Modular categories}\label{ss-21}
A modular category is a
strict monoidal Ab-category $\M$ (see Appendix \ref{a-mod})
with unit object $\one$ and an additional
set of data obeying a system of axioms. The data are
\begin{enumerate}
\item 
A finite set $I$ of simple objects containing $\one$.
\item 
For each pair of objects $V,W$, a braiding morphism
$c_{V,W}\In\Hom(V\OTimes W,W\OTimes V)$.
\item 
For each object $V$, a twist $\theta_V\In \Hom(V,V)$.
\item 
A duality: for each object $V$ there is a unique
dual object $V^*$ and morphisms $b_V\In \Hom(\one,V\OTimes V^*)$
and $d_V\In\Hom(V^*\OTimes V,\one)$. 
\end{enumerate}

These data obey a number of axioms, which we describe
in  Appendix \ref{a-mod}. See \cite{T} for more details.
In general modular categories, $\Hom(\one,\one)$
is a general ring. Here we assume
that $\Hom(\one,\one)\eq\C$. The general case may be dealt with
similarly, as we only use the axioms and results in the general
theory of \cite{T}. However, in order to simplify
certain normalizations, we do use the fact that
$\C$ is an algebraically closed field.

The axioms can be best understood in the language of ribbon graphs, i.e.\
finite collections of disjoint ribbons, annuli and coupons. {\em Ribbons} 
are oriented rectangles $[-1/10,1/10]\Times [0,1]$ embedded in $\R^2\Times [0,1]$,
so that $[-1/10,1/10]\Times (0,1)$ is entirely contained in $\R^2\Times (0,1)$.
{\em Annuli} are oriented annuli $[-1/10,1/10]\Times S^1$, embedded in 
$\R^2\Times (0,1)$.  Ribbons and annuli are labeled by objects of the category.
The {\em core} $\{0\}\Times [0,1]$ or $\{0\}\Times S^1$ of each ribbon or annulus
is also given an orientation. The {\em coupons} are
oriented rectangles embedded in $\R^2\Times(0,1)$
with two preferred opposite sides, the top
and the bottom, and are labeled by morphisms of the category.
The ends $[-1/10,1/10]\Times\{0\}$, $\,[-1/10,1/10]\Times\{1\}$ of the
ribbons are glued to the top or the bottom of coupons, so that the
orientations match to give an oriented (topological) 2-manifold with boundary,
 or are contained in $\R^2\Times\{0,1\}$. The coupons are labeled by morphisms
from the tensor product of the objects labeling the ribbons glued to the bottom
boundary, or their dual objects, to the tensor product of the objects labeling the
ribbons glued to the top boundary, or their dual objects. The ordering
of the tensor product reflects the ordering (from left to right) of the ribbons
meeting at the coupon. The dual object
is taken when the orientation of the core points in the upwards (bottom to top)
direction. For example, if the upper side of the coupon
\[
\ftrentadue
\]
is the top boundary, and the orientation is the standard orientation of the
plane, then $f\in\Hom(U^*\Otimes V,W\Otimes X\Otimes Y^*)$.

Using this correspondence the basic morphisms are represented as follows:
\\[-2.8em]
 \begin{eqnarray*}
 &\alza{0.8cm}{c_{V,W}=\ \ }\ftrecentodue\qquad
 \alza{0.8cm}{\theta_V=\quad \ }\ftrecentotre&
 \\
 &\alza{.5cm}{b_V=\ \ }\ftrecentoquattro\qquad\ \ 
 \alza{.5cm}{d_V=\ \ }\ftrecentocinque \alza{.5cm}{.}\ \ &
 \end{eqnarray*}
Choosing an orientation of a ribbon in an oriented 3-manifold
is the same as choosing a preferred side, which in our drawings will usually
face the reader. The other side is drawn as shaded.

The tensor product of morphisms is represented by the juxtaposition
of the factors, and the composition $fg$ is obtained by drawing
$f$ on top of $g$ and gluing the ends of the ribbons. 
Then the axioms are such that the morphism corresponding to a ribbon
graph depends only on its isotopy class. In particular for every
ribbon graph $\Gamma$ in $S^3=\R^3\cup\infty$ we get an isotopy invariant
$|\Gamma|\in\Hom(\one,\one)=\C$.

In drawing the graphs representing morphisms we will often make
use of the ``blackboard framing'' notation. Instead of drawing
ribbons we will only draw their cores, with the understanding
that the ribbons are contained in the plane of the page (or of the screen)
and inherit the orientation from the standard orientation of the plane.

As a consequence of the axioms one then proves that the objects
in $I$ are pairwise non-isomorphic and are a system of representatives of
all isomorphism classes of simple objects. Also one shows that
there is a canonical isomorphism $V\to(V^*)^*$ for all objects $V$
given in terms of braiding, twist and duality. We will tacitly
identify $(V^*)^*$ with $V$ via this isomorphism below.

Of particular importance is the (quantum) {\em trace} of
an  endomorphism $f\In \Hom(V,V)$. It is defined by the formula
$\mathrm{tr}(f)=d_Vc_{V,V^*} (\theta_Vf\otimes\mathrm{id}_{V^*})b_V\in \C$. 
It obeys $\mathrm{tr}(fg)=\mathrm{tr}(gf)$ (whenever both sides are defined)
and $\mathrm{tr}({\id_\one})=1$.

Out of the trace one then defines the 
{\em quantum dimensions} of simple objects:\,\footnote{~We usually denote 
simple objects by lower case letters $i,j,\dots,$ and general objects by 
capital letters $U,V,\dots$.}
\[
\mathrm{dim}(i)=\mathrm{tr}(\mathrm{id}_{{}i}),
\] 
and the {\em modular matrix} with matrix elements
\[
{\ST}_{i,j}=\mathrm{tr}(c^{}_{{}j,{}i}\,c^{}_{{}i,{}j}).
\]
These numbers depend only on the isomorphism classes of the simple objects 
$i,j$.  One shows that, as a consequence of the axioms,
$\mathrm{dim}(i)=\mathrm{dim}(i^*)\neq 0$, 
$\mathrm{dim}(\zero)=1$, and that the modular
matrix obeys ${\ST}_{i,j}={\ST}_{j,i}={\ST}_{i^*,j^*}$, 
${\ST}_{i,\zero}=\mathrm{dim}(i)$.

On the simple modules ${}i$, the isomorphism $\theta_{{}i}$
acts as a scalar $v_i\,{\neq}\, 0$ times the identity.

A {\em rank} of a modular category is a number
$\mathcal{D}\In \C$ such that 
         \[\mathcal{D}^2=\sum_{i\in I}\dim(i)^2. \]
In the following, we shall assume that a 
rank has been fixed. Related to the rank is the {\em charge}
$\kappa$ of the modular category. It is defined by the formula
\[
\kappa=\mathcal{D}^{-1}\sum_{j\in I} v_j^{-1}\mathrm{dim}(j)^2.
\]
The charge appears in the description of the framing anomaly.

Out of ${\ST}_{i,j}$, $v_i$, $\mathcal D$ one constructs a projective 
representation of the mapping class group of the torus,
see \ref{ss-moca} below.

\medskip
\noindent{\bf Examples.}
\begin{enumerate}
\item 
The category of finite dimensional modules over the group algebra
$\C[\Z_{2N}]$ has simple non-isomorphic objects labeled by $I\eq\Z_{2N}$, 
representing all isomorphism classes of simple objects. All objects
are direct sums of simple objects. This category can be
made into a modular category as follows. Let $\zeta\eq\exp{(\I\pi/2N)}$.
Let for each object $V$, $V^*$ be the dual
vector space to $V$ with action $g\,\alpha=\alpha\circ (-g)$, $g\In\Z_N$,
$\alpha\In V^*$. Thus $j^*\,{\simeq}\,{-}j$ for a simple object $j\In I$.
Let $d_V{:}\ V^*\OTimes V\To \C$, $b_V{:}\ \C\To V\OTimes V^*$, 
are the canonical homomorphisms (evaluation and its dual). For
simple objects let $\theta_j=v_j\,\id_j$ with $v_j=\zeta^{-j^2}$ and
$c_{j,k}=\zeta^{-jk}\,\id_{\C\otimes\C}$. These definitions are extended
to general objects by (bi)linearity with respect to direct sums. Then
\[
v_j=\zeta^{-j^2},\qquad \ST_{j,k}=\zeta^{-2jk},\qquad \mathcal{D}=\sqrt{2N},
\qquad\kappa={\rm e}^{\I\pi/4}.
\]
In particular $\mathrm{dim}(j)=1$ for all simple objects $j$. An equivalent
category appears in conformal field theory and is related to the free field
with values in a circle of radius $1/2N$.

\item
The ``purified'' category of representations of the quantum group $U_q(sl_2)$ 
with $q=\exp(\pi\I/(\ell{+}2))$, which is related to
the SU(2) WZW model at level $\ell$, has $\ell{+}1$ simple objects
${}0,\dots,{}\ell$ up to isomorphism. In this case we have 
\[
v_j=
{\rm e}^{-\pi\I\frac{j(j+2)}{2(\ell+2)}},
\qquad
{\ST}_{j,k}=
\frac
{\sin\left(\frac{\pi(j+1)(k+1)}{\ell+2}\right)}
{\sin\left(\frac\pi{\ell+2}\right)}\,,
\qquad
\mathcal{D}=\frac{\sqrt{\frac{\ell+2}2}}
{\sin\left(\frac{\pi}{\ell+2}\right)}\,,
\qquad
\kappa={\rm e}^{\frac{3\pi\I\ell}{4(\ell+2)}}.
\]
\end{enumerate}

\subsection{Spaces of morphisms}\label{ss-pairing}
Let $U,V,W$ be  objects of a modular category. Then we have linear
isomorphisms $\Hom(U\Otimes V,W)\to \Hom(U,W\Otimes V^*)$,
$\Hom(U\Otimes V,W)\to \Hom(V,U^*\Otimes W)$, given
by
\begin{equation}\label{e-iso}
\phi\mapsto (\phi\otimes\id_{V^*})\circ(\id_U\otimes b_V),
\qquad
\phi\mapsto (\id_{U^*}\otimes \phi)\circ(b_{U^*}\otimes\id_V),
\end{equation}
respectively. In particular we have an isomorphism
\[
\Hom(i_1\otimes\cdots\otimes i_n, j_1\otimes\cdots\otimes j_m)
\to
\Hom(\one,i_n^*\otimes\cdots\otimes i_1^*\otimes j_1\otimes\cdots\otimes j_m),
\]
for any simple objects $i_1,\dots,j_m$.

For $n{+}m\eq3$, it is then sufficient to consider the space 
\[
H^{i,j,k}=\Hom(\one,i\Otimes j\Otimes k).
\]
We have a non-degenerate pairing $\langle\ ,\ \rangle\,{:}\
H^{k^*,j^*,i^*}\otimes H^{i,j,k}\to \C$, given by
\[
\langle\phi,\psi\rangle=d_k\circ(\id\otimes d_j\otimes \id)
\circ(\id\otimes\id \otimes d_i\otimes\id\otimes \id)\circ(\phi\otimes\psi).
\]
It is useful to fix bases $(e_{\alpha}[ijk],\,\alpha\eq1,\dots,N^{i,j,k})$
of the spaces $H^{i,j,k}$ for $i,j,k$ simple objects, so that 
\begin{equation}\label{e-ortho}
\langle e_\alpha[k^*j^*i^*],e_\beta[ijk]\rangle
=\delta_{\alpha,\beta}.
\end{equation}
Graphically, a basis element $e_\alpha[ijk]$ is represented by a coupon
\\[-1.8em]
\[
\fmilleuno
\]
or in the simplified blackboard framing notation by a trivalent vertex with
a  label $\alpha$ drawn where the bottom of the
coupon should be. The relation $\Ref{e-ortho}$ may then be written as
\\[-2.1em]
\[
\fmilledue\alza{0.7cm}{\quad=\quad}
\fmilletre\alza{0.7cm}{\quad=\quad}
\fmillequattro\alza{0.7cm}{\quad=\, \delta_{\alpha,\beta}.}
\]
Using the isomorphisms above, the bilinear pairing may also be formulated as
a pairing between $H_{i,j}^k=\Hom({}i\Otimes {}j,{}k)$ 
and $H_{k}^{i,j}=\Hom({}k,{}i\Otimes {}j)$ given by the trace:
 \[\langle\phi,\psi\rangle=\mathrm{tr}(\phi\psi).\]

The dimensions (Verlinde numbers) of $\Hom(i\otimes j,k)$
will be denoted by $N_{i,j}^k$. Similar notations
are used for the other spaces. Thus $N^{i,j,k}=
\mathrm{dim}(H^{i,j,k})=N_{i^*}^{j,k}$ and so on.

The first computations with these bases are the following two 
identities. The first identity is 
\\[-2.5em]
\begin{equation}\label{e-Mozart}
\ftredici\alza{1.1cm}
{=\;\delta_{\alpha,\beta}\;\frac1{\mathrm{dim}(i)}}\quad\fquattordici
\alza{1.1cm},
\end{equation}
or $e_\alpha[ik^*j^*] e_\beta[jki^*]
=\delta_{\alpha,\beta}\dim(i)^{-1}\id_{{}i}$, 
if $e_\alpha[ik^*j^*]$, $e_\beta[jki^*]$ 
are identified via \Ref{e-iso} with basis elements of $H_{j,k}^i$,
$H_{i}^{j,k}$, respectively. The second identity is
\\[-1.5em]
\begin{equation}\label{e-Bizet}
\alza{.5cm}{\fundici}\alza{1.4cm}
{=\;\sum_{k\in I}\sum_{\alpha=1}^{N_{j,k^*,i}}
\;{\mathrm{dim}(k)}}\quad\fdodici
\alza{1.4cm},
\end{equation}
whose meaning is $\id_{{}i}\otimes\id_{{}j}=\sum_{k\in I,\alpha}
{\mathrm{dim}(k)}e_\alpha[jk^*i]e_\alpha[i^*kj^*]$,
where $e_\alpha[jk^*i]$ is identified with a basis element
of $H_k^{i,j}$ and $e_{\alpha}[i^*kj^*]$ with a basis element
of $H_{i,j}^k$.

The first identity can be obtained by noticing that both sides
of the equations are elements of the one-dimensional vector
space $\Hom({}i,{}i)= \C\,\mathrm{id}_{{}i}$, 
so they are proportional. The constant of
proportionality is computed by taking the trace on both sides.

The second identity follows from the domination axiom
(Appendix \ref{a-mod}, (xi)) which implies
that the left-hand side can be expressed as a linear combination
of the morphisms on the right-hand side, possibly with different
basis elements $e_\alpha,e_\beta$ in the factors. To compute the
coefficients $c_{\alpha,\beta}(k)$ in
$\id_{{}i}\otimes\id_{{}j}=\sum_{k\in I,\alpha,\beta}
c_{\alpha,\beta}(k)e_\alpha[jk^*i]e_\beta[i^*kj^*]$,
we compose both sides of the equation with $e_\gamma[i^*lj^*]
\in H_{i,j}^l$. Since $\Hom({}k,{}l)=0$ for $l\neq k$,
the only term contributing to the sum over $k$ is the one with $k=l$.
Using the first identity, we get 
$e_\gamma\eq\sum_{\alpha,\beta} c_{\alpha\beta}(l)
\delta_{\alpha\gamma}\dim(l)^{-1}e_\beta$, from which the result 
$c_{\alpha,\beta}(l)\eq\mathrm{dim}(l)\delta_{\alpha,\beta}$ follows.
\subsection{Frobenius--Schur indicators}\label{ss-fs}

A {\em self-dual object} in a modular category is an object
isomorphic to its dual object. To each simple self-dual object one
associates a scalar squaring to one, called the
Frobenius--Schur indicator. Its role in conformal field theory was
first emphasized in \cite{B}. It is a generalization of a classical notion in
group representation theory: if $V$ is an irreducible representation
of a group $G$ then there is an action of $\Z_2$ on
the space of invariants $(V\Otimes V)^G$ by permutation of factors.
This space is at most one-dimensional. If it is non-trivial, the generator
of $\Z_2$ acts by multiplication by the Frobenius--Schur indicator.

In the case of a general modular category, the Frobenius--Schur indicator
is defined as follows. Suppose $V$ is a simple self-dual
object and $\phi{:}\ V^*\To V$ is an isomorphism. Then the
Frobenius--Schur indicator is the factor
of proportionality $\nu(V)$ in the identity
\[
(\theta_{V^*}\Otimes \id_V)\,c_{V,V^*}\,b_V=\nu(V)\,(\phi^{-1}\otimes\phi)\, b_V,
\]
between two non-zero elements of the one-dimensional space
$\mathrm{Hom}(\one,V^*\Otimes V)$. Since $V$ is simple, $\phi$ is unique
up to non-zero scalar, so $\nu(V)$ is
independent of the choice of isomorphism $\phi$.

\begin{lemma} Let $V$ be a simple self-dual object.
Then
\begin{enumerate} 
\item[(i)]  $\nu(\one)=1$.
\item[(ii)] $\nu(V)^2=1$.
\item[(iii)]
If $V$ is isomorphic to $W$ then $W$ is simple and self-dual
and $\nu(V)\eq\nu(W)$. In particular $\nu(V^*)\eq\nu(V)$.
\end{enumerate}
\end{lemma}

\noindent{\it Proof:}
(i) follows from $\theta_\one=\id_\one$, $c_{\one,\one}=\id_\one\Otimes\id_\one$.
To prove (ii), let us act on the equation defining $\nu$ with the morphism
$(\theta_V\Otimes\id_{V^*})c_{V^*,V}$. By the naturality of the braiding,
the left-hand side becomes $c_{V^*,V}c_{V,V^*}(\theta_{V}\Otimes\theta_{V^*})b_V$,
which, by the twist axiom (Appendix \ref{a-mod}, (iv)), is equal
to $\theta_{V\otimes V^*}b_{V}=b_V\theta_\one=b_V$. The right-hand side is
\begin{eqnarray*}
(\theta_V\Otimes\id_{V^*})\,c_{V^*,V}\,\nu(V)\,(\phi^{-1}\Otimes\phi)\, b_V 
 &=&\nu(V)\,(\phi\Otimes\phi^{-1})\,(\theta_{V}\Otimes \id_{V^*})\,c_{V,V^*}\,b_V\\
&=&\nu(V)^2 \,b_V.
\end{eqnarray*}
Thus $\nu(V)^2=1$. 

To prove (iii), let us introduce the dual morphism $f^*{:}\ W^*\To V^*$ 
of a morphism $f{:}\ V\To W$: $f^*\eq(d_W\Otimes\id)(\id\Otimes f\Otimes 
\id)(\id\Otimes b_V)$. It is easy to see (by drawing the corresponding 
graphs) that $\id_{V}^*=\id_{V^*}$ and that $(fg)^*=g^*f^*$, whenever the 
composition of the morphisms $f$, $g$ is defined.
In particular $f^*$ is an isomorphism if and only if $f$ is an isomorphism.
Moreover $b_W=f\Otimes (f^{*})^{-1}\,b_V$ if $f{:}\ V\To W$ is an isomorphism.
Using the naturality of twist and braiding, we act by $(f^*)^{-1}\Otimes f$
on the equation defining $\nu(V)$ and get 
\begin{eqnarray*}
(\theta_{W^*}\Otimes \id_W)c_{W,W^*}b_W
&=&
\nu(V)\,({f^*}^{-1}\phi^{-1}\Otimes f\phi)\, b_V\\
&=&
\nu(V) \left((f\phi f^*)^{-1}\Otimes f\phi f^{*}\right) b_W,
\end{eqnarray*}
Since $f\phi f^*{:}\ W^*\To W$ is an isomorphism, it follows that $W$ is 
self-dual (it is clear that $W$ is simple) and that $\nu(W)\eq\nu(V)$.
$\square$

\subsection{Fusing matrices (6j-symbols)}
A modular category is in principle determined by a set of numerical
data. Additionally to  the modular  matrix and the scalars
$v_i$, one has to specify the 6j-symbols. Let $i,j,k,l,m,n$ be simple 
objects (in the applications they are
either elements of $I$ or duals of elements of $I$). Then
one shows that the linear homomorphisms
\[
\Phi:\quad\bigoplus_{l\in I}
H_{l}^{i,j}\otimes H_{m}^{l,k}\to \Hom({}m,{}i\Otimes {}j\Otimes {}k),
\]
given by $\phi\otimes\psi\mapsto (\phi\Otimes\id_{{}k})\circ\psi$, and 
\[
\Phi':\quad\bigoplus_{n\in I}
H_{n}^{j,k}\otimes H_{m}^{i,n}\to \Hom({}m,{}i\Otimes {}j\Otimes {}k),
\]
given by $\phi\otimes\psi\mapsto (\id_{{}i}\Otimes\phi)\circ\psi$,
are isomorphisms. Therefore we have an isomorphism $(\Phi')^{-1}\circ
\Phi$. Its components are the {\em 6j-symbols}:
\[
\sixj ijklmn{}{}:\quad
H_{l}^{i,j}\otimes H_{m}^{l,k}\,\to\, H_{n}^{j,k}\otimes H_{m}^{i,n}.
\]
The matrix elements of the 6j-symbols with respect to the above bases
are defined by
\[
\sixj ijklmn{}{} e_{\alpha}[l^*ij] \otimes e_\beta[m^*lk]
=
\sum_{\gamma,\delta} \sixj ijklmn{\alpha\beta}{\gamma\delta}
e_{\gamma}[n^*jk]\otimes e_\delta[m^*in].
\]
Graphically, these matrix elements are defined by
\\[-1.7em]
\[
\fduecentouno
\;\alza{1.5cm}{=\;\sum_{n,\gamma,\delta}}
\quad\fduecentodue
\;\alza{1.5cm}{\sixj ijlkmn{\alpha\beta}{\gamma\delta}.}
\]
We also introduce inverse 6j-symbols as the components of the
inverse map $\Phi^{-1}\circ\Phi'$:
\\[-1.5em]
\[
\fduecentodue\;\alza{1.5cm}{\,=\,\sum_{l,\alpha,\beta}}
\quad\fduecentouno
\;\alza{1.5cm}{\barsixj ijlkmn{\alpha\beta}{\gamma\delta}\ .}
\]

\subsection{The 3-dimensional topological field theory}\label{ss-mf}
To every modular category there is an associated 3-dimensional topological
field theory. It is a formalization and generalization
of the Chern-Simons path integral of \cite{W}. The TFT associates 
a finite dimensional vector space $\HH(X)$ (the space of conformal
blocks) to each surface $X$ with marked points and some additional structure,
and an element of $\HH(X)$ to each 3-dimensional manifold
with a graph of Wilson lines bounding $X$.

To deal properly with the ``framing anomaly'', we need to endow
surfaces with additional structures and use ribbons 
instead of Wilson lines. We start by introducing the definitions,
following \cite{T}.

An  {\em extended surface} is an oriented closed 2-manifold $X$
with a finite set of disjoint oriented embedded arcs labeled by 
simple objects,  and a lagrangian subspace 
$\lambda(X)$ of the first homology
group $H_1(X,\R)$. A  homeomorphism of extended surfaces
$f{:}\ X\To Y$ is an orientation preserving  homeomorphism 
mapping arcs to arcs with the same label and
the same orientation. A homeomorphism $f{:}\ X\to Y$ of extended 
surfaces will be called {\em strong} if it also maps $\lambda(X)$ to
$\lambda(Y)$\,\footnote{~As we rarely use strong homeomorphisms we 
departed here slightly from the notation of
\cite{T}: there a homeomorphism is called weak
homeomorphism and a strong homeomorphism is called homeomorphism.}.
The {\em opposite} $-X$ of an extended surface is the surface
$X$ with opposite orientation and the same arcs, 
so that if an arc of $X$ is labeled by a simple object $i$ then it has 
opposite orientation and it is labeled by $i^*$ in $-X$.

\medskip

A {\em cobordism of extended surfaces} is a triple 
$(M,\partial_-M,\partial_+M)$ such that:
\begin{enumerate}
\item
$M$ is a 3-dimensional manifold with boundary containing a
ribbon graph\,\footnote{~Instead of ribbon graphs one often considers framed graphs,
whose edges (assumed to be smoothly embedded in a smooth manifold)
come with a normal vector field. A ribbon graph can be made into a framed graph
by taking a vector field normal to the ribbons. The present approach \cite{T} works
also in the topological category.}.
 A ribbon graph consists of ribbons, annuli and coupons 
as in \ref{ss-21}, but the ribbons and annuli are labeled by
simple objects only. Ribbons ends are glued to coupons or are contained in
the boundary $\partial M$.
\item
$\partial_\pm M$ are disjoint disconnected subsets of the boundary
$\partial M$ so that $\partial M=\partial_+M\cup(-\partial_-M)$,
endowed with lagrangian subspaces of their first homology groups. The 
marked arcs at which the ribbons in $M$ end are given the label of the
ribbons whose core is oriented inwards, and the dual label otherwise. The
lagrangian subspaces and the oriented labeled arcs
give $\partial_\pm M$ the structure of extended surfaces.
\end{enumerate}

We say that $(M,\partial_-M,\partial_+M)$
is a cobordism from $\partial_-M$ to
$\partial_+M$.

\medskip

The TFT\,\footnote{~The notation in \cite{T} for this TFT is 
$(\tau^e,\mathcal{T}^e)$} $( Z,\HH)$ associated to a modular category 
$\M$ over $\C$ consists of the following data.
\begin{enumerate}
\item[(i)]
For each extended surface $X$ there is a finite dimensional complex 
vector space $\HH(X)$, the space of states (or of conformal blocks), 
such that $\HH(\emptyset)=\C$ and $\HH(X\sqcup Y)=\HH(X)\otimes \HH(Y)$.
\item[(ii)]
To each homeomorphism of extended surfaces $f{:}\ X\To Y$ there is
an isomorphism $f_\sharp{:}\ \HH(X)\To\HH(Y)$.
\item[(iii)]
If $(M,\partial_-M,\partial_+M)$ is a cobordism of extended surfaces,
then the TFT associates to it a homomorphism
\[
 Z(M,\partial_-M,\partial_+M):\quad\HH(\partial_-M)\to\HH(\partial_+M)
\]
depending linearly on the labels of the coupons.
\end{enumerate}

These data obey the following axioms.
\begin{enumerate}
\item (Naturality) Let $(M,\partial_-M,\partial_+M)$, 
$(N,\partial_-N,\partial_+N)$ be cobordisms of extended surfaces. 
Let $f{:}\ M\to N$ be a degree one homeomorphism mapping the ribbon graph 
in $M$ onto the ribbon graph in $N$, restricting
to homeomorphisms $f_\pm{:}\ \partial_\pm M\To
\partial_\pm N$ preserving the lagrangian subspaces. Then
\[
(f_+)_\sharp\circ Z(M,\partial_-M,\partial_+M)
= Z(N,\partial_-N,\partial_+N) \circ(f_-)_\sharp
\]
\item (Multiplicativity)
If $M_1,M_2$ are two cobordisms of extended surfaces, then under the 
identification $\HH(\partial_\pm M_1\sqcup\partial_\pm M_2)
=\HH(\partial_\pm M_1)\otimes \HH(\partial_\pm M_2)$ we have 
$ Z(M_1\sqcup M_2)= Z(M_1)\Otimes Z(M_2)$.
\item (Functoriality)
Suppose a cobordism $M$ is obtained from the disjoint union of $M_1$ and 
$M_2$ by gluing $\partial_+M_1$ to $\partial_-M_2$ along 
a degree one homeomorphism $f{:}\ \partial_+M_1\to\partial_-M_2$ preserving
marked arcs with their orientation. Then
\[
 Z(M,\partial_-M_1,\partial_+M_2)
=\kappa^m\,
 Z(M_2,\partial_-M_2,\partial_+M_2)
\circ f_\sharp\circ
 Z(M_1,\partial_-M_1,\partial_+M_1),
\]
for some integer $m$.
\item (Normalization)
Let $X$ be an extended surface. Let the {\em cylinder over $X$} be the
3-manifold $X\times[-1,1]$, with the ribbon graph 
consisting of the ribbons $z\times [-1,1]$, where $z$ runs over the 
marked arcs of $X$.  Their orientation is such that they induce
the orientation of the arcs on $X\Times \{1\}$.
Their core is oriented from $1$ to $-1$. Then
\begin{equation}\label{e-Strauss}
 Z(X\Times [-1,1],X\Times \{-1\}, X\Times \{1\})=\id_{\HH(X)}.
\end{equation}
\end{enumerate}

The homomorphism $ Z(M,\partial_-M,\partial_+M)$ is called the {\em invariant}
of the cobordism of extended surfaces $(M,\partial_-M,\partial_+M)$.
By the naturality axiom it is invariant under degree one homeomorphisms 
that restrict to the identity on the boundary.

Moreover the invariant does not change if we remove an edge with
label $\one$ or we replace the label of an edge by its dual and
reverse the orientation of its core.

The TFT gives the system of vector spaces $\HH(X)$
the structure of a {\em modular functor}.

Next we list some of the properties of the modular functor $\HH$ which
we shall use.

\subsubsection{Duality}
The space $\HH(-X)$ associated to the opposite of the extended 
surface $X$ is canonically isomorphic to the dual space to $\HH(X)$. The 
isomorphism is induced by the pairing
\begin{equation}\label{e-Vienna1}
 Z(X\Times [-1,1],X\sqcup(- X),\emptyset):\quad
\HH(X)\otimes\HH(-X)\to \C.
\end{equation}
Here $X$ is identified with $X\Times\{1\}$ and $-X$ with $X\Times \{-1\}$.

\subsubsection{Mapping class group}\label{ss-moca}
The action $f\mapsto f_\sharp$ of homeomorphisms may be expressed in 
terms of the TFT.
Namely, let $f{:}\ X\to Y$ be a homeomorphism of extended surfaces. Then
the 3-manifold obtained by gluing the cylinder over $X$ to the cylinder
over $Y$ defines a cobordism $(M_f,X,Y)$.
The normalization and functoriality axioms then imply that $f_\sharp
\eq Z(M_f,X,Y)$. Moreover, it can be shown, using the naturality axiom,
that if $f,g$ are homotopic in the class
of homeomorphism of extended surfaces, then
$f_\sharp\eq g_\sharp$. In particular, if $X\eq Y$,
$f\mapsto f_\sharp$ defines a projective
representation of the mapping class group of $X$.

\medskip

\noindent{\bf Example.} Let $X$ be a torus with no marked arcs. View
$X$ as the boundary of a solid torus $H=D^2\times S^1$ and take $\lambda=
\lambda(X)$
to be the kernel of the map induced by the inclusion $X\hookrightarrow H$.
Then a basis of $\HH(X)$ is given by
         \[\chi_j^{}(X)= Z((H,j),\emptyset,X),\]
$j\In I$ where $(H,j) $ is $H$ with a ribbon graph consisting of an untwisted
annulus $[-\epsilon,\epsilon]\times S^1$ with label $j$. 
Let $S(z,w)\eq(w^{-1},z)$, $T(z,w)\eq(zw,z)$ be the standard generators of the
mapping class group $\mathrm{SL}(2,\Z)$ 
of $X=S^1\Times S^1$. Then $S_\sharp$, $T_\sharp$ are
represented in this basis by the matrices $\SR\eq
(\mathcal{D}^{-1}{\ST}_{i,j})$ and $\T\eq (v_i^{-1}\delta_{i,j})$ respectively. 
The map $f\,{\mapsto}\,f_\sharp$ is a projective representation of the mapping 
class group: the matrices $\SR$ and $\T$ obey the relations
$\SR^4\eq1$, $(\T\SR)^3\eq\kappa \SR^2$. Moreover $\SR^2$ is the matrix 
$(\delta_{i,j^*})$. Note that if we choose a third root of $\kappa$,
a genuine representation may be obtained
by replacing $\T$ by $T\eq(\kappa^{-1/3}v_i^{-1} \delta_{i,j})$.

\subsubsection{Gluing homomorphisms}\label{sss-gl}
If $X$ is an extended surface with arcs $\gamma$, $\gamma'$ labeled
by $i$, $i^*$, let $X'$ be an extended surface obtained as follows: 
Let $\phi,\phi'{:}\ D^2\to X$ be orientation preserving 
disjoint embeddings of the unit disk $D^2\,{\subset}\,\C$ such that
their restriction to $[-1,1]$ are parametrizations of the oriented
arcs $\gamma,\gamma'$. Then $X'$ is obtained from
$X$ by removing the interiors of the disks $\phi(D^2)$, $\phi'(D^2)$ and gluing
their boundaries by identifying $\phi(z)$ with $\phi'(-\bar z)$, for $z\In S^1$.
The arcs of $X'$ are the remaining arcs of $X$ and the lagrangian subspace 
$\lambda(X')$ consists of images in $X'$ of homology classes of cycles in 
$X-(\phi(D^2)\cup\phi'(D^2))$ which are
mapped by the inclusion to cycles in $\lambda(X)$.

Then one has a {\em gluing homomorphism} $g_{X,X'}{:}\ \mathcal{H}(X)\To
\mathcal{H}(X')$.

The gluing homomorphism is obtained from
the TFT: Let $M$ be the 3-manifold obtained from
$X\Times [-1,1]$,  the cylinder over $X$, by gluing $\phi(D^2)\Times \{1\}$
to $\phi'(D^2)\Times \{1\}$ via the identification
$(\phi(z),1)=(\phi'(-\bar z),1)$, $z\In D^2$. Let the
ribbon graph in $M$ be obtained from the
ribbon graph in $X\Times [-1,1]$ by replacing the part of the
ribbon through the glued disks by a narrower one, so as to fit inside $M$.
Then $M$ has boundary $-X\sqcup X'$ and defines a cobordism of extended 
surfaces from $X$ to $X'$. Then
\[
g_{X,X'}=Z(M,X,X').
\]
It is then known that the gluing homomorphism
has the following completeness property.

If $X_j$ is the extended surface $X$ as above, but with $\gamma$ labeled
by $j$ and $\gamma'$ labeled by $j^*$, then the sum of gluing homomorphisms
\[
\bigoplus_{j\in I}\mathcal{H}(X_j)\to \mathcal{H}(X')
\]
is an isomorphism.

\subsubsection{Description of $\HH(X)$ as a vector space}
Let $X$ be a 2-sphere with $n$ marked arcs
labeled by simple objects $j_1,\dots,j_n$. Let $M$ be
a 3-ball with boundary $X$ and a ribbon graph
consisting of one vertex connected to the
marked arcs by $n$ ribbons. Then $Z(M,\emptyset,X)$
depends linearly on the label of the vertex and is thus a linear map
\[
Z(M,\emptyset,X):\quad\mathrm{Hom}(\one,{}{j_1}\Otimes
\cdots\Otimes {}{j_n})\to \HH(X)
\]
By construction of the TFT out of the modular category, this map is
an isomorphism. Combining this result with the
completeness of the gluing map, one deduces, by
attaching handles to the sphere, that
$\HH(X)\simeq \oplus_{k_1,\dots,k_g\in I}
\mathrm{Hom}(\one,{}{j_1}\otimes
\cdots\otimes {}{j_n}\otimes \,\otimes_{s=1}^g(k_s\otimes {}{k_s^*}))$ 
for a surface of genus $g$ with $n$ marked points. 
Under this identification, the invariant of a cobordism changes
in a covariant way if we replace the labels of the edges by 
equivalent simple objects.

\subsection{Trace formula}
One important variant of the functoriality axiom is a formula
\cite{T} for the invariant of a closed 3-manifold of the form $M=X\Times S^1$,
for a closed oriented surface $X$, with some ribbon graph $\Gamma$. 
We may obtain $M$ by gluing the two components of the
boundary of $N=X\Times [0,1]$. Then the ribbon graph intersects
the boundary along arcs and $\partial_+N=X\Times \{1\}$ becomes an
extended surface. $\partial_-N=X\Times\{0\}$ is then canonically strongly
homeomorphic to the same extended surface, and the {\em trace formula} holds:
\begin{equation}\label{e-traceformula}
Z(M,\emptyset,\emptyset)=\mathrm{Tr}_{\HH(\partial_+N)}(Z(N,\partial_-N,\partial_+N)).
\end{equation}

\subsection{Framing anomaly}
We give here the formula for the integer
$m$ appearing in the functoriality axiom, following Sect.\ IV.7 of \cite{T}.
It is given in terms of Maslov indices, which we proceed to define.

Let $H$ be a symplectic real vector space with
symplectic form $\omega$, and $\lambda_1,\lambda_2,
\lambda_3\subset H$ be lagrangian subspaces.
Then on the subspace $(\lambda_1+\lambda_2)\cap
\lambda_3$ we have a quadratic form $Q(x)=\omega(x_2,x)$, where 
$x=x_1+x_2$ with $x_1\in\lambda_1$, $x_2\in \lambda_2$ ($Q(x)$
does not depend on the choice of the decomposition
of $x$). The {\em Maslov index} $\mu(\lambda_1,\lambda_2,\lambda_3)$
is by definition  the signature of $Q$. It is 
          a 
function which is antisymmetric under permutations of its three
arguments, and in particular vanishes if any two arguments coincide.

If $X$ is an oriented closed
 2-manifold, the intersection form
on  $H_1(X,\R) $ is symplectic. Moreover, if $M$ is a 3-manifold
with boundary and $\partial M=\partial_+M\sqcup\partial_-M$ is
a decomposition of the boundary into closed disjoint subsets,
then we have a map $N_*$ from the set of the lagrangian subspaces
of $H_1(\partial_-M,\R)$ to the set of lagrangian subspaces of
$H_1(\partial_+M,\R)$: $x\in N_*(\lambda)$ if and only if there
exists an $x'\In\lambda$ so that $x-x'$ is homologous to zero as
a cycle in $M$. Similarly we have a map $N^*$ sending lagrangian
subspaces of $H_1(\partial_+M,\R)$ to the lagrangian subspaces of
$H_1(\partial_-M,\R)$. Then the integer $m$ appearing in the 
functoriality property is
\[
m=\mu(f_*N_*\lambda(\partial_-M_1),f_*\lambda(\partial_+M_1),
N^*\lambda(\partial_+M_2))
+ \mu(f_*\lambda(\partial_+M),\lambda(\partial_-M_2),N^*\lambda
(\partial_+M_2)).
\]
The following property is useful for surfaces with orientation
reversing involutions, such as doubles.

\begin{lemma}\label{l-maslov}
Let $H$ be a real symplectic vector space with symplectic form
$\omega$. Suppose $\sigma\in\mathrm{End}_\R(H)$ is an involution
such that $\sigma^*\omega=-\omega$. If $\lambda_1,\lambda_2,\lambda_3$
are lagrangian subspaces invariant under $\sigma$, then
\[
\mu(\lambda_1,\lambda_2,\lambda_3)=0.
\]
\end{lemma}

\medskip

\noindent{\it Proof:}
Let $x$ be an element of the invariant
space $(\lambda_1+\lambda_2)\cap\lambda_3$. 
If $x=x_1+x_2$ with $x_i\in\lambda_i$, then $\sigma(x)
=\sigma(x_1)+\sigma(x_2)$ is a decomposition of $\sigma(x)$
into a sum of elements of $\lambda_1$, $\lambda_2$. Thus
\[
Q(\sigma(x))=\omega(\sigma(x_2),
\sigma(x))=-\omega(x_2,x)=-Q(x).
\]
On the other hand, the signature is an invariant,
so the signature of $Q$ is equal to the signature
of $Q\circ\sigma=-Q$. Thus the signature vanishes.
$\square$

\section{Boundary conditions and correlation functions}
\label{s-3}
\subsection{The double of a surface}
Suppose that $X$ is a two-dimensional compact manifold with boundary,
possibly non-orientable. Then $X$ may be identified with
$\hat X/\Z_2$ for a closed oriented manifold $\hat X$, the 
        {\em   double }
of $X$, with an orientation reversing action of the generator of 
$\Z_2=\Z/2\Z$.  
  The double
is constructed by taking the total space of the orientation
bundle $p{:}\ \mathrm{Or}(X)\to X$ (the $\Z_2$-bundle over $X$ whose
fiber at $x$ consists of the two orientations of the
tangent plane at $x$) and identifying the two points
of each fiber over the boundary: $\hat X= \mathrm{Or}(X){/}\!\!\sim$ 
with $x\,{\sim}\, x'$ iff $p(x)\eq p(x')\In\partial X$.
The double comes with a projection $p{:}\ \hat X\To X$ and an orientation 
reversing involution $\sigma{:}\ \hat X\To\hat X$ exchanging
the two sheets and defining the action of $\Z_2$.

Here are some examples. If $X$ is closed and orientable,
then $\hat X$ consists of two copies of $X$ with opposite
orientations. If $X$ is orientable with non-empty boundary, 
$\hat X$ is obtained by taking two copies of $X$ with opposite 
orientations and gluing the two copies along the boundary. So 
           if 
$X$ is a disk, then $\hat X$ can be
viewed as the unit sphere $S^2$ in $\R^3$ with $\Z_2$ action
generated by the reflection at the $x$-$y$ plane.
           If 
$X$ is the real projective plane $\R P^2$ then 
$\hat X$ is $S^2$ with $\Z_2$ action given by
the antipodal map $x\mapsto -x$ of $S^2\subset\R^3$. The
annulus $X=S^1\times [-1,1]$ is $\hat X/\Z_2$ with
$\hat X=S^1\times S^1$ and involution $(\theta_1,\theta_2)
\mapsto (\theta_1,-\theta_2)$, $\theta_i\In \R/2\pi\Z\eq S^1$.
Taking $\hat X=S^1\Times S^1$ with involution 
$(\theta_1,\theta_2)\mapsto (-\theta_1,\theta_2+\pi)$ gives
the Klein bottle. The involution
$(\theta_1,\theta_2)\mapsto (\theta_2,\theta_1)$ gives the M\"obius strip.

\subsection{The case of closed orientable surfaces}
Let us first consider the case of closed orientable surfaces.
Suppose $X$ is closed and orientable, and choose an orientation
of $X$. Then the double of $X$ is $\hat X=X\sqcup(-X)$, the disjoint
union of two copies of $X$ with opposite orientations. The
involution exchanges the two copies. Let $X$ be endowed with
$n$ distinct points $z_1,\dots,z_n$ on it, labeled by simple objects
$i_1,\dots, i_n$. To these data one associates
a correlation function $C(X)\in \HH(\hat X)$.

To be more precise, we should take care of the framing: so 
$z_1,\dots,z_n$ should be taken as disjoint arcs rather than points.
Also $\HH(\hat X)$ is only unambiguously defined if
$\hat X$ is given a lagrangian subspace $\lambda$ in its first 
homology group with real coefficients. 
As will be clear below, a convenient choice is to  take $\lambda$
to consist of $a\oplus(-a)\in H_1(\hat X,\R)=
          H_1(X,\R)\oplus H_1(X,\R)$, 
where $a$ runs over $H_1(X,\R)$.
We call this lagrangian subspace {\em canonical
lagrangian subspace} and denote it by $\lambda_-(\hat X)$.

The natural candidate for $C(X)$ is then
the element of $\HH(\hat X)$ associated to the
3-manifold $X\Times [-1,1]$, with ribbon graph consisting
of $z_i\Times [-1,1]$, where $z_i$ runs over the marked arcs on $X$:
\[
C(X)= Z(X\Times [-1,1],\emptyset,\hat X).
\]
See \ref{ss-mf} for the orientations of the ribbons
in $X\Times [-1,1]$. Note that the canonical lagrangian
subspace is the kernel of the map induced by inclusion of the boundary.

Let us check that this ansatz obeys the modular and factorization 
properties one expects for correlation functions.

Let $f{:}\ X\To X$ be a degree one homeomorphism
and let $\hat f{:}\ \hat X\to \hat X$ be equal to $f$ on each of the two 
copies of $X$.  The lift $\hat f$ of $f$ is the unique degree one
homeomorphism of $\hat X$ commuting with the involution and projecting to $f$.

\begin{thm}\label{t-closedmodular}
(Modular invariance)
Let $X$ be a closed oriented surface with labeled arcs $z_1,\dots,z_n$. 
Let $f{:}\ X\To X$ be a degree one homeomorphism preserving the marked arcs.
Let $\hat f$ be its lift to $\hat X$. Then
\[
\hat f_\sharp\, C(X)=C(X).
\]
\end{thm}

\medskip

\noindent{\it Proof:}
Let $F{:}\ X\Times [-1,1]\To X\Times [-1,1]$ be defined by $F(x,t)\eq(f(x),t)$.
$F$ is a homeomorphism that restricts to $\hat f$ on
the boundary. Moreover, $\hat f_*$ clearly preserves the 
lagrangian subspace. Thus the naturality axiom applies and we get
\[
\hat f_\sharp\,
 Z(X\times[-1,1],\emptyset,\hat X)
=
 Z(X\times[-1,1],\emptyset,\hat X),
\]
proving the claim.
$\square$

\medskip

It is useful to express the correlation function for 
more general lagrangian subspaces. Let us say that
a lagrangian subspace $\lambda$ of $H_1(\hat X,R)$ is {\em symmetric} if
$\sigma_*\lambda\eq\lambda$. The lagrangian subspace $\lambda_-(\hat X)$
has this property. Let us define, for any symmetric lagrangian subspace
$\lambda$, 
\[
C_\lambda(X)=(\mathrm{id}_{\lambda,\lambda_-})_\sharp\,C(X).
\]
Here $\mathrm{id}_{\lambda_1,\lambda_2}$ denotes the identity
map between extended 
surfaces which differ only in their distinguished lagrangian subspaces.
Then we get the more general modularity property:
\[
\hat f_\sharp\,C_\lambda(X)= C_{\lambda}(X).
\]
This formula follows from the functoriality formula, except
that we have to check that the framing anomaly term is trivial.
The reason for this is that all lagrangian subspaces appearing
in the calculation of the Maslov indices are invariant under $\sigma_*$.
Therefore the Maslov indices  vanish  by Lemma \ref{l-maslov}.

In particular we may choose the symmetric lagrangian subspace 
$\lambda\,{\oplus}\,\lambda\In H_1(X\Sqcup (-X),\linebreak[0]\R)=
H_1(X,\R)\,{\oplus}\, H_1(X,\R)$, for any lagrangian subspace $\lambda$ 
of $H_1(X,\R)$. Under these circumstances we may identify
$\HH(\hat X,\lambda\oplus \lambda)=\HH(X,\lambda)\Otimes\HH(-X,\lambda)$ and 
write 
\[
C_{\lambda\oplus\lambda}(X)=\sum_{j}b_j(X,\lambda)\otimes b_j(-X,\lambda),
\]
for any basis $b_j(X,\lambda)$ of $\HH(X,\lambda)$ and dual basis
$b_j(-X,\lambda)$ with respect to the pairing \Ref{e-Vienna1}.
In this form, the modular invariance is less apparent.

\medskip

\noindent{\bf Example.} Let $X\eq S^1\Times S^1$ 
be a torus with no marked arcs. Let $D^2\,{\subset}\,\C$ be
the unit disk and view $X$ as the boundary of a solid torus 
$H=D^2\times S^1$. Take $\lambda=\lambda(X)$
to be the kernel of the map induced by the inclusion $X\hookrightarrow H$.
Then a basis of $\HH(X)$ is given by $\chi_j^{}(X)= Z((H,j),\emptyset,X)$, $j\In I$, 
see \ref{ss-moca}.
Let $S(z,w)=(w^{-1},z)$, $T(z,w)=(zw,z)$ be the standard generators of the
mapping class group $\mathrm{SL}(2,\Z)$ 
of $X=S^1\Times S^1$. Then $S_\sharp$, $T_\sharp$ are
represented in this basis by the matrices $s\eq(\mathcal{D}^{-1}{\ST}_{i,j})$
and $t\eq(v_i^{-1}\delta_{i,j})$, respectively. 

The map $\theta(z,w)=(\bar z,w)$ is a homeomorphism $X\to -X$, preserving
orientation and lagrangian subspace. Therefore a basis of $\HH(-X)$ is 
$\chi_j^{}(-X)=\theta_\sharp\,\chi_{j^*}^{}(X)$. This basis is dual to the basis
$(\chi_j^{})_{j\in I}$, as the pairing between $\chi_j^{}(X)$ and
$\chi_{k}^{}(-X)$ is the invariant of $S^2\Times S^1$ with two annuli running
along $S^1$. This invariant is $\delta_{j,k}$, by the trace formula. Thus
\[
C_{\lambda\oplus\lambda}(X)=\sum_{j\in I}\chi_j^{}(X)\otimes \chi_j^{}(-X).
\]
As $S\circ\theta=\theta\circ S^{-1}$ and $T\circ\theta=\theta\circ T^{-1}$,
$S_\sharp$, $T_\sharp$ are represented in the basis $(\chi_{j}^{}(-X))_{j\in I}$
by the matrices $\SR_\theta=C\SR^{-1}C$, $\T_\theta=C\T^{-1}C$, 
with 
\begin{equation}\label{e-cc}
C_{i,j}=\left\{\begin{array}{rl} 1, & i\simeq j^*,\\ 0,& {\rm otherwise.}
\end{array}\right.
\end{equation}
 (The anomaly does not contribute
as in each Maslov index there are two coinciding arguments.)  
The statement of modular invariance thus reduces to
\[
{}^{\mathrm{t}}\T\;\T_\theta=
{}^{\mathrm{t}\!}\SR\,\SR_\theta=1.
\]
These identities may also be checked directly, using the relations of
$\SR$, $\T$ of \ref{ss-moca}, the fact that $\SR$ is a
symmetric matrix,
and the relations $v_i=v_{i^*}$, $\SR^2=C$.

\medskip

We turn to the factorization properties of correlation functions. Let 
$X$ be a closed oriented surface with marked oriented labeled arcs.
Suppose two of the marked arcs carry label $j$, $j^*$ respectively. Let 
$X'$ be obtained by removing from $X$ a small open disk
around each of the two arcs, and gluing
their boundaries by an orientation reversing
homeomorphism. We say that we obtain $X'$
by gluing $X$ at the two marked arcs. Then $\hat X'$ is obtained
from $\hat X$ by performing this operation twice, at each of the inverse 
images of the two arcs. Therefore we have a gluing homomorphism
$g_{\hat X,\hat X'}{:}\ \HH(\hat X)\To \HH(\hat X')$,
which is the composition of the two gluing homomorphisms (in either 
order) with $(\id_{\lambda',\lambda_-(\hat X')})_\sharp$. 
Here $\lambda'$ is the symmetric lagrangian subspace 
of $H_1(\hat X',\R)$ obtained form the canonical lagrangian subspace 
$\lambda_-(\hat X)$ of $\hat X$ by the gluing prescription
of \ref{ss-mf}, and $\id_{\lambda',\lambda_-(\hat X')}$ is the identity
map from $\hat X'$ with lagrangian subspace $\lambda'$ to $\hat X'$ with
canonical lagrangian subspace.

\medskip

\begin{thm}\label{t-closedfactorization}
(Factorization)
Let $X$ be as in Theorem \ref{t-closedmodular}.
Let $X'$ be obtained from $X$ by gluing $X$ at
two marked arcs with labels $j$, $j^*$. Let $X_j$ be the surface $X$,
          with $j\in I$ arbitrary, and with
all other labels fixed. Then
\[
C(X')=\sum_{j\in I}\mathcal{D}^{-1} \mathrm{dim}(j)\,
g_{\hat X_j,\hat X'}C(X_j).
\]
\end{thm}

\medskip

\noindent{\it Proof:}
Let $M_j=X_j\times[-1,1]$. Then $g_{\hat X_j,\hat X'}C(X_j)$
is, by \ref{sss-gl}, the invariant $ Z(M_j',\emptyset,\hat X')$ of
a cobordism. The 3-manifold $M_j'$ is obtained from $M_j$ by gluing two 
disks around the two chosen marked arcs on $X\Times \{1\}$, and also 
on $X\Times \{-1\}$. The ribbons ending at the marked arcs are then glued
together to form an annulus.

It is not too hard to see that $ X' \times [-1,1]$ is homeomorphic
to the manifold we obtain from $M_j'$ by performing a surgery at this 
annulus. The surgery is, by definition, the following construction:
we first parametrize a tubular neighborhood $U$
of the annulus by a homeomorphism $\phi{:}\ D^2\Times S^1\To M'_j$
in such a way that the annulus is contained in $\phi([-1,1]\Times S^1)$. 
Then we glue $S^1\Times D^2$ to $M_j'-\mathrm{int}(U)$ via   
the map $\phi$ restricted to $S^1\Times S^1$.
\begin{figure}
\scalebox{0.4}{\includegraphics{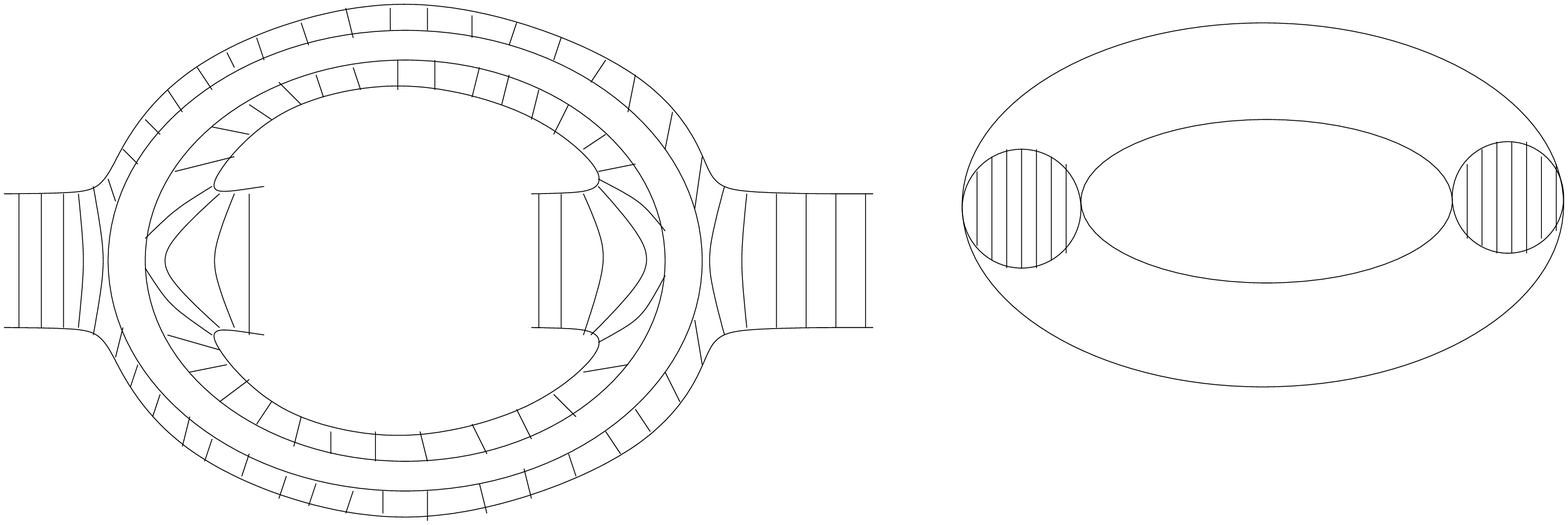}}
\caption{}\label{fig12}\end{figure}
Fig.~\ref{fig12} is supposed to illustrate
the homeomorphism of the resulting manifold
with $ X'\times[-1,1]$: on the left, we embed the
region of interest of $M_j'-\mathrm{int}(U)$ in $\R^3$
and on the right we draw $S^1\Times D^2$. The
fibers $\{x\}\Times [-1,1]$ in a plane section are drawn.

Let $M''\eq M_j'-\mathrm{int }(U)$ and $\chi_j^{}(S^1\Times S^1)
\eq Z((H,j),\emptyset,S^1\Times S^1)$, as in the example above. 
By the functoriality axiom,
\[
 Z(M_j',\emptyset,\hat X')=
 Z(M'',\partial U,\hat X')
\circ\phi_\sharp\,
\chi_j^{}(S^1\times S^1).
\]
The Maslov indices vanish by Lemma
\ref{l-maslov}. Indeed, the
involution of the double extends to an
involution of the boundary of $M''$ and
of $S^1\times S^1$ so
that  all lagrangian subspaces
involved are symmetric under the involution.

On the other hand, using the surgery presentation
of $X'\times [-1,1]$, we have
\begin{eqnarray*}
C(X')&=& Z(X'\Times [-1,1],\emptyset,\hat X')
        \\
     &=& Z(M'',\partial U,\hat X')
        \circ\phi_\sharp\circ S_\sharp\chi_\zero^{}(S^1\Times S^1).
\end{eqnarray*}
Again, the Maslov indices vanish by symmetry.
The claim then follows from the modular property
\[
S_\sharp\,\chi_\zero^{}(S^1\TImes S^1)=
\sum_{j\in I}{\SR}_{j,\zero}\,\chi_j^{}(S^1\TImes S^1).
\]
$\square$

\subsection{The case of the $(n,m)$-point function on the disk}
We  consider the case of the correlation function for $n$ interior
points and $m$ boundary points on the disk
(the $(n,m)$-point function). The double of the
disk is a 2-sphere $S^2$, which we view as the unit sphere in $\R^3$. 
The projection $p{:}\ S^2\To D^2$ is the orthogonal 
projection onto the $x$-$y$ plane and the involution $\sigma$
is the reflection at the $x$-$y$ plane.

Let $X$ be the disk $D^2$ with $n$ distinct labeled marked points
in its interior and $m$ labeled points on its boundary. Let the interior
points be labeled by $i_1,\dots,i_n$, and the boundary points, 
cyclically ordered along the orientation of the circle,
be labeled by $j_1,\dots,j_m$. The segment between the
$r$th and the $r{+}1$st point is given 
a boundary condition, a simple object $a_r$.

Moreover, the correlation function depends also
on an additional datum at each boundary point:
the correlation function $C(X)$ is a linear map
from a tensor product of multiplicity spaces $\bigotimes_{r=1}^m
W_{a_{r-1},a_r}(j_r)$ to $\HH(\hat X)$ (we set $a_0\eq a_m$).
 The presence of these multiplicity spaces reflects the
multiplicities of boundary fields. 
From the physical point of view, one understands these
multiplicities as a consequence of the
field-state correspondence of the conformal field theory,
implying that they can be read off the annulus multiplicities, which for
the boundary conditions of our interest coincide with
fusion rule coefficients. 

Accordingly, we assume 
that the multiplicity spaces are identified
with the space of conformal blocks on the sphere with three points:
\[
W_{a,b}(j)=\Hom({}b,{}j\Otimes {}a).
\]
Using the identification $\Hom({}b,{}j\Otimes{}a)\eq\Hom(\one,{}b^*
\Otimes{}j\Otimes{}a)$, we have a basis $(e_\alpha[b^*ja]$,
$\alpha\eq1,\dots,N_{b}^{j,a})$ of each multiplicity space obeying
the orthogonality properties of \ref{ss-pairing}.

Our ansatz for the $C(X)$ evaluated on a product $e_{\alpha_1}\Otimes
\cdots\Otimes e_{\alpha_m}$ of basis elements is the element of 
$\HH(\hat X)$ associated to the graph in the 3-ball in Fig.~\ref{fig1}. 

To give the precise definition we should again
take the framing into account. So the marked interior
points on $D^2$ should be taken to be disjoint oriented arcs. The boundary 
points are replaced by arcs on the boundary, oriented along the orientation 
of the boundary. The modular invariance property proven below will 
imply that these choices are irrelevant up to canonical isomorphisms.

The graph in Fig.~\ref{fig1} is made into a ribbon graph
as follows: the vertical lines are the cores of ribbons whose
sides are the inverse images by $p$ of the marked
arcs and vertical lines connecting the endpoints
of the marked arcs. The orientation of the ribbons
is chosen so that they induce the orientation of the arcs on the 
upper hemisphere. The part of the graph connected to the equator
is the core of an annulus lying in the $x$-$y$ plane to which ribbons 
also lying in the $x$-$y$ plane are glued along a side. The
opposite sides of  these ribbons coincide with the
marked arcs on the boundary. The orientation
of the ribbon graph is such that it induces
the orientation of the boundary arcs.

Let us call {\em equatorial graph} the component of
the ribbon graph connected to the equator. The remaining
components we call {\em vertical ribbons}.

Let $M$ be the unit ball with this ribbon graph. Then we set 
\[
C(X)= Z(M,\emptyset,\hat X)\in
\Hom_\C(\otimes_{r=1}^m W_{a_{r-1},a_r}(j_r), \HH(\hat X)).
\]
In this formula we regard $ Z(M,\emptyset,\hat X)$
as a multilinear function of the labelings of vertices.

Let us check that this ansatz is modular invariant. 
If $f{:}\ D^2\To D^2$ is a degree one homeomorphism of the disk 
preserving the marked arcs, then there exists a unique
degree one homeomorphism $\hat f$ of $S^2$, the {\em lift}
of $f$, so that $p\circ \hat f\eq p\circ f$.
It preserves the inverse images of the arcs.

\begin{thm}\label{t-diskmodular}
(Modular invariance)
Let $X$ be a disk with $n$ marked arcs
in the interior and $m$ marked arcs on the boundary. Let
$f{:}\ X\To X$ be a degree one homeomorphism
preserving the marked arcs and their orientations.
Let $\hat f$ be its lift to $\hat X$. Then
\[
\hat f_\sharp\, C(X)=C(X).
\]
\end{thm}

\medskip

\noindent{\it Proof:}
Denote by $x,y,z$ the standard coordinates of $\R^3$. View $D^2$ as 
the unit disk in the $x$-$y$ plane. Let $F(x,y,z)=(f(x,y),r(x,y)z)$,
with $r(x,y)=(1-x^2-y^2)^{-1/2}(1-|f(x,y)|^2)^{1/2}$.
Then $F$ is a degree one homeomorphism of the ball 
$D^3$ whose restriction to $S^2$ is 
$\hat f$. The image of the ribbon graph $\gamma$ in $D^3$ by
$F$ is isotopic to $\gamma$.  Therefore, by the naturality axiom,
$\hat f_\sharp\, Z(M,\emptyset,\hat X)= Z(M,\emptyset,\hat X)$. 
$\square$

\medskip

We turn to the factorization properties. In the case of surfaces 
with boundary there are two kinds of factorizations. One may either
cut the surface along a loop in the interior,
as in the case of closed surfaces, or along
a path joining points on the boundary.
In the first case, factorization is analyzed analogously
as for closed surfaces. Thus we consider here only the latter possibility.

Let $X'$, $X''$ be two oriented disks with marked labeled arcs as 
above.  Let $n',n''$ denote the numbers of interior arcs and $m',m''$ 
the numbers of boundary arcs. Suppose that a marked arc $x'$ on the 
boundary of $X'$ has label $j$ and that a marked arc
$x''$ on the boundary of $X''$ has dual label $j^*$. Then we may
glue $X'$ to $X''$ along an orientation
reversing homeomorphism from $x'$ to $x''$.
We assume that the labels of the boundary
segments on the sides of these two arcs
match, so that the gluing results in a
disk $X$ with $n'+n''$ interior marked arcs
and $m'+m''-2$ boundary marked arcs. We
say that $X$ is obtained by gluing $X'$, $X''$ at $x'$, $x''$.
Then the double $\hat X$ may be obtained by
gluing the extended surface $\hat X'\sqcup \hat X''$ 
at the inverse image of $x'$, $x''$, and we have a gluing homomorphism
$g_{\hat X'\sqcup \hat X'',\hat X}{:}\ \HH(\hat X')
\Otimes\HH(\hat X'') \To \HH(\hat X)$.

We need not care about lagrangian subspaces
since the first homology groups are trivial in this case.

\begin{thm}\label{t-diskfactorization}
(Factorization)
Let $X$ be a disk obtained by gluing disks $X'$, $X''$ at the marked arcs
$x'$, $x''$ with labels $j$, $j^*$. Let the labels
of the boundary segment preceding and following $x'$ be $a$, $b$ 
respectively. Let $X'_j$ be the surface $X'$, with $j$ running
over $I$ with all other labels fixed. Similarly, let $X''_{j^*}$ 
be the surface $X''$, with $j$ running
over $I$ with all other labels fixed. 
Let us order the boundary arcs in a way compatible
with the cyclic ordering, so that $x'$ is the last
arc of $X'$ and $x''$ is the first arc of $X''$.
Then for any choice of basis $e_\alpha[b^*ja]$ of $H^{b^*,j,a}$
and dual basis $e_\alpha[a^*j^*b]$ of $H_{a^*,j^*,b}$ as in 
\ref{ss-pairing},
\begin{eqnarray*}
\lefteqn{C(X)(u_1\Otimes\cdots\Otimes u_{m'+m''-2})}
\\[2mm]&=&\sum_{j\in I,\alpha}
\mathrm{dim}(j)\, g_{\hat X'_j\sqcup \hat X''_{j^*},\hat X}
\bigl[
C(X'_j)(u_1\Otimes\cdots\Otimes u_{m'-1}\Otimes e_\alpha[b^*ja])
\\
&&\hspace{2.5cm} \otimes \,C(X''_{j^*})(e_\alpha[a^*j^*b]\Otimes 
u_{m'}\Otimes \cdots\Otimes u_{m'+m''-2})\bigr].
\end{eqnarray*}
\end{thm}

\medskip

\noindent{\it Proof:}
By the gluing construction (see \ref{sss-gl}), the summand labeled by
$j,\alpha$ on the right-hand side is $\mathrm{dim}(j)
\, Z(M_{j,\alpha},\emptyset,\hat X)$, where $M_{j,\alpha}$ is a ball 
obtained by gluing  two balls defining $C(X_j')$ 
and $C(X_{j^*}'')$. $M_{j,\alpha}$ contains a ribbon
graph $\gamma_{j,\alpha}$ obtained by the gluing prescription. 
Thus $\gamma_{j,\alpha}$ has vertical lines
and a piece lying in the $x$-$y$ plane obtained by gluing two 
equatorial graphs. We have to show that the sum over 
the labelings $j$ and $\alpha$  of the invariant of the cobordism with this 
graph gives the same as if we replace it by an equatorial graph.

In the vicinity of the point at which the gluing was performed the
ribbon graph looks like
\\[-2.7em]
\[
\fvuno
\]
Applying \Ref{e-Bizet}, after summing over $j$ and $\alpha$,
we can replace this part of the graph by two horizontal bands. In
this way $\gamma_{j,\alpha}$ is replaced by  the equatorial graph 
appearing on the left-hand side as desired.
$\square$

\subsection{The general case}

We turn now to the general case of a compact surface, possibly with
boundary, possibly non-orientable. There is a subtlety that arises
when one considers non-orientable surfaces. Namely a label of a
point by a simple object is only defined if one chooses a local
orientation. This is formalized by the following definitions of 
labeled surfaces and their doubles. The correlation functions will then 
be defined for {\em labeled surfaces} and they will take
values in the space of states of their doubles,
which are {\em extended surfaces} (see \ref{ss-mf}).

\medskip

\subsubsection{Labeled surfaces and their doubles}
A {\em labeled surface} is a compact
2-dimensional manifold $X$ with (possibly empty) oriented boundary 
and with  marked disjoint arcs (embedded closed intervals). 
The arcs lie either in the interior or on the boundary and carry labels. 
Boundary arcs are labeled by simple objects. 
The connected components of the complement in $\partial X$
of the boundary arcs also carry labels, called 
           {\em boundary conditions}, which are also simple objects.
 The label of an interior arc $z$ is an equivalence class of 
triples $(i,or,or')$ where $i$ is a simple object, $or$ is a local orientation
of the surface at $z$, and $or'$ is an orientation of the arc.
Two triples are equivalent if they are equal or if
one is obtained from the other by taking the dual object and 
reversing the orientations $or$, $or'$.

We call {\em boundary segments} the connected components 
of the complement in $\partial X$ of the boundary arcs.
If a boundary arc $x$ lies between two boundary segments
labeled by boundary conditions
$a$, $b$ in the order given by the orientation of the boundary,
we say that $x$ {\em changes the boundary conditions from $a$ to $b$}.
\medskip

The {\em double} $\hat X$ of a labeled surface
is an extended surface associated to a labeled surface.
It is, as an oriented 2-manifold,
the double $\hat X$ of $X$, with projection $p{:}\ \hat X\To X$
and orientation reversing involution $\sigma{:}\ \hat X\To \hat X$.
The double is made into an extended surface, by taking as arcs
the inverse images of the arcs of $X$. The boundary arcs have one
inverse image and are labeled by the labels in $X$. Their orientation
is inherited from the orientation of the boundary of $X$. Each interior
arc $z$ of $X$ has two inverse images. They are labeled and oriented
by  the two labels in the equivalence class labeling $z$,
in such a way that the local orientation $or$
appearing in the label agrees with the orientation of $\hat X$. The
lagrangian subspace of $H_1(\hat X,\R)$ 
is the eigenspace of $\sigma_*$ to the eigenvalue $-1$. It is called 
canonical lagrangian subspace and denoted by $\lambda_-(\hat X)$. 

This definition makes sense because of the
\begin{lemma}
Let $\sigma{:}\ \hat X\To \hat X$ be an orientation reversing
homeomorphism of a surface $\hat X$ such that $\sigma\circ\sigma=\mathrm{id}$. 
Then the induced map $\sigma_*{:}\ H_1(\hat X,\R)\To H_1(\hat X,\R)$ 
is diagonalizable and its eigenspaces are lagrangian.
\end{lemma}
\medskip
\noindent{\it Proof:}
The induced map $\sigma_*$ is a linear involution of a real vector space.
Thus $p_\pm=\frac12(1\pm\sigma_*)$ are projections onto the eigenspaces 
$\lambda_\pm$ corresponding to the eigenvalues $\pm1$. Since
$p_++p_-=\mathrm{id}$, $\sigma_*$ is diagonalizable. Let $\omega$ denote 
the intersection pairing on $H_1(\hat X,\R)$. Since $\sigma$ reverses 
the orientation, we have $\omega(\sigma_*a,\sigma_*b)=-\omega(a,b)$
for all $a,b\in H_1(\hat X,\R)$. Therefore $\omega$ vanishes identically 
on $\lambda_+$ and $\lambda_-$. Since $\lambda_+\oplus\lambda_-=H_1(\hat X,\R)$, 
the subspaces $\lambda_\pm$ are of
maximal dimension with this property, i.e., lagrangian. $\square$

\medskip

\begin{proposition}\label{p-Aut}
Let $\hat X$ be a double of the labeled surface $X$.
Let $\mathrm{Aut}(\hat X,\sigma)$ be the group of degree one
homeomorphisms of $\hat X$ preserving the marked arcs with their orientation
and commuting with the involution $\sigma$.
Then $f\mapsto f_\sharp $ defines a representation (not just a projective 
representation) of $\mathrm{Aut}(\hat X,\sigma)$ on $\HH(\hat X)$.
\end{proposition}
\medskip

\noindent{\it Proof:} Let $f\In\mathrm{Aut}(\hat X,\sigma)$.
 Since $f$ and $\sigma$ commute, the
induced maps $f_*$, $\sigma_*$ also commute. It follows that
the eigenspaces of $\sigma_*$ are preserved by $f_*$. In
particular the lagrangian subspace of $\hat X$ is
preserved by elements of $\mathrm{Aut}(\hat X,\sigma)$. 
Under these circumstances the anomaly is trivial and
we have the representation property $(f\circ g)_\sharp =f_\sharp 
\circ g_\sharp$.
$\square$.

\medskip

\subsubsection{Connecting 3-manifolds} To each compact surface $X$
we associate a connecting 3-manifold $M_X$.
The connecting 3-manifold $M_X$ is a 3-dimensional oriented manifold with
boundary $\hat X$. It is used to construct the correlation
functions and reduces to the cylinder $X\times[-1,1]$
if $X$ is closed and orientable and to the ball if $X$ is a  disk.

We first describe $M_X$ as an oriented manifold.
           If 
$X$ has no boundary, $M_X$ is $(\hat X\times [-1,1])/\Z_2$,
where $\Z_2$ acts on the first factor by the involution $\sigma$
and on the second by $t\mapsto -t$. This action preserves the
product orientation, so that $M_X$ is naturally an oriented manifold.
It comes with a projection $[(x,t)]\mapsto     p(x)$ to $X$. The fiber of
this projection over $y\In X$ is an interval, the {\em connecting interval over
$y$}, connecting the two inverse images of $y$ in $\hat X$.
If $X$ has a boundary, $M_X$ is obtained from $(\hat X\Times [-1,1])/\Z_2$
by contracting the fibers over the boundary to single points. 

Alternatively, let $\rho{:}\ X\To [0,\infty)$ be any non-negative
function such that $\rho(x)\eq 0$ if and only
if $x\In \partial M$. Then we may define $M_X$ to consist of
$[(x,t)]\In (\hat X\Times \R)/\Z_2$ such that $t^2\leq \rho(p(x))$. 
The points with $t^2\eq \rho(p(x))$ form the boundary which is obviously 
homeomorphic to $\hat X$. Connecting manifolds corresponding to
different choices of $\rho$ are canonically
homeomorphic. The homeomorphism commutes with
$p$ and reduces to the identity on $\hat X\simeq \partial M_X$.
\begin{proposition}\label{p-ctm} Let $\hat X$ be the double of $X$, 
$p{:}\ \hat X\To X$ the projection, $\lambda_-(\hat X)$ the canonical 
lagrangian subspace of $H_1(\hat X,\R)$.
\begin{enumerate} 
\item[(i)] $M_X$ is a compact manifold with boundary $\partial M_X=\hat X$.
\item[(ii)] The restriction of $\pi{:}\ M_X\to X$, $[(x,t)]\mapsto p(x)$
 to $\pi^{-1}(X-\partial X)$ is a fiber bundle whose fiber
over $y$ is an interval with boundary $p^{-1}(y)$.
\item[(iii)] $\lambda_-(\hat X)$ is the kernel of the homomorphism
$H_1(\hat X,\R)\to H_1(M_X,\R)$ induced by inclusion.
\item[(iv)] The involution $\sigma:\hat X\to\hat X$ extends
to the involution $[(x,t)]\mapsto[(x,-t)]$ of $M_X$. Its fixed
point set is the image of $X$ under the embedding $i:y\mapsto
[(x,0)]$ for any $x$ with $p(x)=y$.
\end{enumerate}
\end{proposition}

\medskip

\noindent{\it Proof:}  Choose a function $\rho$ as above.
Let $\{\phi_\alpha{:}\ U_\alpha\to \R^2\}$ be an atlas
of $X$ with connected charts $U_\alpha$. 
Let $\epsilon_{\alpha,\beta}$ be the sign of the Jacobian of
$\phi_\alpha\circ\phi_\beta^{-1}$. Then $M_X$ is homeomorphic to
\[
\bigsqcup_\alpha\,\{(y,t)\In U_\alpha\TImes \R\,|\, 
t^2\,{\leq}\,\rho(y)\}\,{/}\!\sim
\]
with equivalence relation
$(x\In U_\alpha,t)\sim (x\In U_\beta,\epsilon_{\alpha,\beta}\,t)$. 
The projection to $X$ is $p(y,t)=y$ and the involution is 
$\sigma(y,t)=(y,-t)$. Therefore we have a surjective
map $\hat X\Times [-1,1]$ onto $M$ given by $((y,t),s)\mapsto (y,ts)$.
The fibers of this map consist of two points related by the $\Z_2$
action, except if $y\in\partial X$ where the fiber is an interval.

This presentation of $M_X$ implies (i) and (ii). 

To prove (iii), notice that if $a$ is a loop on $\hat X$, then 
$a-\sigma\circ a$ is the boundary of a surface in $M_X$ consisting
of connecting intervals ending at points of $a$. Thus if
$c$ is a cycle on $\hat X$
such that $\sigma_*c$ is homologous to $-c$, then $c$ is homologous
to zero in $M_X$. This shows that $\lambda_-(\hat X)$ is contained
in the kernel $K$ of the homomorphism induced by inclusion. On the
other hand, it is a general fact that 
the intersection form vanishes on $K$. Therefore the dimension of
$K$ cannot be larger than the dimension of the lagrangian subspace
$\lambda_-(\hat X)$.

(iv) is obvious.
$\square$

\subsubsection{Multiplicity spaces}
Suppose that $X$ is a labeled surface. If $x$ is a marked arc on the 
boundary labeled by a simple object $j$ and changing the boundary condition $a$ 
to the boundary condition $b$, the {\em multiplicity space}
of $x$ is $W_{a,b}(j)\eq \Hom({}b,{}j\Otimes {}a)$. The {\em multiplicity space}
$W_{\partial X}$ of a labeled surface $X$ is the (unordered)
tensor product of the multiplicity spaces of its boundary arcs. If 
there are no boundary arcs, we set $W_{\partial X}=\C$.

\subsubsection{Construction of correlation functions}
We are ready to define correlation functions for general labeled
surfaces. Let $X$ be a labeled surface, $M_X$ be its connecting
manifold and $\hat X\eq\partial M_X$ the double of $X$, with its structure
of extended surface. Let $i{:}\ X\To M_X$ be the inclusion of $X$ as zero
section (Prop.~\ref{p-ctm}, (iv)). We construct
a ribbon graph in $M_X$. It consists of vertical ribbons and an equatorial 
graph for each connected component of $\partial X$. 

The {\em vertical ribbons}
are associated to interior arcs of $\hat  X$: if $z$,
$z'$ are interior arcs projecting to an interior arc of $X$,
the corresponding vertical ribbon is the union of the connecting intervals
ending at $z$ and $z'$. It is an embedded rectangle with two sides equal to
$z$, $z'$. The orientation of the vertical ribbon is chosen so as to induce
the orientations of $z$, $z'$. If we orient the core from $z$ to $z'$, the
label of the ribbon is equal to the label of $z$.
The {\em equatorial graphs} consist of annuli and joining ribbons.
The annuli lie in the zero section $i(X)$ of $M_X$ and their
cores are obtained by moving $i(\partial X)$ into $M_X$ by a short 
        amount, where `short' means 
away from the vertical ribbons. The joining ribbons are short ribbons in
$i(X)$ connecting boundary arcs to the annuli at trivalent vertices.
They are labeled by the label of the corresponding boundary arcs and their 
      cores are
oriented inward.
The labels of the parts of the annuli between trivalent vertices are the boundary
conditions between the corresponding arcs.
The orientation of the equatorial graphs is chosen so as to induce the orientation
of the boundary arcs. This does not fix the orientation of the annuli that are
not connected to the boundary; but this does not matter since the correlation
function will not depend on the choice of that orientation.

Then the correlation function of the labeled surface $X$ is
\[
C(X)= Z(M_X,\emptyset,\hat X):\quad W_{\partial X}\to \HH(\hat X),
\]
considered as a multilinear function of the labels of the trivalent vertices.
Thus if $u=\otimes_{x}u_x\in W_{\partial X}$ with $x$ running over the boundary
arcs, $C(X)u= Z(M_{X,u},\emptyset,\hat X)$, where $M_{X,u}$ is the connecting
3-manifold with its ribbon graph, such that the trivalent vertex connected
to $x$ by a joining ribbon is labeled by $u_x$.

\subsubsection{Modular invariance}
If $f{:}\ X\to X$ is a homeomorphism of the labeled surface $X$, preserving 
the orientation and the marked arcs of the boundary and
mapping interior marked arcs to interior marked arcs with the same label,\,%
\footnote{~Recall that a label of an interior arc 
is an equivalence  class of triples $[(i,or,or')]$.
The condition for an interior arc $z$ means that $f$ maps $z$ to an arc $z'$,
and if $z$ has label $[(i,or,or')]$, $z'$ has label $[(i,f_*or,f_*or')]$.}
then there exists a unique degree one homeomorphism
$\hat f$ of $\hat X$, the {\em lift} of $f$, so that $p\circ \hat f=f\circ p$.
It preserves the inverse images of the arcs and their orientation. $f$
commutes with $\sigma$ and therefore preserves 
the canonical lagrangian subspace of $H_1(\hat X,\R)$.

\begin{thm}\label{t-genmodular}
(Modular invariance)
Let $X$ be a labeled surface. Let $f{:}\ X\To X$ be a  
homeomorphism preserving the orientation of the boundary and
mapping marked arcs to marked arcs with the same label and boundary segments
to boundary segments with the same boundary condition.
Let $\hat f$ be its lift to $\hat X$. Then
\[
\hat f_\sharp\, C(X)=C(X).
\]
\end{thm}

\medskip

\noindent{\it Proof:}
Let $F{:}\ M_X\To M_X$ be the map $[(x,t)]\mapsto[(\hat f(x),t)]$ of $M_X\,
{\subset}\,(\hat X\Times [-1,1])/\Z_2$. It is clear that $F$ is a well-defined 
degree one homeomorphism
of $M_X$. It maps vertical ribbons to vertical ribbons with the same label.
The equatorial graphs are mapped to slight deformations of the equatorial graphs.
As the boundary arcs are fixed, we may compose $F$ with a homeomorphism $G$ of 
$M_X$ with support in the vicinity of $i(\partial X)$ and restricting to the
identity on the boundary, in such a way that the equatorial graphs are also kept
fixed. Then $G\circ F$ preserves the ribbon graph and restricts to $\hat f$ on
the boundary. Therefore, by the naturality axiom,
$\hat f_\sharp\, Z(M,\emptyset,\hat X)= Z(M,\emptyset,\hat X)$. $\square$

\medskip

\noindent{\bf Remark.} We may relax the condition that $f$ preserves the
boundary arcs and the orientation of the boundary. We may just assume that
$f$ maps boundary arcs to boundary arcs with the same label, or the dual
label, depending on whether $f$ preserves the local  orientation of the boundary.
Similarly $f$ should be compatible with the labeling of boundary segments.
Then the modular invariance reads $\hat f_\sharp\,C(X)=C(X)\rho(f)$, for
a suitable action on the multiplicity spaces. We leave the details to the
reader.

\subsubsection{Factorization}

If $X$ is a surface, we can obtain
a new surface $X'$ by cutting and pasting
in two basic ways: either we can cut out  disks around two interior 
marked arcs and glue their boundaries together or
we can glue two boundary arcs. 

In both cases we want to relate the correlation
functions on $X'$ to the correlation functions
on $X$. In the first case the relation between
correlation functions is called bulk factorization.
In the second case it is called boundary factorization.

\medskip

Let $z_1,z_2$ be two interior arcs of a labeled
surface. We say that the labels of $z_1$, $z_2$ {\em match}
if they are of the form $[(i,or_1,or_1')]$
and $[(i^*,or_2,or_2')]$, respectively. In this case we construct a 
labeled surface $X'$ as follows. Choose representatives $(i,or_1,or_1')$, 
$(i^*,or_2,or_2')$ so that 
the arcs are oriented and we have local orientations around the arcs.
Let $\phi,\phi'{:}\ D^2\To X$ be orientation preserving 
disjoint embeddings of the unit disk $D^2\,{\subset}\,\C$ such that
their restriction to $[-1,1]$ are parametrizations of the oriented
arcs $z_1,z_2$. Then $X'$ is obtained from $X$ by removing the 
interiors of the disks $\phi(D^2)$ and $\phi'(D^2)$, and gluing
their boundaries by identifying $\phi(z)$ with $\phi'(-\bar z)$, for 
$z\In S^1$.  The arcs of $X'$ are the remaining arcs of $X$.

We say that $X'$ is obtained from $X$ by gluing $z$ to $z'$. 

The double of $X'$ is then obtained from the
double of $X$ by gluing the inverse images
of $z_1$ to the inverse images of $z_2$:
the inverse image of $z_1$ with orientation
$or_1$ is glued to the inverse image of $z_2$ with
orientation $or_2$ and the inverse image
of $z_1$ with the opposite orientation $-or_1$ is glued to the
inverse image of $z_2$ with the orientation $-or_2$.

Then we have a gluing homomorphism
$g_{\hat X,\hat X'}{:}\ \HH(\hat X)\To \HH(\hat X')$, which is 
the composition of the two gluing homomorphisms (in either order) with
$(\id_{\lambda',\lambda_-(\hat X')})_\sharp$. Here $\lambda'$ is the
symmetric lagrangian subspace of $H_1(\hat X',\R)$ obtained 
form the canonical lagrangian subspace $\lambda_-(\hat X)$ 
of $\hat X$ by the gluing prescription
of \ref{ss-mf}, and $\id_{\lambda',\lambda_-(\hat X')}$ is the identity
map from $\hat X'$ with lagrangian subspace $\lambda'$ to $\hat X'$ with
canonical lagrangian subspace.

\medskip

\noindent{\bf Example.} 
Let $z_1$, $z_2$ be 
two marked arcs on a sphere $X$. Choose an
orientation $or$ of the sphere and let
$(i_1,or,or_1')$, $(i_2,or,or_2')$ be representatives
of the labels of these two arcs chosen to agree
with the global orientation. If $i_2=i_1^*$,
the labeled surface obtained by gluing
$z$ to $z'$ is a torus. If $i_2=i_1$, we may take 
the other representative $(i_2^*=i_1^*,-or,-or_2')$ and obtain
a Klein bottle as a result of gluing.  In the first case, $\hat X'$ 
is obtained from the disjoint union of the two spheres by 
        gluing pairs of arcs on the same connected component.
In the second case, $\hat X'$ is obtained 
by gluing arcs on one connected component to arcs of the other.

\medskip

\begin{thm}\label{t-bulkfactorization}
(Bulk factorization)
Let $X$ be a labeled surface.
Let $X'$ be obtained from $X$ by gluing $X$ at
two interior marked arcs 
with labels $[(j,or_1,or_1')]$, $[(j^*,or_2,or_2')]$. 
Let $X_j$ be the surface $X$, with $j$ running
over $I$ with all other labels fixed.
Then
\[
C(X')=\sum_{j\in I}\mathcal{D}^{-1}
\mathrm{dim}(j)\,
g_{\hat X_j,\hat X'}C(X_j).
\]
\end{thm}

The proof of this theorem is the same as the
proof of Theorem \ref{t-closedfactorization}.

We now turn to the boundary factorization. Let $x$, $x'$ be boundary 
arcs with the orientation induced by the orientation of the boundary.
Suppose that the label of $x$ is $j$ and
that the label of $x'$ is $j^*$. Assume
that $x$ changes the boundary conditions 
from $a$ to $b$ and that $x'$ changes the
boundary conditions from $b$ to $a$.
Under these circumstances, we may glue
$x$ to $x'$ via an orientation reversing
homeomorphism and obtain a surface $X'$.
We say that $X'$ is obtained from $X$ by gluing $x$ to $x'$.
 \dontprint{The Euler characteristic $\chi$,
the number $n$ of interior arcs and the
number $m$ of boundary arcs are related by
\[
\chi(X')=\chi(X)-1, \qquad n(X')=n(X), \qquad m(X')=m(X)-2.
\]
}

The double $\hat X'$ of $X'$ may then be identified
with the surface obtained from $\hat X$ 
by gluing the inverse image in $\hat X'$
of $x$ to the inverse image in $\hat X'$ of $x'$.
The lagrangian subspace of $H_1(\hat X',\R)$ obtained
by the gluing prescription coincides in this
case with the canonical lagrangian subspace $\lambda_-(\hat X')$.
We thus have a gluing homomorphism 
$g_{\hat X,\hat X'}:\HH(\hat X)\to \HH(\hat X')$. 

To formulate the boundary factorization properties
of correlation functions, we need to compare
the multiplicity spaces of $X$ and $X'$.

Note that $W_{\partial X}=W_{\partial X'}\Otimes
W_{a,b}(j)\Otimes W_{b,a}(j^*)$, and that
$W_{a,b}(j)$ is dual to $W_{b,a}(j^*)$. We thus
have a natural map $\gamma_{X',X}{:}\ W_{\partial X'}\,{\to }\,
W_{\partial X}$ obtained by taking the tensor product with the 
canonical tensor. In terms of the bases of \ref{ss-pairing},
\[
\gamma_{X',X}(w)=w\otimes\sum_\alpha e_\alpha[b^*ja]
\otimes e_\alpha[a^*j^*b].
\]
\begin{thm}\label{t-boundaryfactorization}
(Boundary factorization)
Let $X'$ be obtained by gluing two marked boundary arcs
$x'$, $x''$ with labels $j$, $j^*$ of a labeled surface $X$. 
Let $X_j$ be the surface $X$, with $j$ running
over $I$ with all other labels fixed. Then
\[
C(X')=\sum_j\mathrm{dim}(j)\,
g_{\hat X_j,\hat X'}\circ C(X)\circ \gamma_{X',X}
\]
\end{thm}

This theorem is proved in the same way as
Theorem \ref{t-diskfactorization}.

\section{Structure constants}\label{s-4}

It is clear that using the factorization property of correlation
functions (Theorems \ref{t-bulkfactorization} and 
\ref{t-boundaryfactorization}) the calculation
of any correlation function can be reduced to four
basic cases: the sphere with three points,
the disk with three boundary points, the
disk with one interior point and one boundary point,
and the real projective plane with one point.

We compute the correlation functions in these four cases.
The special cases of two points on the sphere and on the
disk and one interior point on the disk
can be in principle deduced by setting one of the labels
to $0$. But since the results are particularly simple
we compute them separately.

The calculation of the two-point functions also explains
the appearance of the factors of $\mathrm{dim}(j)$
and $\mathcal{D}^{-1}$ in the factorization formulae.

\subsection{Two-point functions}\label{ss-2p}
We calculate the two-point functions on the sphere and on the disk.

Let $X\eq(S^2,j,j^*)$ be the unit sphere in $\R^3$ with two points,
say at the north pole with label $j$ and at the south pole with label 
$j^*$. We give $S^2$ the standard orientation.
As usual, we rather have to specify two arcs than two points.
Let the arc at the north pole be a short arc in the $x$-$z$ plane 
oriented in the positive $x$ direction, and let the arc at the
south pole be a short arc in the $x$-$z$ plane, pointing in the
negative direction. $X$ may be both viewed 
as a labeled surface and as an extended surface.
There are no lagrangian subspaces here
since the first homology of the sphere is trivial.

The correlation function $C(X)$ on the sphere
takes values in $\HH(X\sqcup(-X))=
\HH(X)\otimes \HH(-X)$. 
A basis of the one-dimensional
vector space $\HH(X)$ is given by 
\[
b(X)= Z((D^3,j),\emptyset,X),
\] 
associated to the unit ball $M_X=D^3$ endowed with a ribbon
$D^3\cap([-\epsilon,\epsilon]\Times \{0\}
\Times \R)$. The ribbon has label $j$; it runs vertically along the
$z$-axis and its core is oriented from
top to bottom. The orientation of the ribbon
is such that it induces the orientation of the arcs.

A basis of $\HH(-X)$ is given by $ b(-X)=\theta_\sharp b(X)$, where
$\theta$ is the reflection with respect to the $x$-$z$ plane.

To compute the two-point function on the sphere,
we have to compute the proportionality
constant $c$ in $C(S^2,j,j^*)=c\, b(X)\Otimes b(-X)$.
This can be done by using the functoriality of the invariant
$ Z$. If we glue two balls, each with a ribbon inside, to $S^2\times [-1,1]$ 
we get $S^3$ with an unknot labeled by $j$. These two balls may be
viewed as a cobordism from $S^2\sqcup S^2$ to the empty set. Applying
its invariant to $C(S^2,j,j^*)$, we get $\mathcal{D}^{-1}\mathrm{dim}(i)$,
the invariant of $S^3$ with the unknot. Applying the invariant of the same
cobordism to $b(X)\otimes b(-X)$, we get the invariant $\mathcal{D}^{-2}
\mathrm{dim}(j)\mathrm{dim}(j^*)$ of a closed manifold with two
connected components, each of which is a 3-sphere with an unknot labeled
by $j$ and $j^*$ respectively. As $\mathrm{dim}(j)=\mathrm{dim}(j^*)\neq 0$,
we get $c=\mathcal{D}\,\mathrm{dim}(j)^{-1}$, with the result
\[
C(S^2,j,j^*)=\frac{\mathcal{D}}{\mathrm{dim}(j)}
\;b(S^2,j,j^*)\otimes b(-(S^2,j,j^*)).
\]

Let us turn to the case of the disk with two boundary points labeled
by $j$, $j^*$, and boundary
conditions $a,b$. The two-point correlation function $C(D^2,j,j^*;a,b)$ is a map
from $W_{ab}(j)\otimes W_{ba}(j^*)$ to the space $\HH(S^2,j,j^*)$ of
conformal blocks on the sphere. Evaluating the correlation function
 on basis vectors $e_\alpha\otimes e_\beta=
e_\alpha[jba^*]\otimes e_{\beta}[j^*ab^*]$
we get the invariant of $D^3$ with an equatorial graph with two outgoing lines.
This graph may be replaced by a single ribbon using \Ref{e-Mozart}. The
result is
\[
C(D^2,j,j^*)\,e_\alpha\otimes e_\beta=\frac{\delta_{\alpha,\beta}}{\mathrm{dim}(j)}
\;b(S^2,j,j^*).
\]
\noindent{\bf Remark.} Our two-point correlation functions have a somewhat non-standard
normalization, which avoids square root ambiguities. For any choice of square
roots one may define ``normalized correlation functions''. If $X$ is a labeled surface
with interior arcs labeled by $i_1,\dots,i_n$ and $m$ boundary arcs labeled by
$j_1,\dots,j_m$, let
\begin{equation}\label{e-Beethoven}
C_\mathrm{norm}(X)=\prod_{\nu=1}^{n}\sqrt{\SR_{i_\nu,\zero}}\,
\prod_{\nu=1}^{m}\sqrt{\frac{\SR_{j_\nu,\zero}}{\SR_{\zero,\zero}}}\,C(X).
\end{equation}
(Recall that $\mathcal{D}^{-1}\eq\SR_{\zero,\zero}$ and that $\mathrm{dim}(j)
\eq\SR_{j,\zero}/\SR_{\zero,\zero}$.) For these correlation
functions there are no factors of $\mathrm{dim}(j)$ or $\mathcal{D}$ in the
two-point functions or in the factorization theorems.
\subsection{The 3-point function on the sphere}
Let $X\eq(S^2,i,j,k)$ be the sphere with three marked arcs labeled
by $i$, $j$ and $k$. To fix the conventions let us take the sphere as
the unit sphere in $\R^3$, with standard orientation
and have the three marked arcs on the equator. We orient the
equator counterclockwise and orient the arcs in the same direction.
The correlation function then takes values in 
$\HH(X\sqcup(-X))=\HH(X)\otimes \HH(-X)$.
A basis of $\HH(X)$ is given by $ e_\alpha(X)= Z(M_{X,\alpha},\emptyset,X)$,
$\alpha\eq1,\dots,N_{k,j,i}$, where the connecting 3-manifold $M_{X,\alpha}$ is
the 3-ball with a ribbon graph in the $x$-$y$ plane with one trivalent vertex 
at the origin. The vertex is labeled by the basis element $e_\alpha[kji]$
of $H^{k,j,i}$. A basis of $\HH(-X)$ is $e_\alpha(-X)\eq\theta_\sharp e_\alpha(X)$,
where $\theta$ is the reflection at the $x$-$y$ plane.

As in the case of the two-point function, to compute the correlation function
$ Z(X\times[-1,1],\emptyset,X\sqcup(-X))$ in terms of the basis
$e_\alpha(X)\otimes e_\beta(-X)$ of $\HH(X\sqcup(-X))=\HH(X)\otimes \HH(-X)$,
we use the functoriality of $ Z$ and act on $C(X)$ 
with $ Z(M_{X,\alpha,\beta},X\Sqcup(-X) ,\emptyset)$.
Here, $M_{X,\alpha,\beta}$ is the disjoint union of two 3-balls, each with a 
ribbon graph with one vertex labeled by $e_\alpha,e_\beta$ respectively.
The result is the invariant of $S^3$ with the ``theta graph'', which has two
vertices connected by three ribbons. By the orthogonality relations of
basis elements this invariant is $\mathcal{D}^{-1}\delta_{\alpha,\beta}$.

On the other hand, if we act with $ Z(M_{X,\alpha,\beta},X\Sqcup(-X),\emptyset)$
on $e_\gamma(X)\otimes e_\delta(-X)$, we obtain the invariant of $S^3\sqcup S^3$
with a theta graph in each copy of $S^3$, which is $\mathcal{D}^{-2}
\delta_{\alpha,\gamma} \delta_{\beta,\delta}$. The result is thus
\[
C(S^2,i,j,k)=\mathcal{D}\,
\sum_{\alpha=1}^{N_{k,j,i}}e_\alpha(S^2,i,j,k)\otimes
e_\alpha(-(S^2,i,j,k)).
\]
As usual, $N_{k,j,i}$ is the dimension 
of $\Hom(\one,{}k\otimes {}j\otimes {}i)$. 
With the normalization \Ref{e-Beethoven} we then have
\[
C_{\mathrm{norm}}
(S^2,i,j,k)=\frac{\sqrt{\SR_{i,\zero}\SR_{j,\zero}\SR_{k,\zero}}}
{\SR_{\zero,\zero}}
\,
\sum_{\alpha=1}^{N_{k,j,i}}e_\alpha(S^2,i,j,k)\otimes
e_\alpha(-(S^2,i,j,k)).
\]

\subsection{The $(1,0)$-point function on the disk}
Let $X\eq(D^2,i;a)$ be the unit disk with an arc labeled
by $i$ at the origin and boundary condition $a$. The correlation
function $C(X)$ takes values in $\HH(\hat X)$ with $\hat X=(S^2,i,i^*)$,
the sphere with two marked arcs as in \ref{ss-2p}. We have, by construction,
\[
C(X)= Z(M,\emptyset,\hat X).
\]
The 3-manifold $M$ is a 3-ball with a vertical ribbon and an equatorial
graph consisting of an annulus labeled by $a$. We want to express
$C(X)$ in terms of the basis $b(S^2,i,i^*)$ (the ``Ishibashi boundary state'')
of the one-dimensional vector space $\HH(\hat X)$. We do this as 
above, by gluing a ball with a vertical ribbon inside
onto $M$ and using the functoriality of $ Z$. The resulting
closed 3-manifold is a 3-sphere with two unknots with linking number
$1$. It has invariant $\mathcal{D}^{-1}{\ST}_{i,a}$. This has to be compared
with the 3-manifold obtained by gluing the same ball to
$D^3$ with a vertical ribbon, which is $S^3$ with an unknot and has invariant
$\mathcal{D}^{-1}{\ST}_{i,\zero}=\SR_{i,\zero}$. The result is
\[
C(D^2,i;a)=\frac{{\SR}_{i,a}}{{\SR}_{i,\zero}}\,b(S^2,i,i^*).
\]
Implementing the normalization as given in eq.\ \Ref{e-Beethoven},
     the corresponding normalized correlation function is
    \[
    C_{\rm norm}(D^2,i;a)=\frac{{\SR}_{i,a}}{\sqrt{{\SR}_{i,\zero}}}\,b(S^2,i,i^*).
    \]
    In conformal field theory, this formula for the one-point functions
     on the disk was first obtained in \cite{C}. 
 
\subsection{The $(1,1)$-point function on the disk}
We calculate the correlation function in the case
of $n=1$ point in the interior of the disk $D^2$ and
$m=1$  point on its boundary. The label of the interior
point is $i$ and the label of the boundary point is $j$.
The double $\hat X$ of the disk with these points is the 2-sphere 
$S^2$ with one point on the equator labeled by $j$ and two points,
say the north and south pole, labeled by $i$, $i^*$.
Then the correlation function with boundary condition $k$ maps
$W_{k,k}(j)=\Hom({}k,{}j\Otimes {}k)$
to $\HH(\hat X)=\Hom(\one,i\Otimes i^*\Otimes j)$. Its matrix
elements $R_{\alpha\beta}$ with respect to the chosen
bases are given by the formula
\[
\funo
\quad\alza{1.5cm}{\;=\;\sum_{\beta}R_{\alpha\beta}}
\quad\fdue\alza{1.5cm}.
\]
To calculate $R_{\alpha\beta}$, we compose this morphism
with the basis element $e_\gamma[j^*ii^*]$ of 
$\Hom({}j\otimes {}i,{}i)\simeq \Hom(\one,j^*\Otimes i\Otimes i^*)$ and obtain
\\[-.8em]
\[
\ftre
\;\alza{1.5cm}{\;=\;\sum_\beta R_{\alpha\beta}}
\quad\fquattro
\;\alza{1.5cm}{=\,\frac1{{\mathrm{dim}(i)}}\,R_{\alpha\gamma}}
\quad\fcinque\alza{1.5cm}.
\]
The left-hand side may be evaluated in terms of 
fusing matrices, with the result:
\\[-.4em]
\[
\;\alza{1.5cm}{\frac1{{\mathrm{dim}(i)}}\,R_{\alpha\gamma}
}\quad\fcinque
\;\alza{1.5cm}{\;=\;\sum_{l,\delta,\varepsilon}}\quad\fsei
\;\alza{1.5cm}{\sixj i{i^*}jkkl{\gamma\alpha}{\delta\varepsilon}}\
\alza{1.5cm}{.}
\]
The expression appearing on the right may be further simplified:
\\[.4em]
\begin{eqnarray*}
&\fsei
     \alza{1.5cm}{=}\quad\fsette
\alza{1.5cm}{=}\quad\fotto&
\\{}\\[.9em]
&\alza{1.5cm}{=\; \frac{v_l}{v_iv_k}}\quad\fnove
\alza{1.5cm}{\quad =\; \delta_{\delta\varepsilon}\,\frac1{{\mathrm{dim}(i)}}
\;\frac{v_l}{v_iv_k}
}\quad\fcinque\alza{1.5cm}.&
\end{eqnarray*}
Putting everything together, we arrive at the result
\[
R_{\alpha\beta}
\,=\, \sum_{l,\varepsilon} \,\frac{v_l}{v_iv_k}
\,\sixj i{i^*}jkkl{\beta\alpha}{\varepsilon\varepsilon}.
\]

\subsection{Three boundary points on the disk}
The correlation function for three points on the
boundary of a disk can be computed in an analogous
way. In this case $\hat X$ is the 2-sphere with
three points on the equator. Let us orient the
boundary of the unit disk counterclockwise and
denote the labels of the three points $i$, $j$, $k$.
The boundary conditions are labeled by $a,b,c$. Then $C( X)$ is a map 
from $W_{c,a}(i)\otimes W_{a,b}(j)\otimes W_{b,c}(i)$
to $\HH(\hat X)=\Hom(\one,{}k\otimes {}j\otimes {}i)$.
The structure constants $C_{\alpha\beta\gamma}^\delta$
are in this case defined by
\[
C(X)\,e_\alpha[a^*ic]\Otimes e_\beta[b^*ja]\Otimes e_\gamma[c^*kb]
=\sum_{\delta}C_{\alpha\beta\gamma}^\delta\, e_\delta[kji],
\]
in terms of the bases of $W_{c,a}(i)\simeq \Hom(\one,{}a^*\Otimes{}i
\Otimes{}{c})$,
etc. The connecting 3-manifold is in this case a ball with an equatorial
graph. The computation of the structure constants then goes as follows. 
\\[-1.2em]
\[
\fventuno
\;\alza{1.5cm}{\;=\;\sum_{d,\delta,\varepsilon} }
\quad\fventidue
\!\!\alza{1.5cm}
{\barsixj jidcba{\delta\varepsilon}{\alpha\beta}\ .}
\]
The diagram appearing on the right-hand side can be deformed to
\\[-.4em]
\[
\fventitre\;\alza{1.5cm}{=\;\delta_{\varepsilon\gamma}
\delta_{dk}\,\frac1{{\mathrm{dim}(k)}}}\quad\fventiquattro\alza{1.5cm},
\]
with the result
\[
C_{\alpha\beta\gamma}^\delta\,=\,
\frac1{{\mathrm{dim}(k)}}\,\barsixj jikcba{\delta\gamma}{\alpha\beta}\;.
\]
The normalized correlation function is then
\[
C_{\mathrm{norm}}(X) \,e_\alpha[a^*ic]
\Otimes e_\beta[b^*ja] \Otimes e_\gamma[c^*kb]
=
\sqrt\frac{\SR_{i,\zero}\SR_{j,\zero}}
          {\SR_{k,\zero}\SR_{\zero,\zero}}
    \sum_{\delta}
\barsixj jikcba{\delta\gamma}{\alpha\beta} e_\delta[kji],
\]
cf.~\cite{R,BPPZ,FFFS1}.

\subsection{One point on the projective plane}
We consider the one-point correlation function
on the projective plane. Thus our labeled surface $X\eq(\R P^2,i)$ is 
$\R P^2\eq S^2/\Z_2$ with a marked arc labeled by $i\In I$ and some
local orientation. The connecting 3-manifold is here $(S^2\Times 
[-1,1])/\Z_2$. View $S^2$ as the unit sphere in $\R^3$, and let us put 
the marked arc $z$ at the north (= south) pole $\pm(0,0,1)\In S^2/\Z_2$
of the projective plane. The local orientation is obtained by identifying,
locally around the arc, the projective plane with the upper hemisphere of
$S^2$. Then the correlation function is 
the element of $\HH(\hat X)$ associated to the ribbon graph given
by the fiber over $z$, i.e., the image in $(S^2\Times [-1,1])/\Z_2$ of the
interval $(0,0,1)\times {[-1,1]}$.
The framing is determined by taking a neighboring point, say
$(\epsilon,0,\sqrt{1{-}\epsilon^2})$ and taking at each point of
the fiber a vector pointing to the fiber over the neighboring point.

\begin{figure}
\scalebox{0.4}{\includegraphics{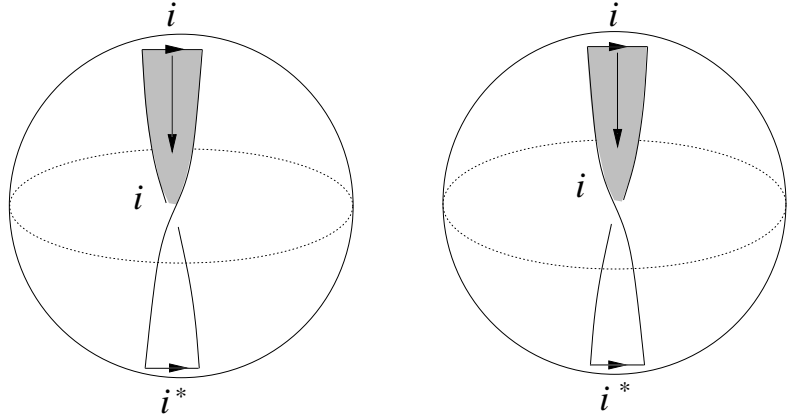}}
   \caption{The vectors $\psi_i^+$, $\psi_i^-$ are associated to the
ribbon graphs in $D^3$ on the left and on the right, respectively}
\label{fig15}
\end{figure}

The space of states $\HH(\hat X)$ is one-dimensional in this case. A basis 
of this space is given by the ``Ishibashi cross cap state'' $\psi_i$.
Before giving its definition we notice that there
are two natural candidates for a basis. 
Namely, we can take any of the two states $\psi_i^{\pm}$ associated
to the ribbon graphs in the 3-ball $D^3$ of Fig.~\ref{fig15}.
The two states differ by a twist, so $\psi_i^+\eq v^{}_i\psi_i^-$. To 
define the cross cap state we choose square roots of $v_i$
and normalize the cross cap state salomonically as
\[
\psi_i=v_i^{1/2}\psi_i^-=v_i^{-1/2}\psi_i^+.
\]
Our task is to express the one-point function $C(X)$ on the 
projective plane in terms of this basis:
\[
C(X)=c_i^{}\,\psi^{}_i=c_i^{}\,v_i^{1/2}\,\psi_i^- .
\]
In other words, we have to compute the constant of proportionality 
$c_iv_i^{1/2}$ between two states in $\HH(\hat X)$ given by ribbon 
graphs in 3-manifolds with the same boundary $S^2$.
This can be done using the functoriality axiom by attaching
a 3-ball with a ribbon graph consisting of a single ribbon connecting
the north pole with the south pole to $S^2$, and
comparing the corresponding (scalar) invariants of ribbon graphs
in closed 3-manifolds.

If we attach a 3-ball to a 3-ball we get a 3-sphere. If we choose
 the ribbon graph in the 3-ball properly, we get in
$S^3$ an unknotted circle with zero framing. Its invariant is
$\mathcal{D}^{-1}\dim(i)$, where $\mathcal{D}^{-1}$ is the invariant of $S^3$.
This proper choice of the ribbon may be described as follows. Suppose
for definiteness that the upper and lower sides
of the ribbon defining $\psi_i^-$ are centered at the poles $(0,0,\pm1)$ of
$S^2$ and lie in the $x_2$-$x_3$ plane. The ribbon
graph in the 3-ball we attach to $S^2$ should be chosen so that it
can be deformed to a ribbon lying on the surface $S^2$ whose midline follows a
``tennis ball pattern'' joining the north pole to the south pole:
this pattern may be parametrized by $\rho\in[0,\pi]$:
\[
x(\rho)=\left(\Frac12\,\sin\,2\rho,\Frac12\,(1-\cos\,2\rho),\cos\,\rho\right).
\]

If we attach a 3-ball to the boundary of $(S^2\Times [-1,1])/\Z_2$
we get the real projective space $\R P^3=S^3/\Z_2$. Indeed 
the map $S^2\Times [-1,1]\hookrightarrow S^3$:
\[(x,t)\mapsto (x\cos\,\pi t/4,\sin\,\pi t/4)\in S^3\subset\R^4,\] 
defines an embedding of the connecting manifold into $\R P^3$. Its image
is the complement of the ball in $\R P^3$ determined by the
equation $x_4^2> 1/2$.

\begin{figure}
\scalebox{0.5}{\includegraphics{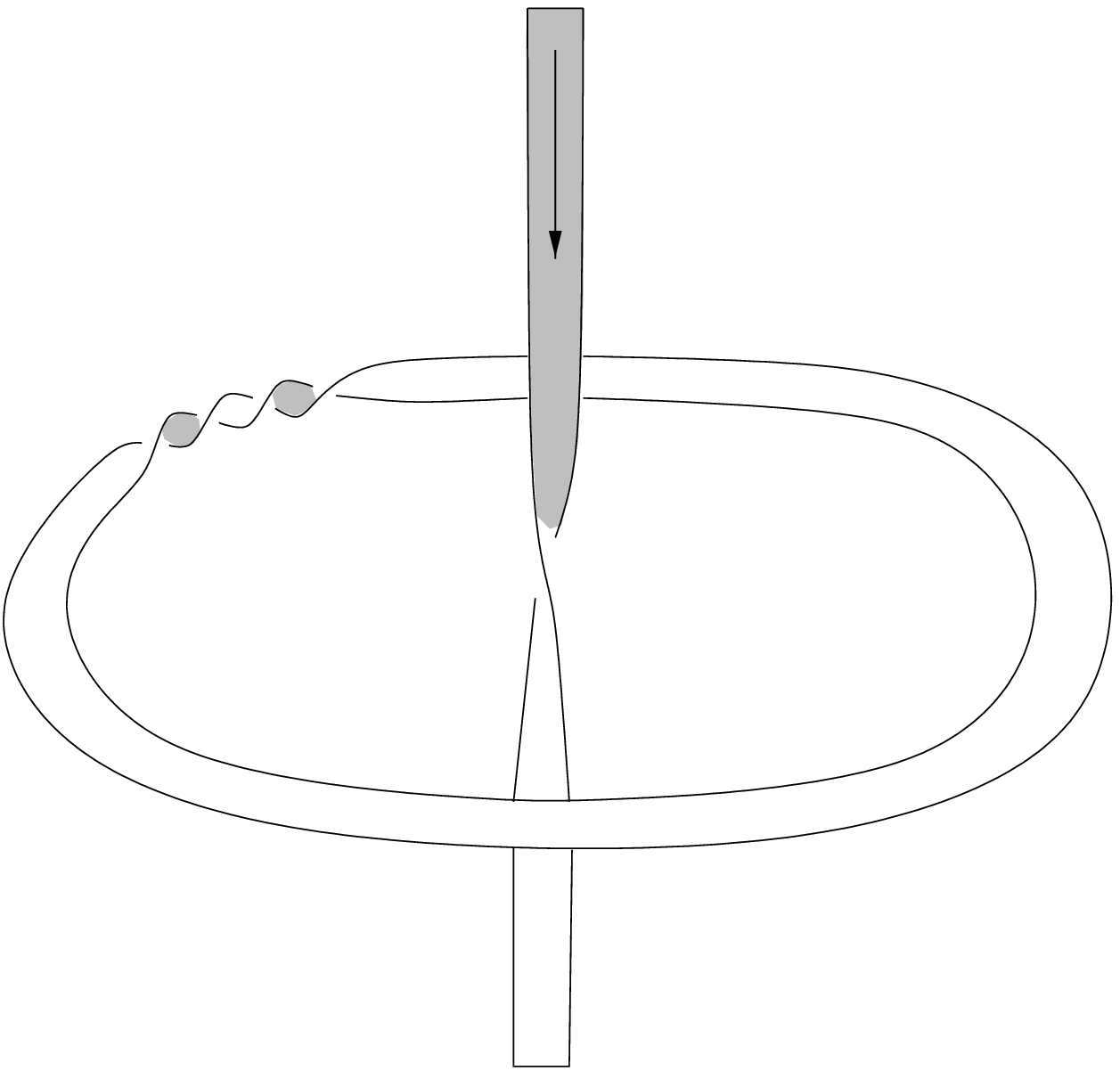}}
\caption{}\label{fig2}
\end{figure}

It is known that $\R P^3$ can be obtained from $S^3$ by surgery
on the unknot with framing $-2$, see Appendix \ref{a-surg}. If we follow how
the ribbon graph is mapped by the surgery and view $S^3$ as the
one point compactification of $\R^3$, we may describe the
situation as in Fig.\ \ref{fig2}. 
The region depicted is contained in a ball in $\R^3$ which after surgery 
on the annulus drawn as horizontal is mapped to the connecting
3-manifold $(S^2\Times [-1,1])/\Z_2$.
The vertical line is mapped to the ribbon graph in the  connecting
3-manifold. If we attach the 3-ball onto $S^2$ and compute with the formulae
of Appendix B the image of the tennis ball pattern, we see that the vertical 
line matches the ribbon graph in the 3-ball to give an unframed unknot
linked to the horizontal unknot. According to the 
Reshetikhin--Turaev surgery formula \Ref{e-surgeryformula}
the resulting invariant is 
\[
\Delta^{-1}
 \mathcal{D}^{-1}\sum_{j\in I}v_j^{-2}\, \mathrm{dim}(j)\,{\ST}_{i,j},
\]
with $\Delta=\sum_{i\in I}v_i^{-1}\dim(i)^2$.
By using $\kappa=\Delta\mathcal{D}^{-1}$ we then get
\[
c_i=v_i^{-1/2}\Delta^{-1} \sum_{j\in I}v_j^{-2}\,
\frac{\mathrm{dim}(j)}{\mathrm{dim}(i)}\, {\ST}_{i,j}
=v_i^{-1/2}\kappa^{-1} \sum_{j\in I}
\, {\SR}_{i,j}\,v_j^{-2}\,\frac{\SR_{j,\zero}}{\SR_{i,\zero}}.
\]
This result can be expressed in terms of the
matrix $P$ \cite{BS}, which is defined in terms
of the representation of $\mathrm{SL}(2,\Z)$:
\[
P=\kappa^{-1}\T^{1/2}\SR\T^2\SR\T^{1/2}=T^{1/2}\SR T^2\SR T^{1/2}.
\]
The square root is defined using the choice of square roots of the $v_i$.
The matrix $P$ is symmetric. Its square is
the conjugation matrix $C=(\delta_{i,j^*})$.

We then have $c_i=P_{i,\zero}/\SR_{i,\zero}$.

Summarizing, our result is 
\[
C(\R P^2,i)=\frac{P_{i,\zero}}{\SR_{i,\zero}}\,\psi_i .
\]
The corresponding normalized correlation function agrees with
the result 
\[
C_{\mathrm{norm}}(\R P^2,i)
=\frac{P_{i,\zero}}{\sqrt{\SR_{i,\zero}}}\,\psi_i 
\]
obtained in conformal field theory, see e.g.\ \cite{PSS1}.

\section{Annulus, Klein bottle, M\"obius strip}\label{s-5}
We consider here the three cases of surfaces whose double is a
torus: the annulus, the Klein bottle and the M\"obius strip. 
The correlation function with no marked points (partition function)
can then be expressed in terms of the basis $\chi_j^{}(S^1\times S^1)$
of invariants of the solid torus. It is then expected on physical
grounds that the coefficients of the partition functions obey certain
integrality conditions. We compute the partition function in these
three cases and show that these conditions are obeyed. Different ways
of computing correlation functions  implies
remarkable properties of the $SL(2,\Z)$ representations arising
from modular categories. The most well-known one is the Verlinde formula
\begin{equation}\label{e-Verlinde}
N_{j,k,l}=\sum_{r\in I}\frac{\SR_{r,j}\SR_{r,k}\SR_{r,l}}{\SR_{r,\zero}},
\end{equation}
which may be understood as the result of two different computation
of the annulus partition function.

\subsection{Annulus partition function}
Let $X=(A,a,b)$ 
be an annulus whose boundary has connected components labeled
by $a$ and $b$. The double $\hat X$ is then a torus and the connecting
3-manifold is a solid torus $D^2\Times S^1$ with two equatorial 
       graphs
without outgoing edges. To be concrete, let us think of $A$ as the
region between two circles centered at the origin of the
$x$-$y$ plane. The two circles forming the boundary are oriented 
counterclockwise. Then $\hat X$ may be thought of as the surface obtained
by revolution around the $z$-axis
of a circle $\mathcal{C}$ in  the $x$-$z$ plane with center on the $x$-axis. Then
the projection $p{:}\ \hat X\To X$ is the orthogonal projection onto the $x$-$y$ 
plane and the involution $\sigma$ is the reflection at the $x$-$y$ plane. 
The connecting 3-manifold $M$ is the solid obtained by revolution of the 
disk in the $x$-$z$ plane with boundary $\mathcal{C}$. It contains two annuli in 
the $x$-$y$ plane oriented counterclockwise and labeled by $a$, $b$.
A well-known calculation using \Ref{e-Bizet} and \Ref{e-Mozart}
shows that 
$Z(M,\emptyset,S^1\TImes S^1) =\sum_k N_{a,b}^k\,\chi_k^{}(S^1\TImes S^1)$. Thus
\[
C(A,a,b)=\sum_{k\in I}N_{a,b}^k\,\chi_k^{}(S^1\Times S^1).
\]
The alternative way of doing this calculation is to glue another solid torus
with an annulus graph to obtain the invariant of $S^3$ with three unknots. The
identity between the two results is the Verlinde formula, see \cite{W}.
Our result can therefore be understood as a three-dimensional version
    of the derivation of the Verlinde formula in \cite{C}, and shows that
    at this level the arguments of \cite{C} are completely equivalent
    to those given in \cite{W}.

A similar reasoning will be used below for the M\"obius strip.

\subsection{The partition function of the Klein bottle}
Let $\Z_2$ act on $S^1\Times S^1$ via the involution
$\sigma{:}\ (z,w)\mapsto (z^{-1},-w)$. The quotient
space is the Klein bottle $K\eq(S^1\Times S^1)/\Z_2$, with double
$\hat K\eq S^1\Times S^1$. We compute the correlation function 
$C(K)\In\HH(\hat K)$ of the Klein bottle with
no marked points. A basis of $\HH(\hat K)$ is given by
\[
\chi_j^{}(S^1\Times S^1)=Z((H,j),
\emptyset,S^1\Times S^1),\qquad j\In I,
\]
with $H\eq D^2\Times S^1$ containing an annulus labeled by $j$, as above. 

So 
\[
C(K)=\sum_{i\in I}c_i(K) \chi_i^{}(S^1\Times S^1),
\] 
for some complex coefficients $c_i(K)$.
The convention we have chosen here is  that the first factor
of $S^1$ in $\hat K\eq S^1\Times S^1$, which becomes a contractible cycle
in the solid torus $H$, generates the kernel of $p_*{:}\ H_1(\hat K,\R)
\To H_1(K,\R)$.

We compute $c_j(K)$ by composing
both sides of the above equation with the invariant
$Z((-H,j^*),S^1\Times S^1,\emptyset)$. If we
compose this invariant with $\chi_i^{}(S^1\Times S^1)$
we get $\delta_{i,j}$, the invariant of $S^2\Times S^1$
with two annuli $z_1\Times S^1$, $z_2\Times S^1$ labeled by $i,\,j^*$. 
Therefore the right-hand side becomes $c_j(K)$.

The left-hand side is then the invariant of
the 3-manifold $M'$ obtained by gluing $(-H,j^*)$ to the connecting 
manifold $M_K$ of the Klein bottle. We claim that this 3-manifold
is homeomorphic to $S^2\Times S^1$ with a certain annulus labeled by $j$.

To see this, let us identify $S^2$ with
$\C P^1=\C\cup\{\infty\}$. The connecting
manifold $M_K$ consists of classes $[(z,w,t)]$ of 
triples $(z,w,t)\in S^1\Times S^1\Times[-1,1]$ modulo
\[
(z,w,t)\sim (z^{-1},-w,-t).
\]
Then we have the embedding $M_K\hookrightarrow \C P^1\Times S^1$ given by
\[
\iota:\quad [(z,w,t)]\mapsto (2w\,\frac{e^tz-1}{e^tz+1},w^2).
\]
The complement of $\iota(K)$ in $S^2\times S^1$ is the interior of a 
solid torus $D^2\times S^1$, embedded
via $(z,w)\,{\mapsto}\,(2w(e^{-1}z{-}1)/(e^{-1}z{+}1),w^2)$.
The image of the ribbon graph $[-\epsilon,\epsilon]
\Times S^1\subset D^2\Times S^1$ in $M'=S^2\Times S^1$ is an annulus.
The intersection of this annulus with the fiber
over $u\in S^1$ consists of two segments centered
at $\pm 2\sqrt{u}$ and is contained in the straight
line connecting these two points. As $u$ runs over
the unit circle, the two points rotate around the
origin by $180$ degrees. By the trace formula
\Ref{e-traceformula}, the invariant of $M'$ is the trace 
\[
Z(M')=\mathrm{Tr}_{\HH(S^2;j,j)}(\Phi_j)
\]
over the space of states for the sphere with two marked arcs of the
morphism $\Phi_j$ represented by the graph
\newcommand{\fzuno} 
{\begin{picture}(3,3)(0,0) 
\scalebox{.3}{\includegraphics{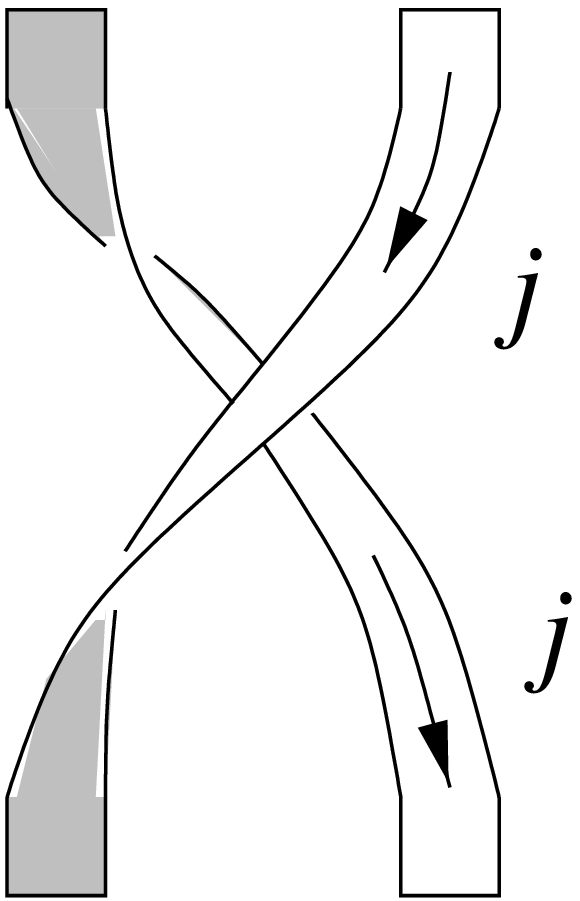}}
\end{picture}}
\newcommand{\fzdue} 
{\begin{picture}(3,4)(3,0) 
\scalebox{.3}{\includegraphics{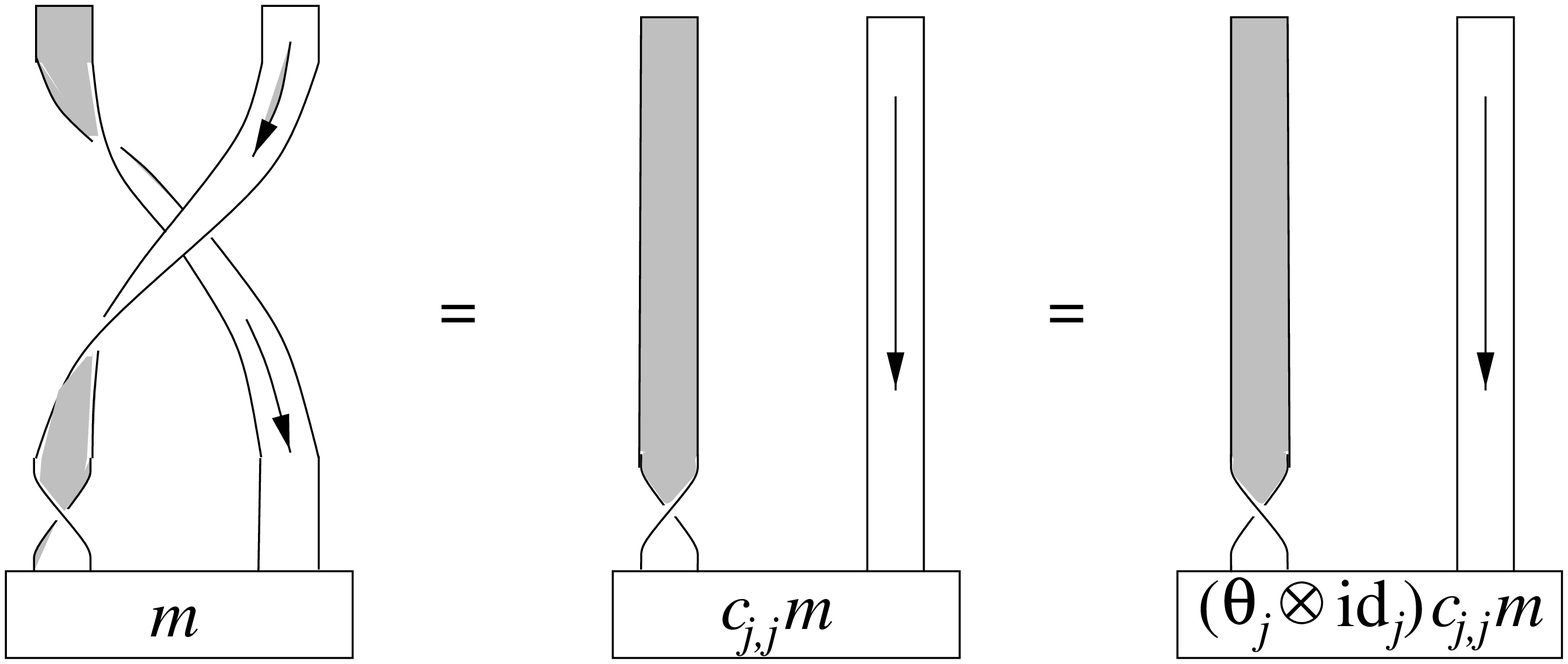}}
\end{picture}}
\[
\fzuno
\]
in $S^2\times[0,1]$, where we think of $S^2$ as the $x-y$ plane in $\R^3$
by stereographic projection. If $j\not\simeq j^*$, $\HH(S^2;j,j)=\{0\}$ and 
$Z(M')$ vanishes. If $j\simeq j^*$, $\HH(S^2;j,j)$ is
one-dimensional with a basis given by the
invariant of a ball with a ribbon graph consisting of ribbons connecting
the two arcs to a two-valent vertex labeled by any non-zero morphism
$m\in\Hom(\one,j\Otimes j)$. Any such morphism may be written as
$m=(\id_j\Otimes\phi)\circ b_j$ for any isomorphism $\phi{:}\ j^*\To j$.
Then we have $(\theta_j\Otimes\id_j)\circ c_{j,j}\circ m=\nu(j) m$,
see \ref{ss-fs}, where $\nu(j)=\pm1$ is the Frobenius--Schur indicator
of $j$. The morphism $\Phi_j$ is then the Frobenius--Schur indicator times the 
identity, as can be seen by acting on the basis element:
\[
\fzdue
\]
Therefore $Z(M')=\nu({}j)$. We conclude that
\[
C(K)=\sum_{j\simeq j^*}\nu({}j)\,\chi_j^{}(S^1\times S^1),
\]
in agreement with \cite{HSS}.
The summation is over all self-dual objects $j\In I$.

\subsection{The M\"obius strip}
\begin{figure}
\scalebox{0.4}{\includegraphics{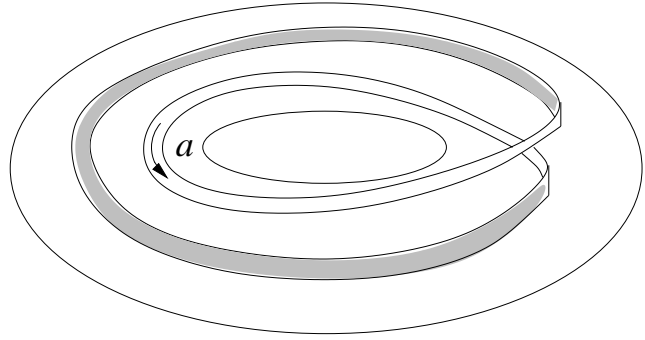}}
\caption{A solid torus with a ribbon graph, whose invariant is the correlation
functions of the M\"obius strip with boundary condition $a$}\label{fig3}
\end{figure}

\begin{figure}
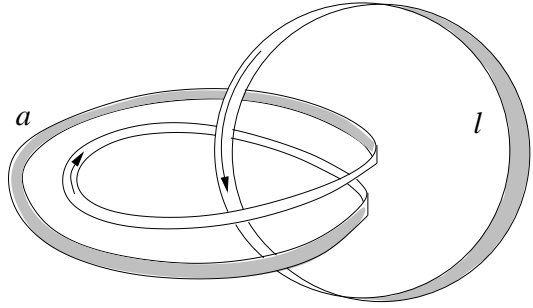

\fquattrocentouno
\caption{A link in $S^3$ used to compute the partition function of the
M\"obius strip}\label{fig4}
\end{figure}

We now consider the correlation function of the M\"obius strip with
no marked point and boundary condition $a\in I$. The double
of the M\"obius strip  is $S^1\Times S^1$ with involution
$\sigma{:}\ (z,w)\mapsto (w/z,w)$. Thus if {\it M\"{o}} is the M\"obius strip,
its connecting manifold is $M_{\mbox{\em\footnotesize M\"o}}
=(S^1\Times S^1\Times [-1,1])/\Z_2$ and is degree one homeomorphic via
\[
(z,w,t)\mapsto \left(\Frac{1+t}2\,z+\Frac{1-t}2\,wz^{-1},w\right),
\]
to $D^2\Times S^1$. The equatorial graph consists of an annulus 
lying in the zero section $t=0$ and running close to the boundary,
see Fig.~\ref{fig3}.
The correlation function has then the form
\[
C(\mbox{\it M\"o};a)=\sum_{j\in I}m_{a,j}\,\chi_j^{}(S^1\Times S^1).
\]
The coefficients $m_{a,j}$ may be computed by composing both 
sides of the equation with the invariant of a solid torus with
a ribbon graph consisting of an annulus labeled by $l\In I$,
in such a way that the manifold obtained by gluing is the
3-sphere. The right-hand side becomes $\sum_jm_{a,j}\mathcal{D}^{-1}{\ST}_{j,l}$,
and the left-hand side is $\mathcal{D}^{-1}$ times the invariant
of the link represented in Fig.~\ref{fig4}.
The invariant may be further simplified, by first flattening the
ribbon onto a plane:
\\[-1.3em]
\begin{eqnarray*}
\fquattrocentodue\qquad 
&\alza{1.1cm}
{=\ \ \sum_{k,\alpha}\mathrm{dim}(k)}&
\fquattrocentotre 
\\{}\\[-.2em]
&\alza{1.3cm}
{=\quad \sum_{k,\alpha}\mathrm{dim}(k)}&
\fquattrocentoquattro \quad \alza{1.3cm}{.}
\end{eqnarray*}
The expression in the sum is then $v_k^2v_l^{-2}\mathrm{dim}(k)\delta_{\alpha,
\alpha}$.
The sum over the $N_{a,l}^k$ possible values of $\alpha$ 
may  be performed, with the result
\[
\sum_j m_{a,j}{\ST}_{j,l}=\sum_{k}v_k^2v_l^{-2}\mathrm{dim}(k)N_{a,l}^k
\]
By using the relation $\sum_l {\ST}_{i,l}{\ST}_{l,j^*}\eq\mathcal{D}^2\delta_{i,j}$,
and expressing our result in terms of the matrix $S\eq\mathcal{D}^{-1}s$, we obtain:
\[
C(\mbox{\it M\"o};a)=\sum_{j\in I}m_{a,j}\,\chi_j^{}(S^1\Times S^1),
\]
with
\begin{equation}\label{e-Moe}
m_{a,j}=
\sum_{k,l\in I}v_k^2v_l^{-2}
{\SR}_{\zero,k}
N_{a,l}^k{\SR}_{l,j^*}.
\end{equation}
An alternative way to do the computation is to cut out a ball
in the solid torus of Fig.~\ref{fig3}
intersecting the ribbon graph in two segments, and use \Ref{e-Bizet}
with $i\eq j\eq a$. Then the ribbon graph may be replaced by a  ribbon
labeled by the summation index $k$ which starts and ends at a two-valent 
vertex after going  once around the solid torus. The two-valent vertex 
is labeled by dim$(k)$ times the morphism  represented by the graph
\[
\ftrentatre
\]
The sum over $\alpha$ is $\sum_{\alpha}e_\alpha[a^*ka^*]\,\Phi_{a,k}\, 
e_\alpha[ak^*a]$, with $e_\alpha[ak^*a]$, $\alpha=1\dots N_{a,k^*,a}$,
regarded as a basis of $\Hom(k,a\Otimes a)$ and $e_\alpha[a^*ka^*]$ as
a basis of $\Hom(a\Otimes a,k)$. $\Phi_{a,k}$ is the linear endomorphism of 
$\Hom(k,a\Otimes a)$ given by $\Phi_{a,k}(x)=(\theta_a\Otimes\id_a)\circ c_{a,a}\circ x$.
Since $e_\alpha[ak^*a]\circ e_\beta[a^*ka^*]=\dim(k)^{-1}\delta_{\alpha,\beta}$ 
(eq.~\ref{e-Mozart}), the quantum dimensions cancel, and we are left with
\[
m_{a,k}=\mathrm{Tr}_{\Hom(k,a\otimes a)}\,\Phi_{a,k}
\]
Moreover $\Phi_{a,k}^2=v_k\,\id$ as may be seen by deforming the
graph representing this morphism put on top of itself or
by using axioms (iv), (v) of Appendix \ref{a-mod}. It follows that
$\Phi_{a,k}$ is diagonalizable with eigenvalues $\pm\sqrt{v_k}$. Therefore
we have
\[
m_{a,k}=M_{a,k}v_k^{1/2},\qquad M_{a,k}\in\Z,\qquad
M_{a,k}\equiv N^k_{a,a}\mod2,\qquad |M_{a,k}|\leq N^k_{a,a}
\]
The two different ways of calculating the M\"obius strip partition function
implies the following result, essentially due to Bantay \cite{B},
on representations of SL$(2,\Z)$ arising from modular categories. 

\begin{thm} Let $\SR\eq(\mathcal{D}^{-1}\ST_{j,k})$, $\hat T\eq(v_j^{-1}\delta_{j,k})$
the matrices defining the projective representation of SL$(2,\Z)$ associated
to a modular category with rank $\mathcal{D}$, see \ref{ss-moca}, and let 
$N^i_{j,k}\eq N_{i^*,j,k} \In\Z_{\geq 0}$ be the corresponding Verlinde 
numbers \Ref{e-Verlinde}. Let $Q\eq S\hat T^2 S^{-1}$. Then the numbers
\[
M_{a,k}=v_k^{-1/2}\sum_{r\in I} (Q^{-1})_{\zero,r}\, 
\frac{{\SR}_{a,r}}{\SR_{\zero,r}}\,Q_{k,r},
\]
are integers and obey $M_{a,k}\equiv N^k_{a,a}\mod 2$, $|M_{a,k}|\leq N^k_{a,a}$.
\end{thm}

The above expression for $M_{a,k}$ was obtained from \Ref{e-Moe} by using
the Verlinde formula \Ref{e-Verlinde}.

\medskip
\noindent{\bf Remarks.}
\begin{enumerate}
\item 
The above formula amounts to not completely trivial identities even for the simple
$\Z_{2N}$ example of \ref{ss-21}. There we have 
\[
\frac{{\SR}_{a,r}}{\SR_{\zero,r}}={\rm e}^{-\pi\I ar/N},\qquad
Q_{k,r}=\left\{\begin{array}{rl}
\frac1{\sqrt N}\,{\rm e}^{\frac{\I\pi}4-\frac{\I\pi}{4N}(k-r)^2},
&\mbox{if $k{-}r{+}N$ is even,}\\
0, &\mbox{otherwise,}
\end{array}\right.
\]
and, with $v_k^{-1/2}=\exp(\pi\I  k^2/4N)$,
\[
M_{a,k}=\left\{\begin{array}{rl}
1,&\mbox{if $k$ is even and $a\equiv k/2\mod 2N$,}\\
0,&\mbox{otherwise.}
\end{array}\right.
\]
Note that being even is well defined in $\Z_{2N}$, and that
if $k$ is even the above choice of square root of $v_k$ is unambiguous.

\item
More generally one expects \cite{PSS1}  the numbers
\[
Y^k_{a,j}=
v_k^{-1/2}v_j^{1/2}\sum_r (Q^{-1})_{j,r}\,
\frac{{\SR}_{a,r}}{\SR_{\zero,r}}\,Q_{k,r}
\]
to be integer for any $j$. This has been recently shown to be
true under some additional assumptions, see \cite{Gan}.
\end{enumerate}

\bigskip

\appendix
\section{Modular categories}\label{a-mod}

We give here the precise definition of modular categories, following \cite{T}.
A monoidal (=\,tensor) category with product $\otimes$ and
     unit $\one$ for the product 
is called {\em strict} if for any objects 
$U,V,W$, we have $(U\Otimes V)\Otimes W=U\Otimes(V\Otimes W)$,
and $V\Otimes \one=\one\Otimes V=V$. A monoidal Ab-category is a monoidal
category such that morphisms between any two objects $U,V$ form  an additive 
abelian group $\Hom(U,V)$, and such that compositions and tensor products
of morphisms are bilinear. In particular the endomorphisms of the unit $\one$ of
the tensor product form a ring with unit.  This ring is called the ground ring. 
The groups of morphisms are naturally modules over the ground ring. An object 
$V$ of a monoidal Ab-category is called simple if
$\mathrm{Hom}(V,V)\simeq K$ as a $K$-module.

A {\em ribbon category} is a strict monoidal
category with additional data: a braiding, a twist and  a duality.

A {\em braiding} associates to any pair of objects
$V$, $W$ an isomorphism $c_{V,W}\In\Hom(V\Otimes W,\linebreak[0]
W\Otimes V)$. A {\em twist} associates to any object $V$ an isomorphism 
$\theta_V\in\Hom(V,V)$. A {\em duality} associates to any object $V$ a
dual object $V^*$ and morphisms $b_V\In\Hom(\one,V\Otimes V^*)$,
$d_V\In\Hom(V^*\Otimes V,\one)$.

These data obey the following set of axioms:
\begin{enumerate}
\item [{(i)}]
$c_{U,V\otimes W}= (\id_V\Otimes c_{U,W}) (c_{U,V}\Otimes \id_W)$.
\item [(ii)]
$c_{U\otimes V,W}= (c_{U,W}\Otimes \id_V) (\id_U\Otimes c_{V,W})$.
\item [(iii)]
$g\otimes f c_{V,W} =c_{V,W}\, f\Otimes g$.
\item [(iv)]
$\theta_{V\otimes W}= c_{W,V}c_{V,W}(\theta_V\Otimes \theta_W)$.
\item [(v)]
$\theta_{V'}f=f\theta_{V}$.
\item [(vi)]
$(\id_V\Otimes d_V) (b_V\Otimes \id_V)=\id_V$.
\item [(vii)]
$(d_V\Otimes \id_{V^*}) (\id_{V^*}\Otimes b_V)=\id_{V^*}$.
\item [(viii)]
$(\theta_V\Otimes \id_{V^*}) b_V= (\id_V\Otimes \theta_{V^*})b_V$.
\end{enumerate}
Here $U,V,\dots$ are arbitrary objects and
$f\in\Hom(V,V')$, $g\in\Hom(W,W')$ are arbitrary morphisms.

Finally a {\em modular category} is
a ribbon Ab-category with a finite family $I$ of
simple objects such that:
\begin{enumerate}
\item[(ix)] $\one\In I$.
\item
[(x)] For every ${}i\In I$,
 ${}i^*$ is isomorphic to an object in $I$.
\item
[(xi)] Every morphism $f{:}\ V\to V'$ may be
decomposed into a finite sum 
$\sum_rg_rh_r$, where $h_r\in \Hom(V,{}i)$ and
$g_r\in \Hom({}i,V')$ for some $i=i(r)$.
\item
[(xii)] The matrix $({\ST}_{i,j})=(\mathrm{tr}(c_{{}j,{}i}
c_{{}i,{}j}))$ indexed by $i,j\in I$ is invertible.
\end{enumerate}

\section{Surgery on the unknot}\label{a-surg}
Here we describe explicitly the homeomorphism between 
$\R P^3$ and the 3-manifold obtained by surgery on
the unknot in $S^3$ with framing $\pm2$.

Let us first fix some conventions about orientations. 
We give an orientation
to the boundary of an oriented manifold $M$ by the ``outward
normal first'' rule. This means that a basis $(b_1,\dots,b_n)$ of
the tangent space $T_x\partial M$ at a boundary point $x$ is
positively oriented if and only if $(b_0,b_1,\dots,b_n)$ is
a positively oriented basis of $T_xM$ for any $b_0$ pointing
outwards. The orientations of the disks $D^n=\{x\In\R^n\,|\,|x|\leq1\}$ 
are inherited from $\R^n$. They define orientations of the
spheres $S^{n-1}=\partial D^n$. We orient $\C^n$ via the isomorphism
$(x_1+\I y_1,\dots,x_n+\I y_n)\mapsto (x_1,y_1,\dots,x_n,y_n)$ with
$\R^{2n}$, and view odd dimensional spheres $S^{2n-1}$ as subsets of $\C^n$.

Recall that if $L$ is the image of a
smooth embedding of $S^1$ in $S^3$, and $n$ is an integer, then
the surgery on $L$ with framing $n$ is the following construction.
Let $U$ be a closed tubular neighborhood of $L$. Fix an embedding 
$j{:}\ S^1\Times D^2\hookrightarrow S^3$  with image $U$ so that
$S^1\Times \{0\}$ is sent to $L$, and $S^1\Times\{1\}$ is sent
to a knot $L'$ whose linking number
with $L$ is $n$. The linking number is calculated using the orientations
of $L$, $L'$ coming from the orientation of $S^1$ via $j$.
We may think that $L$ with framing $n$ 
as an embedded annulus with boundary $L\cup L'$. Let $S^3_{(L,n)}$ 
be the manifold obtained by gluing the complement $S^3-\mathrm{int}\,U$ of
the interior of $U$ to $D^2\Times S^1$ via the restriction to $S^1\Times S^1$
of the map $j$: $S^3_{(L,n)}= (S^3-\mathrm{int}\,U)\sqcup (D^2\Times S^1)
/(x\sim j(x),\, x\In S^1\Times S^1)$. The orientation of $S^3_{(L,n)}$
is defined to be the orientation that extends
the standard orientation of $S^3-U\subset S^3$.

If $L$ is the unknot defined, say, by the embedding $z\mapsto (z,0)$
of $S^1\subset\C$ into $S^3\subset
\C^2$, and the framing is $n$, then we may take, for small $\epsilon>0$,
\[
j(z,w)=\frac1{\sqrt{1+|\epsilon w|^2}}\,
(z,\epsilon wz^n),\qquad (z,w)\in S^1\Times D^2.
\]
The image is $U\eq\{(u,v)\In S^3\,{\subset}\,\C^2\,|\,\epsilon|u|\,{\ge}\,|v|\}$.
The 3-manifold $S^3_{(L,n)}$ obtained by surgery on this link is 
(for $n\,{\ne}\,0$) homeomorphic to the {\em lens space} $S^3/\Z_{|n|}$,
with $\Z_{|n|}$ action generated by $(u,v)\mapsto
(\zeta u,\zeta v)$, $\zeta=\exp(2\pi\I /|n|)$.

Indeed the map
\[
i_n:\quad (u,v)\mapsto \left(\frac v{|v|}\right)^{-1/n} (u,|v|),
\]
is a homeomorphism $S^3-L\to (S^3-L)/\Z_{|n|}$ ($L$ is invariant
under the $\Z_{|n|}$ action), with inverse map
$(u,v)\mapsto (uv^{-1}|v|,v^{-n}|v|^{n+1})$.
This homeomorphism extends to a homeomorphism
from $S^3_{(L,n)}$ onto $S^3/\Z_{|n|}$. This follows from
the fact that $i_n\circ j|_{S^1\Times S^1}{:}\
(z,w)\mapsto \frac{w^{-1/n}}{\sqrt{1+\epsilon^2}}
(1,\epsilon z^{-1})$ extends to the homeomorphism
\[
(z,w)\mapsto \frac{w^{-1/n}}{\sqrt{1+\epsilon^2|z|^2}}\, (1,\epsilon \bar z)
\]
from $D^2\Times S^1$ onto a tubular neighborhood
of $L/\Z_{|n|}$ in $S^3/\Z_{|n|}$.

The sign of $n$ is connected with the orientation.
Since the $\Z_{|n|}$ action on $S^3$ preserves
the orientation, the lens space $S^3/\Z_{|n|}$
inherits an orientation from $S^3$. The map
$i_n$ is then orientation preserving if
$n\,{<}\,0$ and orientation reversing if $n\,{>}\,0$,
as can be seen by linearizing $i_n$ at $(0,1)$.

Let us summarize the results.

\begin{proposition}
Let $L$ be the image of the embedding $z\mapsto (z,0)$ of $S^1\To S^3$. 
Then, for each $n\In\Z-\{0\}$, the map $i_n{:}\ S^3-L\To(S^3-L)/\Z_{|n|}$ 
\[
(u,v)\mapsto \left(\frac{v}{|v|}\right)^{-1/n} (u,|v|),
\]
extends uniquely to 
a homeomorphism from the manifold
$S^3_{(L,n)}$ obtained by surgery on $L$ with
framing $n$ onto the lens space $S^3/\Z_{|n|}$.
The degree of this homeomorphism is $-
\mathrm{sign}(n)$. In particular,
$i_{-2}$ is a degree one homeomorphism from
$S^3_{(L,-2)}$ onto $\R P^3$.
\end{proposition}
\medskip

Moreover we have the Reshetikhin--Turaev formula for the invariant of
a ribbon graph in a manifold obtained from $S^3$ by surgery, see
\cite{T}. It can be deduced from the functoriality axiom applied
to a solid torus. In the case of an unknot with framing $n\,{<}\,0$,
assuming that the ribbon graph does not intersect the solid
torus at which we perform the surgery, it reads
\begin{equation}\label{e-surgeryformula}
Z(S^3_{(L,n)};\Gamma)=\Delta\sum_{j\in I}\mathrm{dim}(j)\,Z(S^3,\Gamma_j),
\end{equation}
with $\Delta=\sum_{i\in I}v_i^{-1}\dim(i)^2$.
The ribbon graph $\Gamma_j$ is obtained from  $\Gamma$
by adding $L$, viewed as an embedded annulus in $S^3$, with label $j$.
This involves a choice of orientation of $L$, but the result does
not depend on this choice.
 
To visualize the results it is useful to stereographically project
$S^3-\{\mathrm{pt}\}$ to $\R^3$. For our application, a useful
projection is 
\[
(x_1{+}\I x_2,x_3{+}\I x_4)
\mapsto \Frac1{1+x_3}\,(x_1,x_2,x_4).
\]
It preserves the orientation.
Then the link $L$ at which we do surgery is mapped to the unit circle
in the $x_1$-$x_2$ plane.

\bigskip

\newcommand\J[5]{{\it #5}, {#1} {#2} ({#3}) {#4}}
\newcommand\Prep[2]{{\it #2}, preprint {#1}}
\def\comp{Com\-mun.\,Math.\,Phys.}
\def\jgap{J.\,Geom.\,and\,Phys.}
\def\invm{Inv.\,Math.}
\def\ijmp{Int.\,J.\,Mod.\,Phys.}
\def\nupb{Nucl.\,Phys.\ B}
\def\phlb{Phys.\,Lett.\ B}

\end{document}